\documentclass[letterpaper]{JHEP3}
\pdfoutput=1

\usepackage{array}

\usepackage{nicefrac}
\usepackage{amsmath}

\usepackage{amssymb,amsfonts,mathrsfs}
\usepackage{graphicx}
\usepackage{epsfig}

\usepackage{amsfonts}

\newcommand{\IZ}{\mathbb{Z}}

\newcommand{\Tr}{\mbox{Tr}}

\def\g{\gamma}

\def\a{\alpha}
\def\b{{\beta}}
\def\e{\epsilon}

\def\h{\eta}

\def\CS{{\cal S}}

\def\CI{{\cal I}}
\def\CD{{\cal D}}

\def\CN{{\cal N}}

\def\IC{\relax\hbox{$\inbar\kern-.3em{\rm C}$}}

\def\IC{{\bf C}}

\def\CN{{\cal N}}

\def\CT{{\cal T}}

\def\bea{\begin{eqnarray}}
\def\eea{\end{eqnarray}}
\def\be{\begin{equation}}
\def\ee{\end{equation}}
\def\ba{\begin{align}}
\def\ea{\end{align}}
\def\bse{\begin{subequations}}
\def\ese{\end{subequations}}
\def\1F1{{}_1\!F_1}
\def\2F0{{}_2\!F_0}

\def\a{\alpha}

\def\h3{$\textrm{H}_3^+$}

\def\IC{{\mathbb C}}

\def\IZ{{\mathbb Z}}

\def\Tr{{\rm Tr}}

\def\ge{{\Gamma_e}}

\def\lbldef#1#2{\expandafter\gdef\csname #1\endcsname {#2}}

\def\href#1#2{#2}

\newcommand{\beq}{\begin{equation}}
\newcommand{\eeq}{\end{equation}}
\newcommand{\ber}{\begin{eqnarray}}
\newcommand{\eer}{\end{eqnarray}}

\def\be{\begin{eqnarray}}
\def\ee{\end{eqnarray}}

\def\({\left(}
\def\){\right)}
\def\[{\left[}
\def\]{\right]}
\def\<{\langle}
\def\>{\rangle}

\def\Z{\mathbb Z}
\def\S{\mathbb S}

\title{${\cal N}=1$ theories of class ${\cal S}_k$}
\author{Davide Gaiotto$^{\alpha}$
 and Shlomo S. Razamat$^{\beta,\gamma}$ 
\\
\\
$^\alpha$ Perimeter Institute for Theoretical Physics,
 Waterloo, Ontario N2L 2Y5, Canada\\
$^\beta$\it NHETC, Rutgers, Piscataway, NJ 08854, USA\\
$^\gamma$\it Department of Physics, Technion, Haifa 32000, Israel\\
}

\bigskip

\abstract{

\

We construct classes of ${\cal N}=1$ superconformal theories elements of which are labeled by punctured Riemann surfaces. Degenerations of the surfaces correspond, in some cases, to weak coupling limits.  Different classes are labeled by two integers $(N,k)$. The $k=1$ case coincides with  $A_{N-1}$ ${\cal N}=2$ theories of class ${\cal S}$ and simple examples of  theories with $k>1$ are $\Z_k$ orbifolds of some of the $A_{N-1}$ class ${\cal S}$ theories. For the space of ${\cal N}=1$ theories to be complete 
in an appropriate sense we find it necessary to conjecture existence of new ${\cal N}=1$ strongly coupled SCFTs. These SCFTs when coupled to additional matter can be related by dualities to gauge theories. We discuss in  detail the $A_1$ case with $k=2$ using the  supersymmetric index as our analysis tool. The index of theories in classes with $k>1$ can be constructed using eigenfunctions of elliptic quantum mechanical models generalizing the Ruijsenaars-Schneider integrable model. When the  elliptic curve of the  model degenerates these eigenfunctions become polynomials with coefficients being algebraic expressions in fugacities, generalizing the Macdonald polynomials with rational coefficients appearing when $k=1$.
}

\begin{document}

\section{Introduction}\label{sect:introd}
There are several known large classes of $\CN=1$ superconformal field theories, which often have some kind of geometric 
interpretation. A classical example is the class of theories associated to dimer models \cite{Hanany:2005ve,Franco:2005rj,Franco:2005sm,Feng:2005gw}, 
labelled by bi-partite graphs drawn on a torus and a choice of rank $N$. These are quiver gauge theories with $SU(N)$ gauge groups and bi-fundamental matter, 
with a topology and choice of superpotential which is determined by the bipartite graph. They are associated to a string theory setup involving 
$N$ D3-branes placed at the tip of a singular toric Calabi-Yau cone. The combinatorial rules for associating quivers to bi-partite graphs can be extended to 
more general two-dimensional geometries \cite{Franco:2012mm,Xie:2012mr,Heckman:2012jh}, possibly including boundaries, such as disks, or higher genus Riemann surfaces. 
These constructions produce quiver gauge theories including both $SU(N)$ gauge groups and $SU(N)$ flavor groups, again with bi-fundamental matter.

There is another natural way to produce large classes of $\CN=1$ superconformal field theories with a geometric origin, by a twisted compactification 
on a Riemann surface of six-dimensional SCFTs. Such constructions generalize the derivation of class $\CS$ of $\CN=2$ 4d SCFTs from the twisted compactifications  of
the six-dimensional $(2,0)$ SCFTs on a Riemann surface decorated with punctures \cite{Gaiotto:2009we,Gaiotto:2009hg}. The class $\CS$ construction may produce both standard gauge theories and 
strongly-interacting SCFTs which lack a known Lagrangian description. Geometric manipulations of the Riemann surface lead to specific manipulations of the 
associated SCFTs, which allow one to derive S-dualities  relating in various ways the standard gauge theories and the strongly interacting SCFTs. 
The class $\CS$ construction has an unexpected computational power, allowing for example to compute the superconformal index of all the four-dimensional class $\CS$ theories,
even though they might lack a Lagrangian description \cite{Gadde:2009kb,Gadde:2010te,Gaiotto:2012xa,Rastelli:2014jja}, and relate $\S^4$ partition functions to $2d$ CFTs \cite{Alday:2009aq,Wyllard:2009hg}.

A straightforward extension of the class $\CS$ construction to $\CN=1$ gauge theories involves alternative twisted compactifications of 
the $(2,0)$ SCFTs. The basic strongly-interacting building blocks remain the same as for the $\CN=2$ class $\CS$, but they are 
glued together using $\CN=1$ gauge multiplets rather than $\CN =2$ ones \cite{Benini:2009mz,Bah:2012dg,Beem:2012yn,Xie:2013gma}.

In this paper we are interested in a more general extension, which involves the twisted compactification of $(1,0)$ SCFTs. 
There is a rather large number of known $(1,0)$ SCFTs, many of which can be built through F-theory constructions \cite{Heckman:2013pva,Heckman:2015bfa}.
In general, they have tensor branches of vacua where they take the appearance of six-dimensional gauge theories, coupled to 
matter fields which may themselves be irreducible SCFTs. A subset of the $(1,0)$ SCFTs become standard gauge theories on their tensor branch
and often admits a D-brane engineering construction \cite{Hanany:1997gh,Brunner:1997gf,Bhardwaj:2015xxa}. 

We focus here on the $(1,0)$ SCFTs $\CT^N_k$ associated to $N$ M5 branes sitting at the tip of an $A_{k-1}$ singularity of M-theory. These theories are somewhat well understood 
and have several features in common with the $(2,0)$ theories. Furthermore, RG flows induced by vevs on the Higgs branch of these theories 
allow one to reach many more $(1,0)$ SCFTs \cite{Gaiotto:2014lca}. It should be possible to extend our analysis to D- and E-type singularities, but we
will not do so here. 

Our main result is the conjectural definition of a new large class $\CS_k$ of $\CN=1$ SCFTs associated to the compactification of the
6d SCFTs $\CT^N_k$ on a Riemann surface with punctures. It is important to observe that the reduced amount of supersymmetry should make 
one very cautious in extending to $\CN=1$ theories the intuition developed with $\CN=2$ class ${\cal S}$ theories. With that concern in mind, 
in this paper we will follow a ``bottom up'' approach: we focus at first on a set of conventional four-dimensional $\CN=1$ theories which 
play a ``data collection'' role analogous to the role of linear quiver gauge theories in class $\CS$ theories. By looking at the properties and S-dualities of these theories 
we find several pieces of evidence connecting them to their six-dimensional avatars and to a larger conjectural set of $\CN=1$ SCFTs labelled by punctured Riemann surfaces. 
In the process, we learn new information about the six-dimensional SCFTs and their compactifications. 

The basic set of ``core'' $\CN=1$ gauge theories which are central to our analysis belongs squarely to the family of bi-partite quivers: they correspond to bi-partite hexagonal graphs 
drawn onto a cylinder. From that perspective, our work selects a subset of bi-partite theories which enjoy a larger than usual set of S-dualities 
and subjects them to a variety of manipulations to embed them into a larger family of non-bipartite $\CN=1$ SCFTs.

The supersymmetric index plays an important role in our analysis. In particular, we are able to recast the index of class $\CS_k$ theories as a 2d TFT, 
built from wave-functions which are eigenfunctions of novel difference operators, which generalize the elliptic RS difference operators 
used in bootstrapping the index of class $\CS$ theories. In principle, that opens the possibility to compute the index of 
any $\CN=1$ SCFT in the class $\CS_k$, even in the absence of a Lagrangian description. 

This paper is organized as follows. In section \ref{sect:brane} we motivate our choice of ``core'' gauge theories by some general considerations 
on brane constructions. In section \ref{sect:basic} we discuss our basic theories and dualities and set the stage for building the class ${\cal S}_k$ models.  
In particular we introduce the the notions of theories corresponding to punctured Riemann surfaces and the basic punctures, maximal and minimal ones. 
This and following sections will be divided into two parts with the first part detailing general physical arguments and the second part, {\it the index avatar}, giving quantitative evidence and details using the supersymmetric index for the particular example of class ${\cal S}_2$ $A_1$ theories.
Next, in section \ref{sect:closemin} we analyze RG flows starting from the basic models triggered by vacuum expectation values for baryonic operators. 
Such RG flows lead to theories in the IR which naturally correspond to surfaces with one of the minimal punctures removed. 
However, unlike the class ${\cal S}$ case here the removal of a minimal puncture leads to new theory and to the notion of 
discrete charge labels attached to the Riemann surface. 
We will also start  encountering new  strongly coupled theories belonging to our putative classes of models.
In section \ref{sect:closemaxi} we discuss closing maximal punctures by giving vacuum expectation values to certain mesonic operators. 
Such vevs lead to more general punctures and we in particular will be interested in closing maximal punctures down to minimal ones.  
One of the outcomes of this analysis will be a derivation of Argyres-Seiberg like duality frames for our basic duality and yet more irreducible strongly coupled SCFTs. In section \ref{sect:surface} we will discuss how to introduce surface defects into our theories by triggering flows with space-time dependent vevs. 
This construction will give us certain difference operators which we expect to characterize completely the wave-functions and the 2d TFT structure of the index. 
We will explicitly derive the difference operators for $A_1$ theories of class ${\cal S}_2$. 
In section \ref{sect:fivedimi} we go back to the higher-dimensional perspective: we study the interplay between class $\CS_k$ theories and the five-dimensional 
gauge theories which arise from circle compactifications of the $\CT^N_k$ theories. We will finish in section \ref{sect:summa} with an outlook of possible further research directions. Several appendices complement the text with further technical details.

\

\section{The $\IZ_k$ orbifold of $\CN=2$ linear quivers}\label{sect:brane}
The analysis of class $\CS$ theories was greatly facilitated by the existence of a core set of class $\CS$ $\CN=2$ SCFTs which 
admit both a Lagrangian description and a six-dimensional engineering construction: linear quiver gauge theories of unitary groups. 
These are theories which can be engineered by configurations of D-branes in IIA string theory, which are then lifted to M-theory 
to make contact with compactifications of the $A_{N-1}$ $(2,0)$ SCFT onto a cylinder geometry.

\begin{figure}
\begin{center}
\begin{tabular}{c}
\includegraphics[scale=0.5]{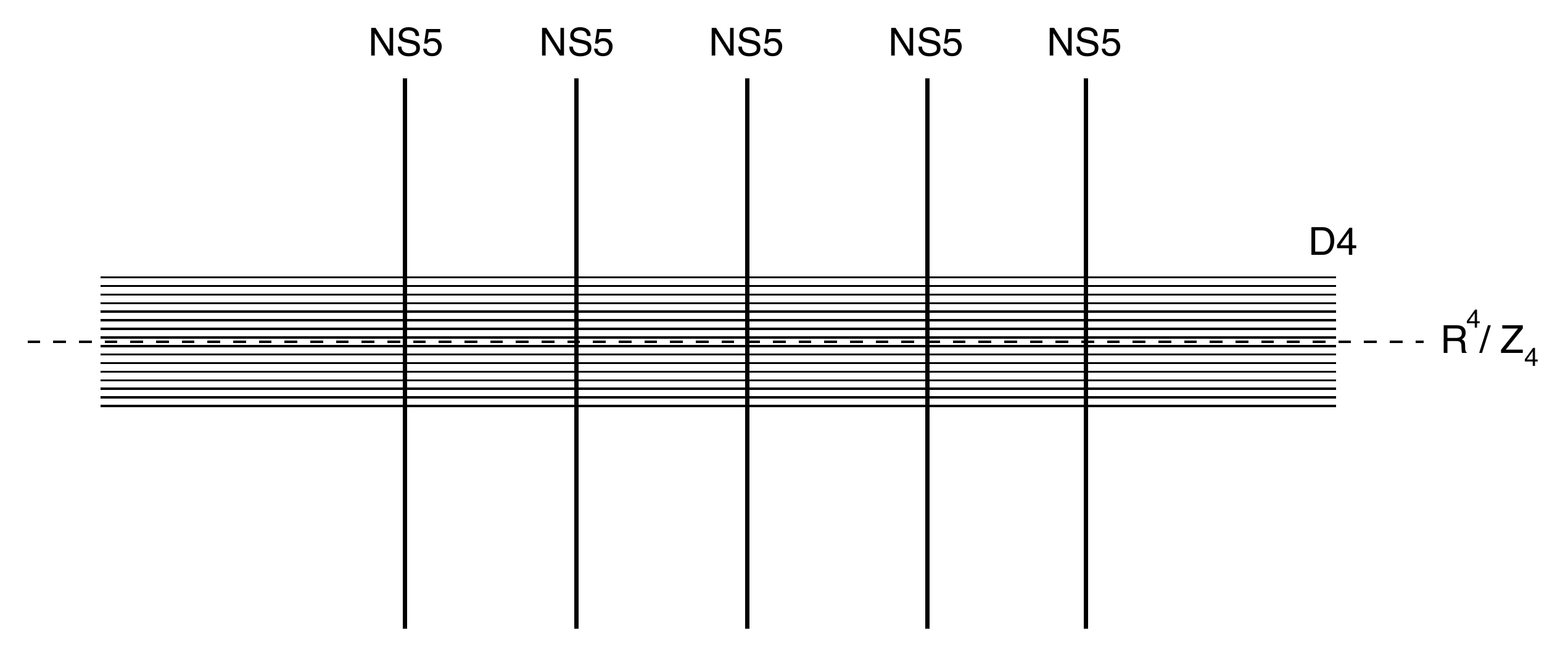} 
\end{tabular}
\end{center}
\caption{ The brane configuration which engineers the core $\CN=1$ gauge theories. We draw a setup with $N=4$ D4 branes (or $kN=16$ fractional D4 branes)
sitting at the locus of a $\Z_4$ singularity, intersected by 5 NS5 branes. The D4 branes in each segment between consecutive NS5 branes engineer the four necklaces of $SU(N)$ gauge groups. 
The 4-4 strings across the NS5 branes engineer the five sets of zig-zag chiral multiplets between the necklaces. The semi-infinite D4 branes can be either associated to 
the two necklaces of flavor groups or to five-dimensional necklace gauge theories coupled to the four-dimensional quiver theory.}
\label{honeyb}
\end{figure}

It is thus natural to seek for some class of $\CN=1$ SCFTs which admit a similar brane engineering construction in IIA and may be lifted to M-theory 
to a cylinder compactification of the $\CT^N_k$ $(1,0)$ SCFTs. The latter arises on the world-volume of $N$ M5 branes in the presence of an 
$A_{k-1}$ singularity. We will thus take the standard NS5-D4 brane system used to engineer $\CN=2$ linear quivers in \cite{Witten:1997sc} and 
super-impose it to an $A_{k-1}$  singularity in IIA string theory. More precisely, the $D4$ branes extend along directions 
$01234$, the NS5 branes along directions $012356$ and the $\IZ_k$ orbifold action rotates in opposite directions the 
$56$ and $78$ planes. See Figure \ref{honeyb}.

As the degrees of freedom of the original $\CN=2$ gauge theory entirely arise from open strings, we can 
apply a standard orbifold construction to arrive at our candidate $\CN=1$ theories \cite{Douglas:1996sw}. 
The analysis implicitly assumes that appropriate B-fields have been turned on so that the orbifold singularity admits a perturbative description. 
We will come back to this point momentarily. 

The orbifold procedure can be implemented in a straightforward way at the level of the gauge theory. Schematically, the orbifold group is embedded both into the 
global symmetries which correspond to the rotations of the internal space-time directions and into the gauge group, and all fields charged non-trivially under the 
orbifold action are thrown away. 

\begin{figure}[htbp]
\begin{center}
\begin{tabular}{c}
\includegraphics[scale=0.45]{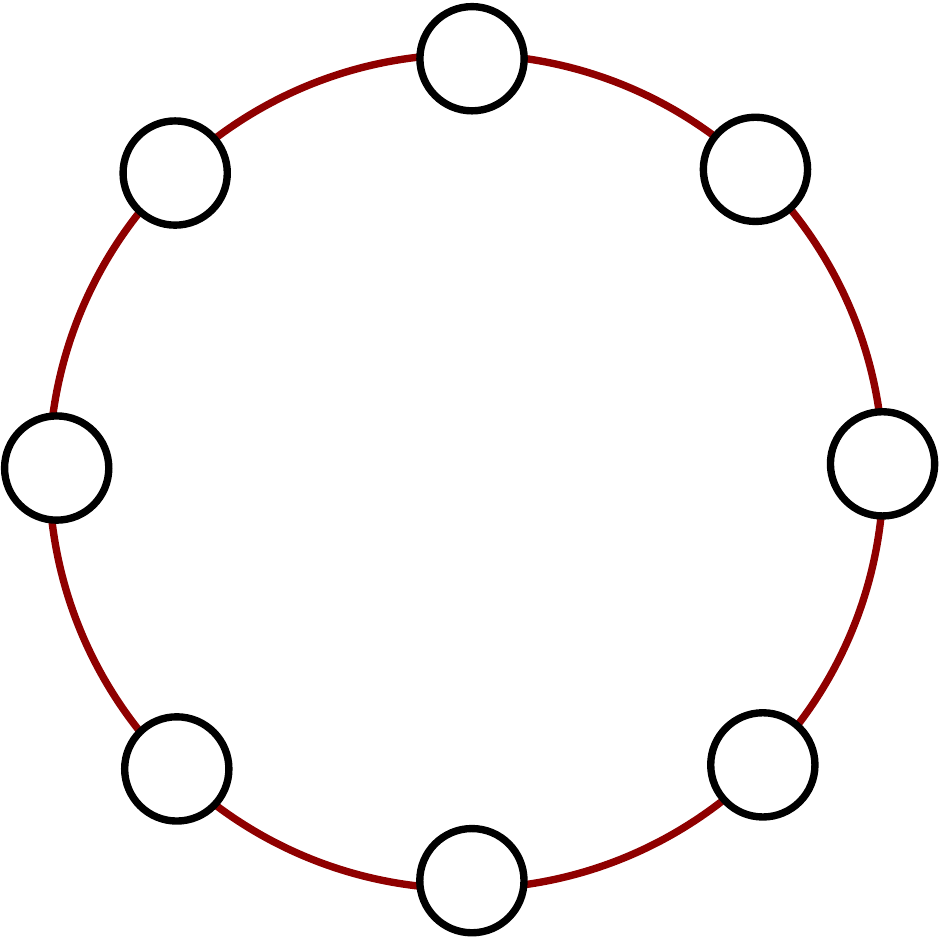}
\end{tabular}
\end{center}
\caption{The necklace quiver ${\cal N}_{N,k}$ in five dimensions. It is a $\Z_k$ orbifold of the maximally supersymmetric YM, with $k$ $SU(N)$ nodes. Notice that each link here represents a full bi-fundamental hypermultiplet and each node a full $SU(N)$ vector multiplet. There is  $U(1)$ global symmetry associated with each link, rotating the bifundamental hypermultiplets. There is also a $U(1)$ symmetry, the instanton symmetry, associated with each gauge group node.
\label{neck}}
\end{figure}

We can focus at first on the D4 branes. In the absence of NS5 branes, they will support a five-dimensional $\CN=1$ gauge theory described by a necklace quiver,
the result of orbifolding $\CN=2$ 5d SYM. The $\IZ_k$ group acts with charges $0,1,-1$ respectively on the real scalar associated to the $9$ direction 
and on the complex scalars associated to the $56$ and $78$ directions.  The embedding of $\IZ_k$ in the gauge group splits the gauge fields and the neutral real scalar into $k$ separate blocks of vector multiplets, while the complex scalar fields with charge $\pm 1$ under the embedding of $\IZ_k$ in the global symmetry will give bi-fundamental hypermultiplets between consecutive nodes of the 5d quiver. See Figure \ref{neck}.

The six-dimensional $(1,0)$ SCFT $\CT^N_k$, compactified on a circle, is expected to give a UV completion of precisely such necklace quiver theory with gauge group $SU(N)$ at each node. 
We will review the properties of such five-dimensional theory in a later section. For now we only need to observe that it has a $U(1)^{2k}$ global symmetry,
which is an Abelian remnant of the $SU(k)_\beta \times SU(k)_\gamma \times U(1)_t$ global symmetry of the 6d SCFT, together with the rotation symmetry of
the compact circle. The non-Abelian gauge symmetry in the UV is broken by Wilson lines for the $SU(k)_\beta \times SU(k)_\gamma$ global symmetry,
which have to be turned on in order to have a perturbative gauge theory description in 5d, rather than an interacting 5d SCFT. In the string theory setup, this corresponds to the B-field which has to be turned on for the orbifold singularity to admit a perturbative description. 

In order to arrive to our $\CN=1$ gauge theories, we need to include the effect of the NS5 branes intersecting the D4 branes and in particular the orbifold action on the $4-4'$ strings stretched across NS5 branes. From the point of view of the four-dimensional gauge theory, we can start with an $\CN=2$ linear quiver of four-dimensional 
$U(k N)$ gauge groups and embed $\Z_k$ into the combination of $SU(2)_R$ and $U(1)$ R-symmetries which preserves an $\CN=1$ sub-algebra,
under which the 4d vector multiplet scalars (not to be confused with the 5d vector multiplets above!) transform with charge $1$ and the 4d hypermultiplets with charge $-1/2$. 

The $a$-th $\CN=2$ vector multiplet will give us a necklace $\CN_a$ of $\CN=1$ $SU(N)$ gauge groups (dropping the overall $U(1)$ which decouple in the IR),
with bi-fundamental chiral multiplets running, say, counter-clockwise along the necklace from the $(i+1)$-th to the $i$-th gauge groups in the necklace. 
Because the $\CN=2$ bi-fundalemental hypermultiplets have charge $-1/2$ under the 
R-symmetry group, we need to embed $\Z_k$ in the gauge groups accordingly, with integral charges at even nodes and half-integral at odd nodes of the original linear quiver. 
Then each $\CN=2$ bi-fundamental hypermultiplet is projected down to a set of bi-fundamental chiral fields which zig-zags back and forth 
between the nodes of consecutive necklaces, say from the $i$-th node of each necklace $\CN_a$ to the $i$-th node of the next necklace $\CN_{a+1}$
and from the $i$-th node of $\CN_{a+1}$ back to the $(i+1)$-th of $\CN_a$. 

\begin{figure}
\begin{center}
\begin{tabular}{cc}
\includegraphics[scale=0.5]{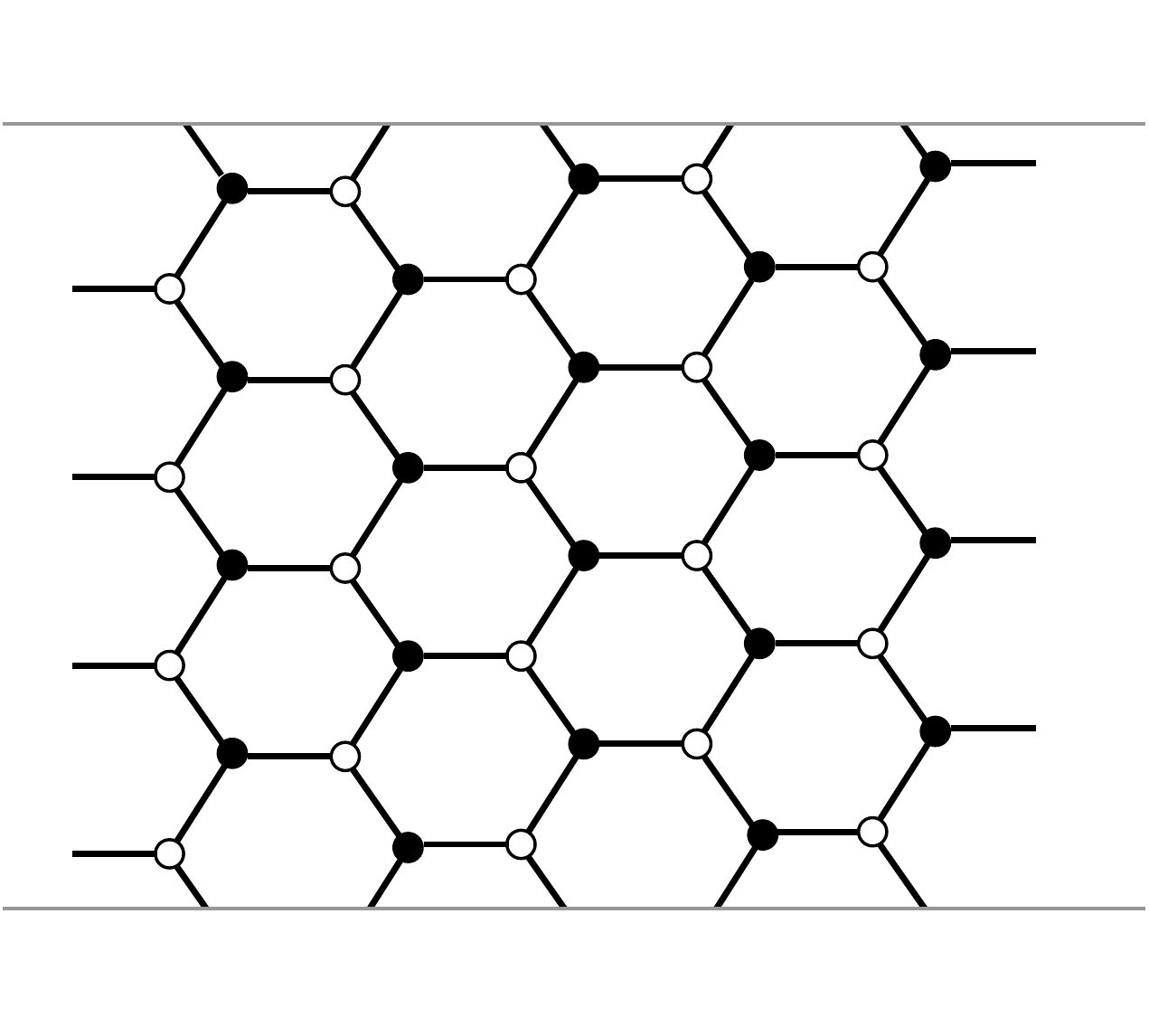} \includegraphics[scale=0.51]{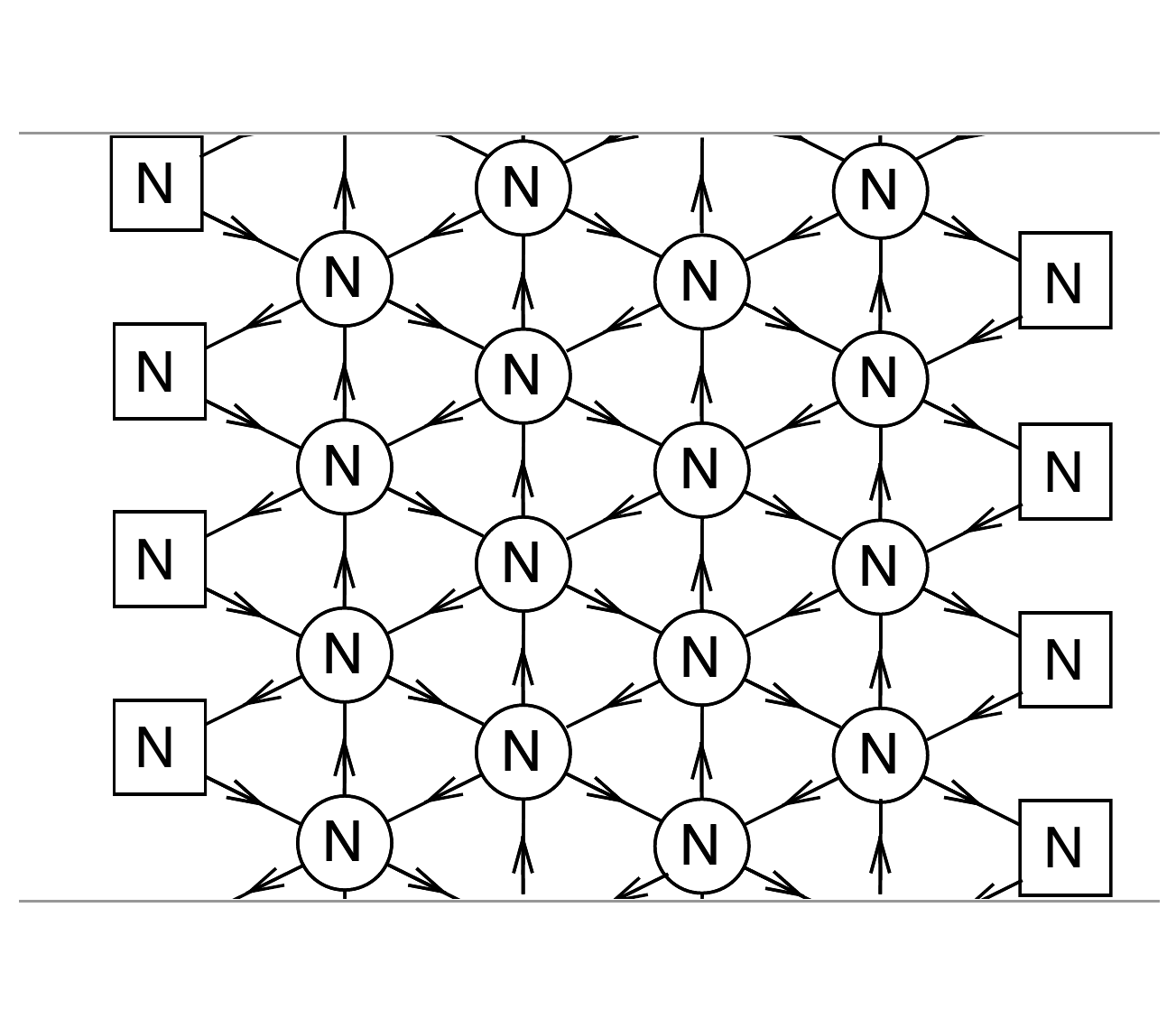}
\end{tabular}
\end{center}
\caption{ The ``honeycomb'' bi-partite graph drawn on a cylinder and the corresponding quiver gauge theory. The top and bottom lines are identified.
Each cell of the bi-partite graph maps to an $SU(N)$ gauge group with $3N$ flavors 
corresponding to the six edges of the cell. Each node of the bi-partite graph indicates a cubic superpotential term, with sign associated to the color of the node}
\label{honeyc}
\end{figure}

Thus if we start with a $\CN=2$ linear quiver of $n$ $U(kN)$ gauge groups, with $kN$ flavors at each end, 
we end up with a $\CN=1$ quiver of $k n$ $SU(N)$ gauge groups with the topology of a tessellation of a cylinder, 
with triangular faces associated to cubic super-potential couplings (arising from the $\CN=1$ superpotential 
coupling vectors and hypers in the original $\CN=2$ theory) and $k$ $SU(N)$ flavor groups at each end. 
This is the theory associated to a bi-partite honeycomb graph drawn on the cylinder, with one side of the hexagons 
aligned with the cylinder's axis. See Figure \ref{honeyc} for an example with $k=4$ and four necklaces.

\begin{figure}
\begin{center}
\begin{tabular}{cc}\includegraphics[scale=0.3]{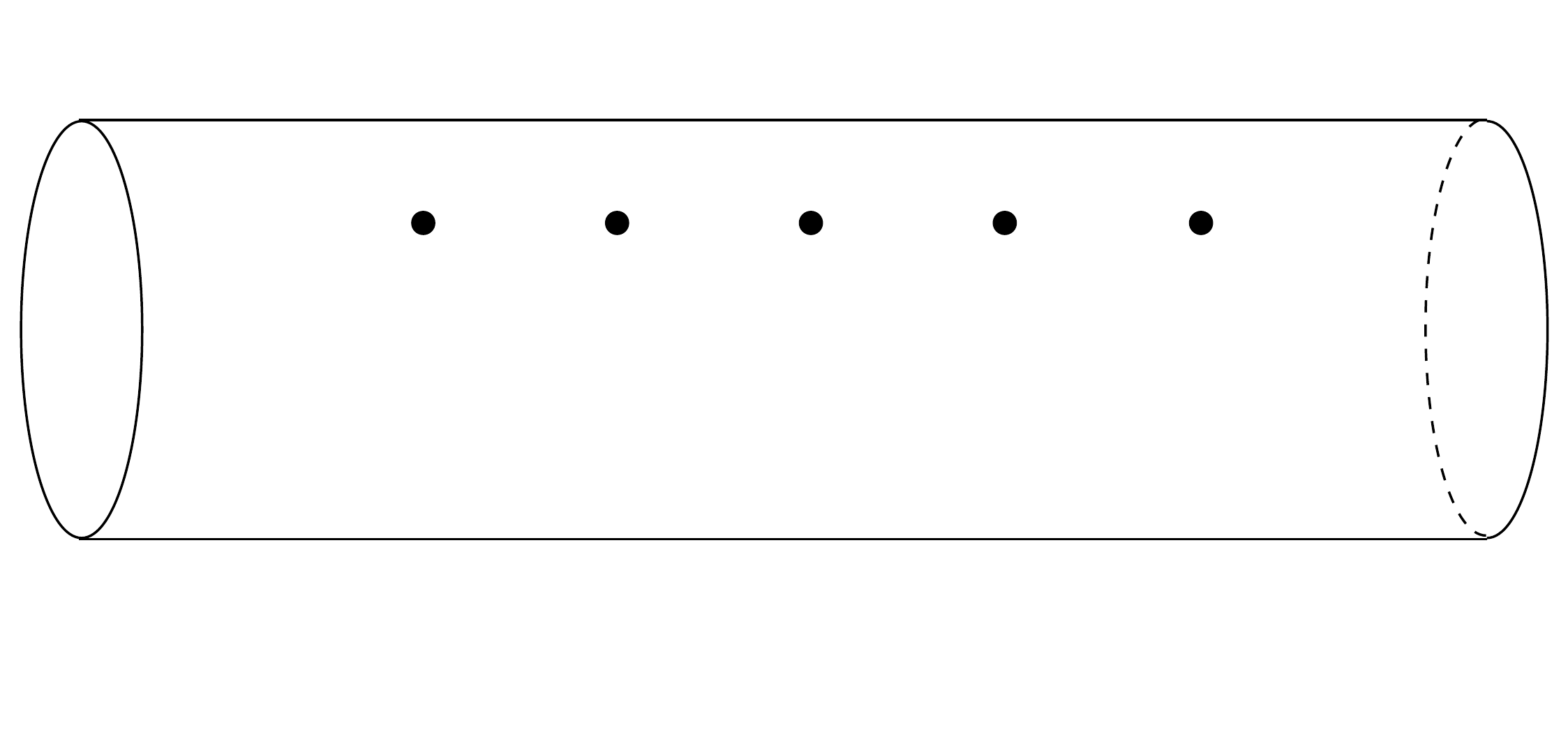} \includegraphics[scale=0.3]{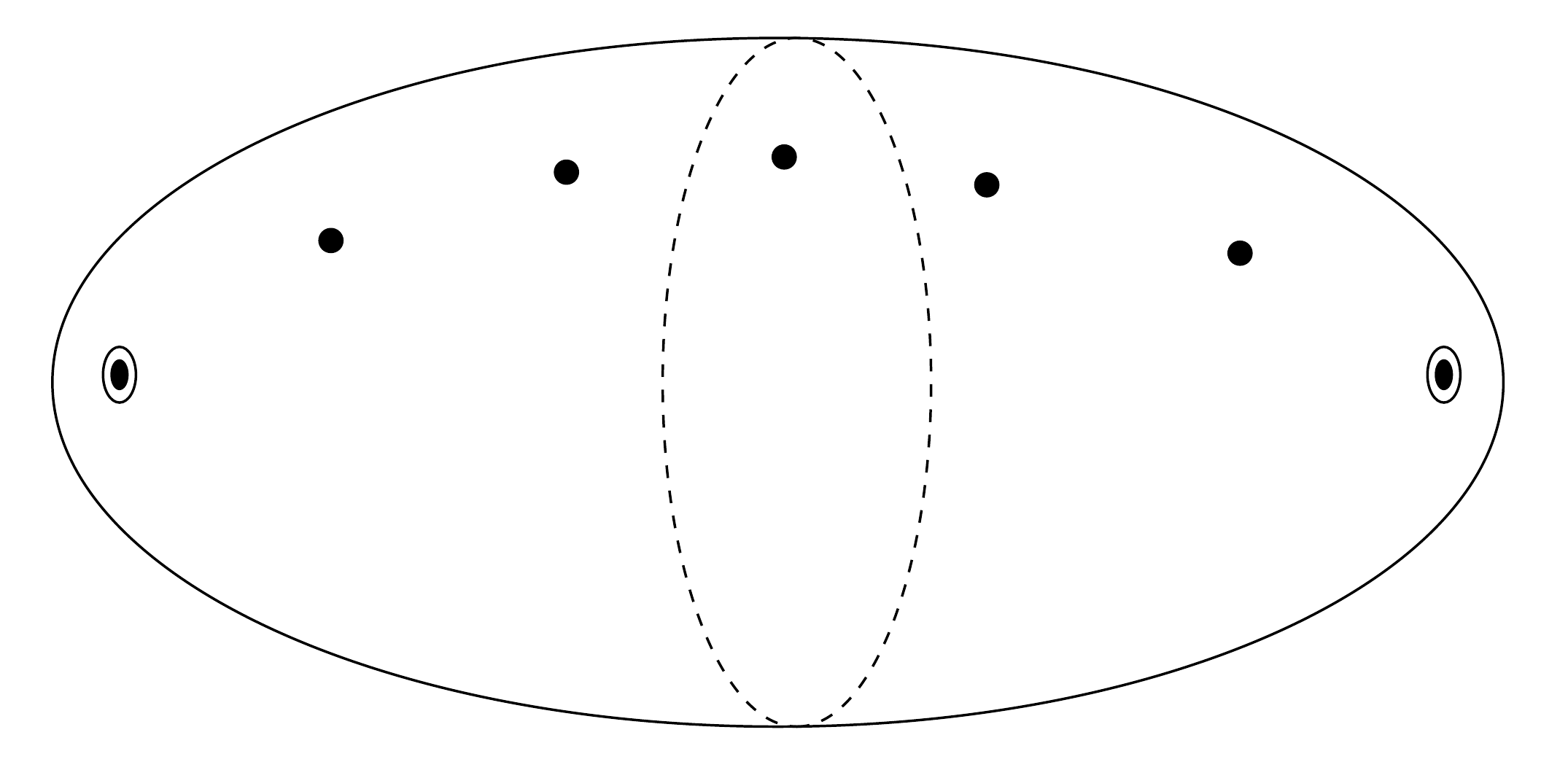}
\end{tabular}
\end{center}
\caption{ The six-dimensional lift of the brane configuration which engineers the core $\CN=1$ gauge theories. We have $N$ M5 branes sitting at the locus of a $\Z_k$ singularity and wrapping a cylinder,
intersected by 5 transverse M5 branes. The world volume theory of the $N$ M5 branes is the $\CT^N_k$ $(1,0)$ SCFT, with $5$ ``minimal'' defects (left). We expect that the de-coupling the four-dimensional degrees of freedom can be implemented by replacing the semi-infinite ends of the cylinder by maximal punctures on a sphere (right). }
\label{honeys}
\end{figure}

This $\CN=1$ gauge theory is our candidate to describe the compactification of the $\CT^N_k$ $(1,0)$ SCFT
on the cylinder, decorated with $(n+1)$ ``minimal punctures'' each associated to a single transverse M5 brane 
wrapping two directions of the $A_{k-1}$ singularity (the M-theory lift of the NS5 branes in the IIA description). 
In analogy with the class $\CS$ analysis, we hope to identify that with a compactification on a sphere, with two extra ``maximal punctures'' 
playing the role of the cylinder's ends. See Figure \ref{honeys}.

Here and in the next sections, we will study these ``core'' $\CN=1$ gauge theories, their global symmetries, exactly marginal deformations and S-dualities, 
in order to find manifestations of their conjectural six-dimensional origin. In particular, we aim to find
\begin{itemize}
\item One exactly marginal deformation parameter for each complex structure modulus of the underlying Riemann surface.
\item Global symmetries matching the six-dimensional description. 
\item S-dualities which manifest the indistinguishability among minimal punctures and imply a 2d TFT-like associativity structure for the index
\item RG flows which relate different types of punctures and, in particular, relate maximal and minimal punctures, thus justifying the picture of a sphere as opposed to a cylinder. 
\end{itemize}

The count of global symmetries and exactly marginal deformation parameters are closely related. As each gauge group has $N_f = 3 N_c$ 
and the super-potentials are cubic, the core theories can be thought of as deformations of a free theory. The $(3n+2)k$ sets of bi-fundamental hypermultiplets 
have each a $U(1)$ global symmetry and there are $3 n k$ marginal couplings. By the arguments of \cite{Green:2010da}, 
the theory will have $x$ exactly marginal couplings iff $2 k + x$ of the global symmetries are unbroken by superpotential terms or 
mixed gauge anomalies. 

\begin{figure}
\begin{center}
\begin{tabular}{cc}
\includegraphics[scale=0.6]{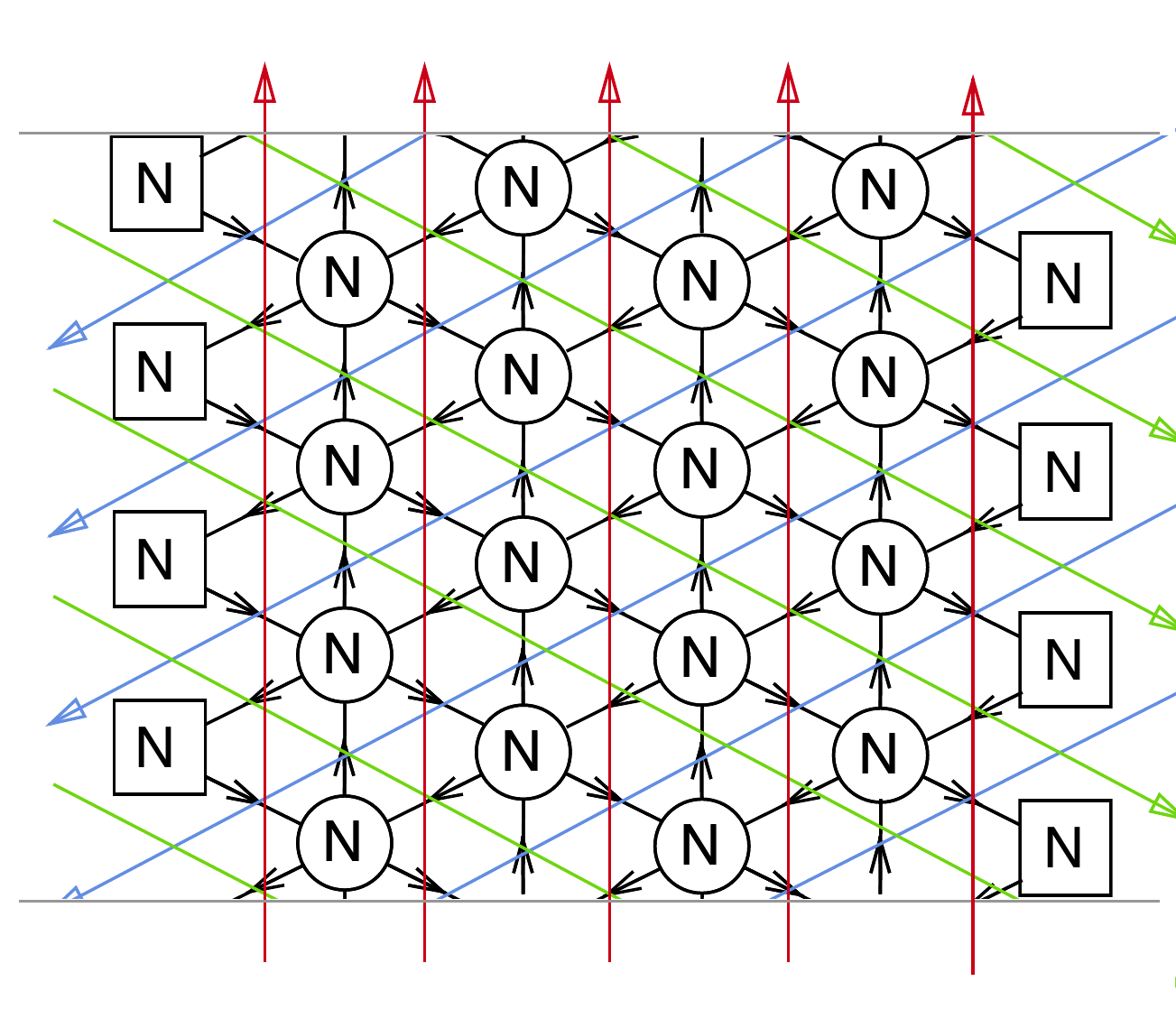}
\end{tabular}
\end{center}
\caption{A simple set of generators for the Abelian global symmetries of the core theories. Each arrow represents a generator, 
acting on the chiral multiplets crossed by the arrow with charge $\pm 1$, depending on the sign of the crossing. The sum of all generators is $0$. 
The vertical (red) arrows are chosen to generate the $U(1)_{\alpha}$ global symmetries associated to minimal punctures. The 
SE pointing (green) arrows, quotiented by their diagonal, generate the $\left[\frac{U(1)^k}{U(1)}\right]_\beta$ ``intrinsic'' symmetries. 
The SW pointing (blue) arrows, quotiented by their diagonal, generate the $\left[\frac{U(1)^k}{U(1)}\right]_\gamma$ ``intrinsic'' symmetries.
Finally, the anti-diagonal combination of blue and green arrow give the generator of the $U(1)_t$ intrinsic symmetry. }
\label{honeyf}
\end{figure}

The non-anomalous Abelian global symmetries of the core theories can be understood graphically as in Figure \ref{honeyf}: each symmetry
is associated to a straight sequence of chiral multiplets with alternating charges $\pm1$. There are $2k + n + 1$ generators which add up to $0$. 
The set of $(n+1)$ $U(1)_{\alpha}$ global symmetries associated to each of the 
$(n+1)$ blocks of chiral multiplets with the same bi-fundalental hypermultiplet ancestor can be thought of arising from the NS5 branes 
world volume gauge symmetry. We thus associate them to the minimal punctures in the six-dimensional description. 
The remaining set of $U(1)^{2k-1}$ global symmetries can be thought of as the Cartan generators of 
the $SU(k)_\beta \times SU(k)_\gamma \times U(1)_t$ global symmetry of the underlying 6d SCFT. Thus we are left with $n$ 
exactly marginal couplings, precisely as in the original $\CN=2$ theory. We associate them to the relative positions of the minimal punctures on the cylinder.

\

\section{Basic building blocks}\label{sect:basic}

In this section we study the core $\CN=1$ theories and their dualities. It is useful to begin by introducing some nomenclature, 
which helps abstracting the properties of the core $\CN=1$ theories to the expected properties of class $\CS_k$ theories. 

Each theory we build will be labelled tentatively by a punctured Riemann surface.
Some set of ``intrinsic'' global symmetries will be always present independently of the choice of punctures: 
we can denote them as ${\frak G}_k\equiv U(1)_t \times \left[\frac{U(1)^k}{U(1)}\right]_\beta \times \left[\frac{U(1)^k}{U(1)}\right]_\gamma$. 
It will be convenient to associate fugacities to these symmetries to keep track of charges of different fields and operators.
We will denote the fugacities for the above mentioned global symmetries as $t$, $\beta_i$ with $\prod_i \beta_i =1$ and $\gamma_i$ with $\prod_i \gamma_i =1$, respectively. We will also have a non-anomalous R-symmetry, whose fugacity we can indicate as $r$ or, with index computations in mind, $\sqrt{pq}$. 

Each puncture will be associated to some specific set of flavor symmetries and a specific set of 't Hooft anomalies involving these flavor symmetries and the intrinsic flavor symmetries. It will also be associated to a set of canonical chiral operators with prescribed charges. 

A ``maximal'' puncture will be associated to an $SU(N)^k$ global symmetry. Maximal punctures will be labelled by a ``color'' $n \in \Z_k$ and an ``orientation'' $o=\pm 1$, positive or negative, 
which determine the pattern of 't Hooft anomalies and the charges of the canonical chiral operators. We will have $N$ units of $U(1)_t SU(N)^2$ anomaly for all $SU(N)$ groups,
$N$ units of $U(1)_{\beta_{i+n-o}} SU(N)_i^2$ and $-N$ units of $U(1)_{\gamma_i} SU(N)_i^2$.
In other words, $SU(N)_i$ has a mixed anomaly with a $U(1)$ symmetry associated to the fugacity $t \beta_{i+n-o} \gamma_i^{-1}$. 
The R-symmetry mixed anomaly should be the same as if the $SU(N)_i$ acted on $N$ fundamental and $N$ anti-fundamental free chiral multiplets with R-symmetry $0$ (the anomaly, of course, is computed from the fermions in the chiral multiplets, which have charge $-1$). 

We also require a set of ``mesons'', chiral operators ${M^{a_{i+1}}}_{a_{i}}$ which transform as fundamental/antifundamental of 
$SU(N)_{i+1}$ and $SU(N)_i$.\footnote{This requirement is for a generic theory. As we will see say in appendix \ref{app:eig}, in some degenerate examples 
some mesons may be naively missing, but can be reinstated by adding pairs of gauge-neutral chiral fields with superpotential masses which allow one to integrate them away. 
One element of each pair play the role of the missing meson and the other is coupled to the rest of the theory by superpotential couplings. }
At a positively oriented puncture, the mesons have fugacity $t \beta_{i+n} \gamma^{-1}_{i}$. A useful mnemonic rule is that they involve the $\beta$ fugacity 
from the $SU(N)_{i+1}$ node, and the gamma fugacity from the $SU(N)_i$ node. At a negatively oriented puncture 
the mesons have fugacity $t \beta_{i+n+1} \gamma^{-1}_{i+1}$. A useful mnemonic rule is that they involve the $\beta$ fugacity  from the $SU(N)_{i}$ node, and the gamma fugacity from the $SU(N)_{i+1}$ node.
Clearly, a cyclic re-definition of the $\beta_i$, with fixed $\gamma_i$, will simultaneously shift the color of all the maximal punctures in a theory. 

\

\subsection{A gluing prescription}

Next, we give a gluing prescription, an operation on four-dimensional theories which is interpreted in six dimensions as 
replacing two maximal punctures of opposite orientation and the same color with a tube. 

Consider a positively oriented maximal puncture of color $0$ and flavor groups $SU(N)_i$ and a negatively oriented puncture of color $0$
and flavor groups $\widehat {SU(N)}_i$. We will gauge the diagonal combinations $SU(N)_i^g$ of $SU(N)_i$ and $\widehat {SU(N)}_{i-1}$.
We will also add $k$ blocks of $N^2$ chiral fields ${\Phi^{a_i}}_{a_{i+1}}$ in the fundamental/antifundamental representation
of $SU(N)^g_i\times SU(N)^g_{i+1}$. These fields also couple to the mesonic operators associated to the maximal punctures, 
through cubic superpotential couplings, 

\be\label{superpotAgen}
W=\lambda \left( \Tr M \Phi - \Tr \widehat M \Phi  \right)\,.\ee
The superpotential couplings determine the charges of ${\Phi^{a_i}}_{a_{i+1}}$ under the intrinsic global symmetry ${\frak G}_k$ such that they have  fugacities $p q t^{-1} \beta_{i}^{-1} \gamma_{i}$.  The mixed anomalies at the $SU(N)^g_i$ gauge node cancel out and the intrinsic global symmetries remain non-anomalous. 

It can be easily seen using Leigh-Strassler-like~\cite{Leigh:1995ep,Green:2010da} arguments that this gauge theory has one exactly marginal coupling which can be continuously tuned to zero. The fact that the exactly marginal coupling  can be switched off is related to the fact that for each $SU(N)$ gauge factor we have matter equivalent to $3N$ fundamental
and $3N$ anti-fundamental chiral fields and thus the one-loop contribution to the NSVZ beta-function
vanishes, and the superpotential couplings are also classically marginal. 

In particular, we can count the difference between marginal parameters and broken $U(1)$ symmetries~\cite{Green:2010da}. We have $3k$ marginal couplings, but we break the $2(2k-2)$ separate intrinsic symmetries of the two theories together with the $k$ 
symmetries acting on the $\Phi_i$ down to a diagonal combination of the intrinsic symmetries. Thus we break $3k-1$ symmetries. 
Each broken symmetry lifts a marginal coupling, leaving only a single exactly marginal one. We expect it to correspond roughly to the length and twist of the 
newly-created tube in the underlying Riemann surface.   

\

\subsection{The free trinion}

Having described some generalities of the setup let us turn our attention to concrete 
and important examples.
Our first ingredient is a free theory which we would like to label by a trinion, a sphere with three punctures.
Two of the punctures are maximal  and one is a  ``minimal'' one. 
We will refer to this theory as the {\it free trinion}. 

\begin{figure}
\begin{center}
\begin{tabular}{c}
\includegraphics[scale=0.41]{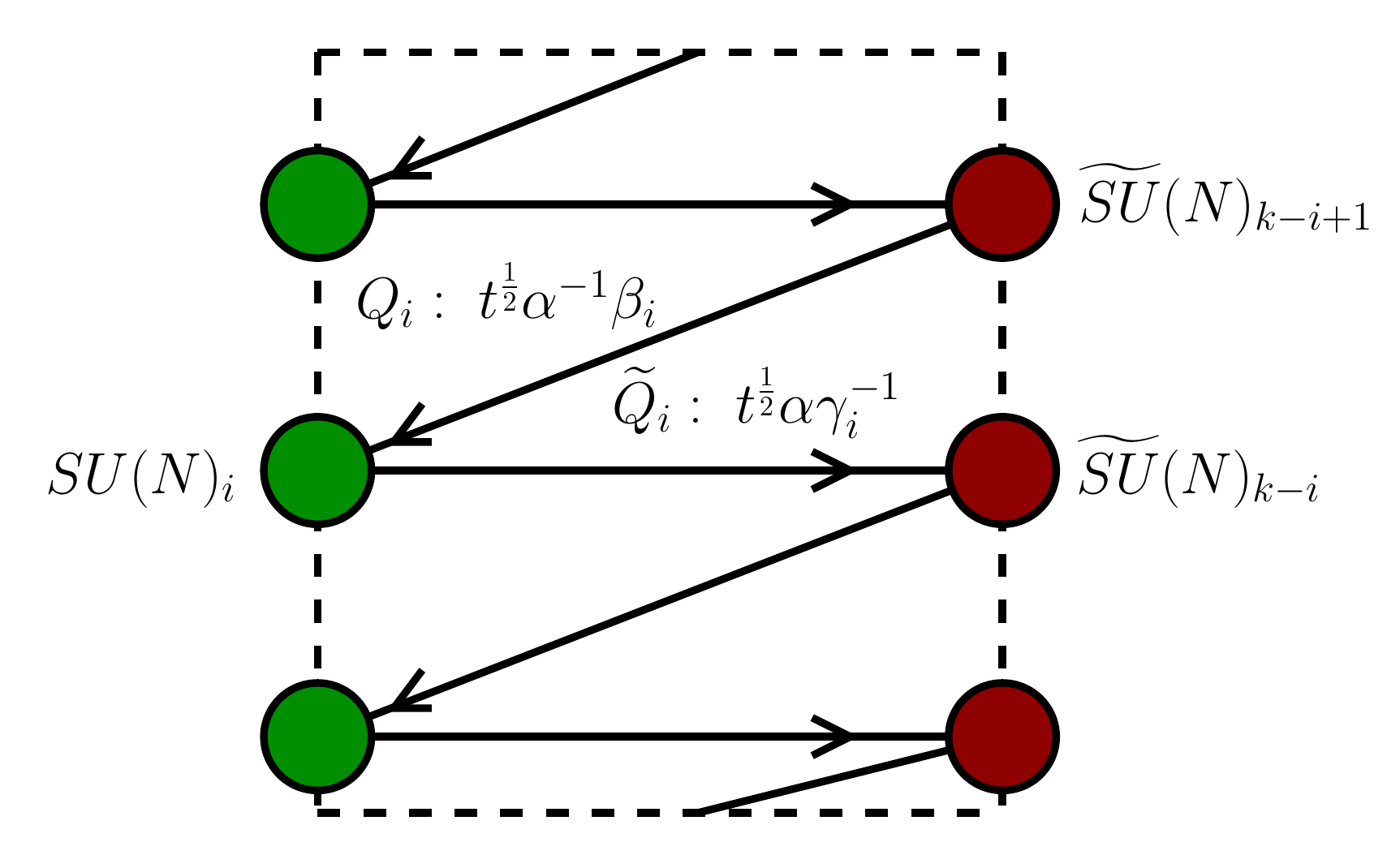}
\end{tabular}
\end{center}
\caption{The free trinion. Dots on the left an on the right represent $SU(N)$ groups associated to the two differnet maximal punctures. The horizontal lines are the bi-fundamentals $\widetilde Q$ and the diagonal lines $Q$. 
\label{freetrinion}}
\end{figure}  

The free trinion is a  collection of $2kN^2$ free chiral fields, equipped with a specific action of the intrinsic and puncture-related global symmetries.  
We organize the fields in two sets of $k$ blocks of $N^2$ chiral fields, denoted respectively as 
${Q^{a_i}}_{b_i}$ and ${\widetilde{ Q}_{a_i}}{}^{b_{i+1}}$: the former transform as fundamental/anti-fundamental representation of $SU(N)_i\times \widetilde{SU}(N)_{k-i+1}$
and the latter in anti-fundamental/fundamental  representation of $SU(N)_{i}\times\widetilde{SU}(N)_{k-i}$. 
The sets of $k$ $SU(N)_i$ and $k$ $\widetilde{SU}(N)_{i}$ global symmetries are associated respectively with the two maximal punctures. 

It is convenient to depict the free trinion as a ring with $2k$ dots connected by arrows. 
The dots at even/odd places correspond to $SU(N)_i/\widetilde{ SU}(N)_i$ flavor groups. See Figure~\ref{freetrinion}. 
We also define the action of the intrinsic ${\frak G}_k=U(1)_t \times \left[\frac{U(1)^k}{U(1)}\right]_\beta \times \left[\frac{U(1)^k}{U(1)}\right]_\gamma$ global symmetries, and of an extra $U(1)_\alpha$ 
associated to the minimal puncture. Expressing the charges in terms of fugacities, we associate fugacity $t^{\frac12} \beta_i \alpha^{-1}$ to ${Q^{a_i}}_{b_i}$ and $t^{\frac12} \alpha \gamma^{-1}_i$ to ${\widetilde Q_{a_i}}{}^{b_{i+1}}$. 

It is straightforward to compute the 't Hooft anomalies of the free trinion. The $U(1) SU(N)^2$ anomalies for each $SU(N)$ 
flavor group receive contributions from two blocks of chirals: we have $N$ units of $U(1)_t SU(N)^2$ anomaly for all $SU(N)$ groups,
$N$ units of $U(1)_{\beta_i} SU(N)_i^2$ and $U(1)_{\beta_i} \widetilde{SU}(N)_{k-i+1}^2$, $-N$ units of 
$U(1)_{\gamma_i} SU(N)_i^2$ and $U(1)_{\gamma_i} \widetilde{SU}(N)_{k-i}^2$.
In other words, $SU(N)_i$ has a mixed anomaly with a symmetry associated to the fugacity $t \beta_i \gamma_i^{-1}$, 
while $\widetilde{SU}(N)_{k-i+1}$ has a mixed anomaly with a symmetry associated to the fugacity $t \beta_i \gamma_{i-1}^{-1}$.

There are various $U(1)^3$ anomalies, such as $k N^2$ units of $U(1)_\alpha^2 U(1)_t$ anomaly, $-N^2$ of 
$U(1)_\alpha U(1)_{\beta_i}^2$ and $N^2$ of $U(1)_\alpha U(1)_{\gamma_i}^2$, and some anomalies involving intrinsic 
symmetries only. 

Finally, we have chiral operators we can associate with each puncture. For example, $M^{a_i}_{a_{i-1}} =\widetilde Q_{a_{i-1}}^{b_{i}} Q^{a_i}_{b_i}$ 
are fundamental/antifundamental of consecutive $SU(N)_i$ groups, with fugacity $t \beta_i \gamma^{-1}_{i-1}$, while 
$\widetilde M^{b_{i+1}}_{b_i} =\widetilde Q_{a_{i}}^{b_{i+1}} Q^{a_i}_{b_i}$ are fundamental/antifundamental of consecutive $\widetilde SU(N)_i$ groups, with fugacity $t \beta_i \gamma^{-1}_{i}$.
We associate them to the respective maximal punctures. On the other hand, baryon operators $B_i = \det Q^{a_i}_{b_i}$ and $\widetilde B_i = \det Q_{a_i}^{b_{i+1}}$ can be associated to the minimal puncture. 

We see that the first maximal puncture, associated to the $SU(N)_i$, has color $1$ and positive orientation. The second maximal puncture is associated to the groups $\widetilde SU(N)_{k-i}$. It has negative orientation and color $0$.  

\

\subsection{Gluing two trinions}
All our core theories are produced by gluing together a sequence of $n+1$ free trinions. 
We can focus on the simplest interacting theory built by gluing  two trinions. 

\begin{figure}
\begin{center}
\begin{tabular}{c}
\includegraphics[scale=0.41]{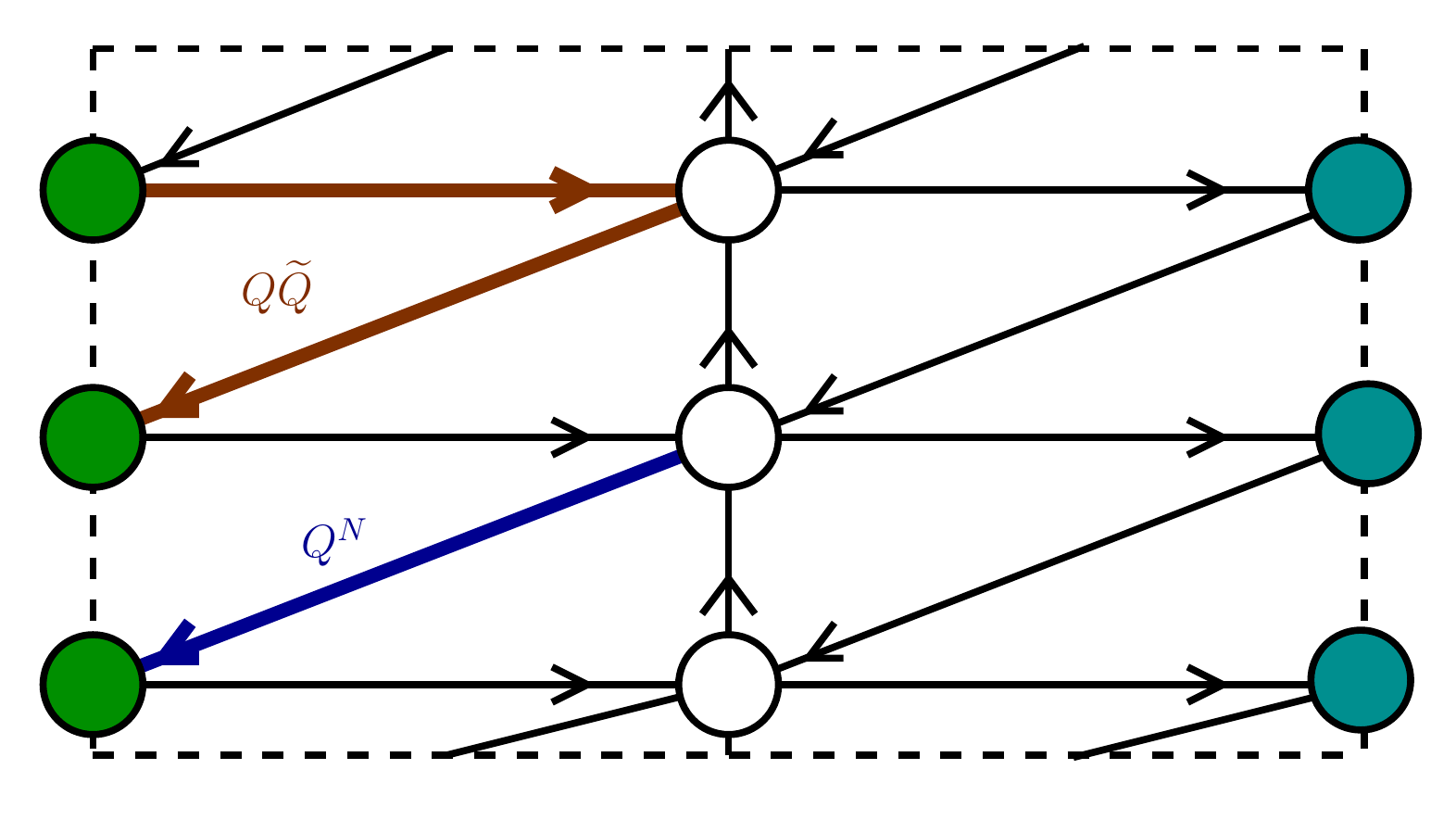}
\end{tabular}
\end{center}
\caption{ The interacting superconformal theory. The circles correspond to $SU(N)$ groups with the the white ones being gauged and colored ones being flavor symmetries. The fat brown line represent the mesonic operator built from the quarks which is charged under the symmetries associated to maximal puncture but is neutral under the symmetry associated to the minimal puncture. Fat blue line represents a quark from which a baryon is built which is charged under the $U(1)$ symmetry associated with the minimal puncture but is a singlet under the symmetry associated with the maximal puncture.
\label{intersphere}}
\end{figure}

We can take two free trinions, shift the definition of the $\beta_i$ in the second trinion so that it 
has a positive maximal puncture of color $0$ and a negative of color $-1$, and glue them together as 
described in the previous subsection. 
The result is our candidate for a theory with two maximal punctures, of opposite orientation and color $1$ and $-1$, 
and two minimal punctures.  Let us denote this theory by $\tilde T_k$ and the quiver description of the theory is in Figure~\ref{intersphere}. Throughout the paper this theory will play an important role and  we will refer to it as our {\it basic interacting theory} or {\it basic four-punctured sphere}.

Our crucial claim is that the interacting theory described here enjoys an S-duality property, which 
corresponds to the exchange of the two minimal punctures. Notice that this is consistent with 't Hooft anomalies, as 
the global symmetries $U(1)_\alpha$ and $U(1)_\delta$ associated to the two minimal punctures have mixed anomalies with the intrinsic global symmetries only, and the anomalies are identical. 
Since we have here an exactly marginal coupling, we can switch on in a correlated manner the gauge couplings and the superpotential coupling such that the beta-functions vanish, the superconformal R-charges of the different fields are the free ones. For example the superconformal anomalies $a$ and $c$ 
are just the ones of this particular collection of free fields,

\be\label{anomacofint}
a=\frac{k}{48}(14N_c^2-9)\,,\qquad\quad\qquad c=\frac{k}{24}(8N_c^2-3)\,.
\ee We will find it convenient  to assign R-charge $0$ to all $Q$ and $\widetilde Q$ and R-charge $2$ for all $\Phi$. 

Notice that in our notation we are focusing on a subset of the full global symmetry of the gauge theory. The 
quarks and anti-quarks which are associated to a given gauge group can be re-assembled into groups of $2N$ 
flavors, so that the theory really has $SU(2N)^k \times U(1)^{k+1}$ global symmetry. The $U(1)$ symmetries 
are associated to fugacities such as $t_i = t \beta_i \gamma_i^{-1}$ and $\eta^2 = \alpha \delta^{-1}$. 
The S-duality transformation should invert the definition of the latter $U(1)$ symmetry, the anti-diagonal combination of $U(1)_\alpha$ and $U(1)_\delta$. 

This structure is obviously similar to what one encounters in $\CN=2$ SQCD, the $k=1$ version of our story.  As we will see momentarily, this is no coincidence: all our conjectural S-dualities can be related recursively to the $\CN=2$ SQCD S-duality by Seiberg dualities.

Our starting point is a theory $\tilde T_k$ defined as a necklace of $SU(N)_i$ gauge groups, connected by bi-fundamental chiral fields 
$\Phi^{a_i}_{a_{i+1}}$ and to fundamental fields $Q_{a_i}^{c_i}$ and $\widetilde Q_{c_{i}}^{a_{i+1}}$,
where the $c_i$ indices transform under $SU(2N)$ global symmetry groups and obvious cubic superpotential. 
We can take the $\Phi^{a_i}_{a_{i+1}}$ to have fugacity $t_i^{-1}$, $Q_{a_i}^{c_i}$ of fugacity $\sqrt{t_i} \eta$, the 
$\widetilde Q_{c_{i}}^{a_{i+1}}$ of fugacity $\sqrt{t_i} \eta^{-1}$. 
If we apply Seiberg duality formally at the $k$-th node, we obtain a new theory ${\tilde T}'_k$ which has a rather simple description: 
it consists of a shorter necklace ${\tilde T}_{k-1}$ where the $(k-1)$-th $SU(2N)$  flavor group has been gauged and coupled to 
$2N$ dual quarks and $2N$ dual anti-quarks. The theory also has a set of $2N \times 2N$ mesons coupled to the latter fields by a cubic superpotential. 
The dual quarks and anti-quarks are rotated by the $SU(2N)_{k-1}$ and $SU(2N)_k$ global symmetries of the original theory. 
Crucially, the $U(1)$ flavor symmetry assignments work out in such a way that 
only fields in $\tilde T_{k-1}$ transform under $U(1)_\eta$.

At this point, we can immediately conclude, recursively, that say the supersymmetric index of ${\tilde T}_k$ is invariant under the transformation $\eta \to \eta^{-1}$:
the Seiberg duality manipulation leaves  the index invariant, and the index of ${\tilde T}'_k$ is built from the index of ${\tilde T}_{k-1}$, 
with $\eta$ only entering in the latter. As ${\tilde T}_1$ is $\CN=2$ SQCD with $N_f = 2 N_c$, which has an S-duality which acts as  
$\eta \to \eta^{-1}$ on the flavor fugacities, the index of ${\tilde T}_1$ is invariant under the transformation $\eta \to \eta^{-1}$.

In order to believe a full S-duality statement, we need to make some assumptions about the RG flows 
associated to the Seiberg duality we employed, as the $SU(2N)$ gauge node is strongly-coupled. 
The RG flow of the theory ${\tilde T}'_k$ defines a map from the conformal manifold of ${\tilde T}_{k-1}$ 
to the conformal manifold of ${\tilde T}_k$. By induction, we can assume that the conformal manifold of 
${\tilde T}_{k-1}$ has two weakly-coupled cusps, where the quiver gauge theory description is good. 
Near these cusps, the RG flow will map us to weakly-coupled ${\tilde T}_k$ quivers, with opposite $U(1)_\eta$
charge assignments. Thus the crucial assumption for the induction to hold is that 
the connected path between the two cusps in the conformal manifold of 
${\tilde T}_{k-1}$ maps to a connected path in the conformal manifold of ${\tilde T}_k$.

\begin{figure}[htbp]
\begin{center}
\begin{tabular}{c}
\includegraphics[scale=0.61]{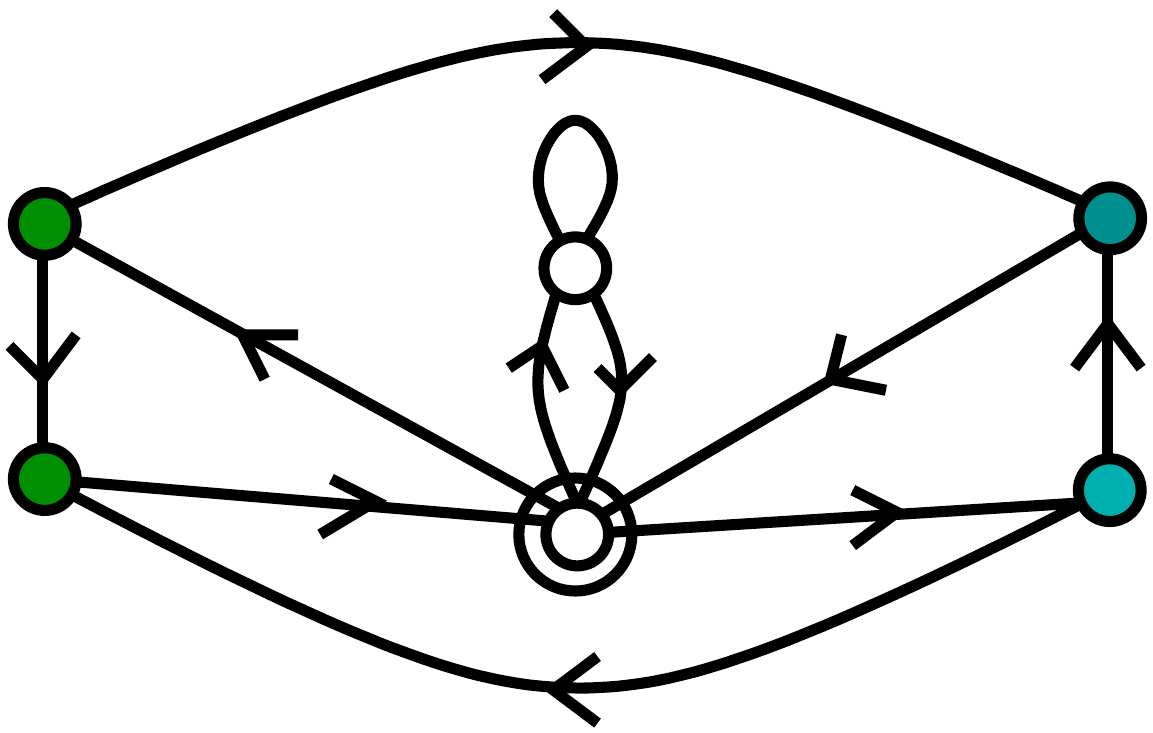}
\end{tabular}
\end{center}
\caption{Seiberg duality of one of the $SU(N)$ gauge groups in the case of $k=2$. 
The node with a circle is an $SU(2N)$ gauge group.  There is a superpotential term for
each triangle in this quiver theory.
\label{seibergdualA}}
\end{figure}

\

\

In analogy with class $\CS$, we can ask if other trinions may exist, in particular trinions which carry three maximal punctures. 
This is of course far from obvious. In the case of class $\CS$, a strong piece of evidence came from the existence of Argyres-Seiberg-like 
dualities~\cite{argyres-2007-0712}. A key step in the analysis was to find operations which ``reduce'' a maximal puncture to a minimal one, a Higgs branch RG flow 
which replaces one end of a linear quiver of $SU(N)$ gauge groups with a ``quiver tail'' which is associated to a set of minimal punctures only,
all related by S-dualities. That suggests the existence of alternative S-duality frames where all minimal punctures 
are produced by quiver tails, attached to a conjectural SCFT with maximal punctures only. 

Due to the intricacies of the quivers we study here, there is a bewildering array of possible RG flows one can trigger by a sequence of 
vevs for chiral operators. Correspondingly, one can modify a maximal puncture to a wide array of ``smaller'' punctures. Our challenge is to 
find a sequence which leads to a minimal puncture.  We will undertake this challenge in the following
sections.

\

\subsection{The index avatar}\label{subsec:basic:index}
 
Let us discuss the basic theories and the dualities in the language of the supersymmetric index setting the nomenclature for the discussion in the following sections. The reader can accustom him/herself with the standard definitions of the index in appendix \ref{app:index}. The index is a very precise tool to test dualities. To be very explicit and keep our formulae readable, 
we will often specialize to the index of $A_1$ theories of class ${\cal S}_2$. All we will say can be rather straightforwardly extended to more general cases.

\

\subsection*{Free trinion}

\begin{figure}[htbp]
\begin{center}
\begin{tabular}{c}
\includegraphics[scale=0.21]{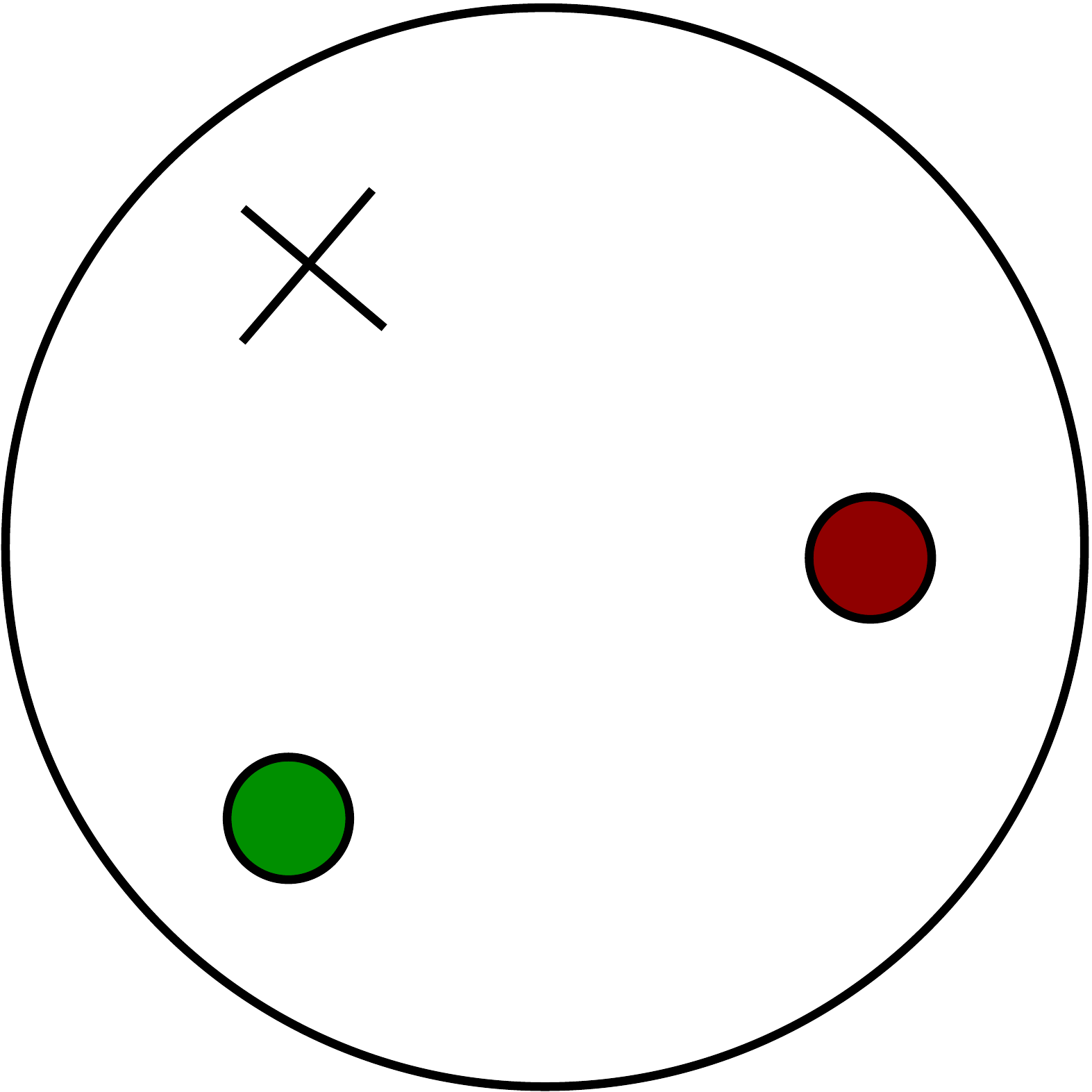}
\end{tabular}
\end{center}
\caption{The sphere with two maximal punctures of different color and one minimal puncture, {\it aka} the free trinion.
\label{spherefP}}
\end{figure}  

Let us start from writing down the index of the free trinion.
As discussed in  previous subsections and depicted in Figure~\ref{spherefP} we associate this trinion  to a sphere with two maximal punctures and one minimal puncture. 
The two maximal punctures are of different colors, and we will return to this 
feature momentarily.  The index can be written by collecting together the contributions of the different free field. For example, in the $A_1$ $k=2$ case the index of the free trinion is,

\be\label{aonetrin}
&&\CI_{\mathrm{ft}}({\bf u}, \alpha, {\bf v}; \beta, \gamma, t)\equiv\;{{\cal I}_{{\bf u}\alpha}}^{\bf v}(\b,\g,t,p,q)\equiv\\
&&\quad\qquad\Gamma_e(t^{\frac12}v_1^{\pm1}u_1^{\pm1}\beta\a^{-1})
\Gamma_e(t^{\frac12}v_1^{\pm1}u_2^{\pm1}\gamma^{-1}\a^{})
\Gamma_e(t^{\frac12}v_2^{\pm1}u_1^{\pm1}\gamma\a^{})
\Gamma_e(t^{\frac12}v_2^{\pm1}u_2^{\pm1}\beta^{-1}\a^{-1})\,.\nonumber
\ee  Here the fugacities ${\bf u}$ and ${\bf v}$ correspond to $SU(2)^2$ and $\widetilde{SU}(2)^2$ flavor symmetries associated to the maximal punctures and $\a$ is the $U(1)$ fugacity associated to the minimal puncture.
Note the index of the trinion is not symmetric under the exchange of the two maximal punctures. 
This is a reason why one should associate an additional parameter, {\it color}, to the maximal punctures.
If we set $\beta=1$ there is no distinction between the colors and indeed the index becomes symmetric under exchanging the two maximal punctures.  In the $k=2$ case we have two colors which are $\Z_2$ valued, $0$ and $1$. We will also refer to the two punctures in this case as lower and upper ones.

We can glue two trinions together and obtain a theory corresponding to two maximal and two minimal punctures, and we will come to details of the gluing at the level of the index momentarily. We can also continue gluing free trinions to obtain theories corresponding to spheres with
many minimal punctures and two maximal ones. The S-duality operation exchanging two punctures surely holds at the level of the index, under the usual assumption that no new $U(1)$ symmetry emerges accidentally in the IR. 

Let us for a moment discuss implications of the S-duality for general $k$  and $N$.
At the level of the index, we can glue a free trinion to a theory $T$ with a positive maximal puncture of color $0$ by the formula:
\begin{align}
\CI_T&({\bf v}; \beta_i,\gamma_i, t) \to \,\CI_{T+\mathrm{min}_\alpha}({\bf v}, \alpha;\beta_i,\gamma_i, t) \equiv \cr &\oint_{{\bf u}} d\mu_n({\bf u};\beta_i, \gamma_i, t) \CI_{\mathrm{ft}}({\bf v}, \alpha, {\bf u}; \beta_i, \gamma_i, t)\CI_T({\bf v}; \beta_i,\gamma_i, t)\,,
\end{align}
where $\CI_{\mathrm{ft}}$ is the free trinion index, $d\mu_n({\bf u},\beta_i, \gamma_i, t)$ the measure which accounts for the gauge fields and $\Phi$ fields involved in the gluing
and the ${\bf u}$ contour integral is the projection on gauge-invariant operators. This can be thought as an integral transformation on the index of $T$, with a kernel which 
depends on the fugacity $\alpha$ of the new minimal puncture. 

The basic S-duality we derived in this section implies that if we apply the transformation twice, the result will not depend on the order of the two transformations.
In other words, the integral operators associated to different values of the fugacity $\alpha$ commute. 
It is reasonable to assume that the transformations can be ``diagonalized'', as in the class ${\cal S}$ case~\cite{Gadde:2011ik,Gadde:2011uv}, 
by expanding the free trinion into ``wavefunctions'' associated to each puncture,
\begin{equation}
\CI_{\mathrm{ft}}({\bf v}, \alpha, {\bf u}; \beta_i, \gamma_i, t)= \sum_\lambda \psi^{[1]}_\lambda({\bf u};\beta_i, \gamma_i, t) \phi_\lambda(\alpha;\beta_i, \gamma_i, t) 
\bar \psi^{[0]}_\lambda({\bf v};\beta_i, \gamma_i, t)
\end{equation}
with $\psi^{[n]}_\lambda({\bf u};\beta_i, \gamma_i, t) \equiv \psi_\lambda({\bf u};\beta_{i+ n}, \gamma_i, t)$ being the wavefunction for positive maximal punctures, 
$\bar \psi^{[n]}_\lambda({\bf u};\beta_i, \gamma_i, t) \equiv \bar \psi_\lambda({\bf u};\beta_{i+ n}, \gamma_i, t)$ being the wavefunction for negative maximal punctures
and $\phi_\lambda(\alpha;\beta_i, \gamma_i, t)$ being the wave function for minimal punctures, invariant under color shift,
\begin{equation}
\phi_\lambda(\alpha;\beta_{i+n}, \gamma_i, t) = \phi_\lambda(\alpha;\beta_i, \gamma_i, t)\,.
\end{equation}
  We take the maximal wavefunctions to be normalized as 
 \begin{align}
\oint_{{\bf v}} d\mu_n({\bf v};\beta_i, \gamma_i, t) \bar \psi^{[n]}_\lambda({\bf v},\beta_i, \gamma_i, t) \psi^{[n]}_{\lambda'}({\bf v},\beta_i, \gamma_i, t) = \delta_{\lambda \lambda'}\,.
\end{align}
Then if we glue together $n$ trinions we get an index 
\begin{equation}\label{linindex}
\CI_{\mathrm{ft}^n}(u_i, \alpha_k, v_i, \beta_i, \gamma_i, t) = \sum_\lambda \psi^{[n]}_\lambda(u_i,\beta_i, \gamma_i, t) \left[\prod_k \phi_\lambda(\alpha_k,\beta_i, \gamma_i, t) \right]
\bar \psi^{[0]}_\lambda(v_i,\beta_i, \gamma_i, t)\,,
\end{equation}
which is explicitly invariant under S-duality.  Note that gluing $n$ free trinions we get a theory of two maximal punctures differing by $n$ units of the $\Z_k$ color.

In the rest of this subsection we will specialize to the case of $k=2$ $A_1$ theories. As already mentioned here we have two colors for maximal punctures. The orientation of the puncture  corresponds to the  ordering  of the two $SU(2)$ groups. We will denote

\be
&&\psi({\bf u};\b,\g,t)=\psi^{[0]}({\bf u};\b,\g,t)\,,\qquad
\widetilde \psi({\bf u};\b,\g,t)=\psi^{[1]}({\bf u};\b,\g,t)\,,\\
&&\psi({\bf u}^\dagger;\b,\g,t)=\bar\psi({\bf u};\b,\g,t)\,,\nonumber
\ee  where  we define $(u_1,u_2)^\dagger=(u_2,u_1)$.
Thus the index of the free trinion will be written as,

\be\label{indtrind}
\CI_{\mathrm{ft}}({\bf u}, \a, {\bf v}; \beta, \gamma, t)={{\cal I}_{{\bf u}\alpha}}^{\bf v}=\sum_{\lambda}  \psi_\lambda(u_1,u_2)\widetilde \psi_\lambda(v_1,v_2) \phi_\lambda(\alpha)\,.
\ee  
The way to determine the functions $\psi_\lambda$, $\widetilde \psi_\lambda$, $\phi_\lambda$, as well as the  possible values of the index $\lambda$ will be determined in section \ref{sect:surface}. 
These functions will turn out to be eigenfunctions of certain difference operators and thus we will refer 
to them as {\it eigenfunctions}.
We stress that it is a rather non trivial assumption that such a representation of the index exists. We will discuss more of the physical implications of this property later on.

We have the following symmetry properties satisfied by the index of the free trinion following from its explicit expression~\eqref{aonetrin},

\be\label{symtrin}
&&{{\cal I}_{{\mathbf u}\alpha}}^{\mathbf v} (\b,\g,t)=
{{\cal I}_{{\mathbf v}\alpha}}^{{\mathbf u}}(\b^{-1},\g,t)\,,\nonumber\\
&&{{\cal I}_{{\mathbf u}\alpha}}^{{\mathbf v}}(\b,\g,t)=
{{\cal I}_{{\mathbf v}^\dagger\alpha}}^{{\mathbf u}^\dagger}(\b^{},\g^{-1},t)\,,\\
&&{{\cal I}_{{\mathbf u}\alpha}}^{{\mathbf v}}(\b,\g,t)=
{{\cal I}_{{\mathbf u}\alpha^{-1}}}^{{\mathbf v}^\dagger}(\g^{},\b,t)\,,\nonumber\\
&&{{\cal I}_{{\mathbf u}\alpha}}^{{\mathbf v}}(\b,\g,t)=
{{\cal I}_{{\mathbf u}^\dagger\alpha^{-1}}}^{{\mathbf v}}(\g^{-1},\b^{-1},t)\,,\nonumber
\ee It is natural to assume that these transformations act on single eigenfunctions and do not act on the labels $\lambda$. Under this assumption ~\eqref{symtrin} implies that,

\be\label{relpsitildepse}
&&\psi_\lambda({\mathbf u};\b,\g,t)=\widetilde \psi_\lambda({\mathbf u};\b^{-1},\g,t)=\widetilde \psi_\lambda({\mathbf u}^\dagger;\b^{},\g^{-1},t)=\psi_\lambda({\mathbf u};\g,\b,t)= \psi_\lambda({\mathbf u}^\dagger;\g^{-1},\b^{-1},t)\,,\nonumber\\
&&\widetilde \psi_\lambda({\mathbf u};\b,\g,t)=\widetilde \psi_\lambda({\mathbf u}^\dagger;\g^{},\b^{},t)=
\widetilde \psi_\lambda({\mathbf u};\g^{-1},\b^{-1},t)\,,\\
&&\phi_\lambda(\a,\beta,\g,t)=
\phi_\lambda(\a,\beta^{-1},\g,t)=
\phi_\lambda(\a,\beta,\g^{-1},t)=
\phi_\lambda(\a^{-1},\g,\beta,t)\,.
\nonumber
\ee We do not have to assume these properties for what follows but assuming them will make some of the considerations simpler and we will state explicitly when such assumption will be made. Moreover, when explicitly computing the eigenfunctions we find that these indeed are satisfied although we will not prove them mathematically. 

\

\subsection*{The gauging}

We gave a prescription for gluing two theories together at maximal punctures of appropriate color and orientation,
by adding extra chiral fields and superpotential couplings to the mesons and gauging diagonal combinations of the flavor symmetries.  
The gauge group in general is  $SU(N)^k$ and in case at hand it is $SU(2)^2$. 
Since there are two types of maximal punctures, we can glue theories along upper {\it or } lower punctures. In both cases the gauge group is the same.
We identify $SU(2)_i$ of one theory with $SU(2)_{3-i}$ of the other one, and we add bifundamental chiral fields of $SU(2)_i\times SU(2)_{i+1}$. The difference between the two gaugings is that when we 
glue two upper punctures the two bifundamental chirals have charges $(+1,+1)$ and $(-1,-1)$ under $U(1)_\beta\times U(1)_\gamma$, whereas when gluing two lower punctures the charges are $(+1,-1)$ and $(-1,+1)$. 

For the index this implies that the functions
$\psi_\lambda$ and $\widetilde \psi_\lambda$ are orthonormal under the following measures,

\be\label{measures}
&&\widetilde\Delta({\mathbf z}) = \frac{{\cal I}_V^2}4\frac{\Gamma_e(\frac{pq}t(\b\g)^{\pm1}z_1^{\pm1}z_2^{\pm1})}{ \Gamma_e(z_1^{\pm2})\Gamma_e(z_2^{\pm2})},\qquad\quad
\Delta({\mathbf z}) = \frac{{\cal I}_V^2}4 \frac{\Gamma_e(\frac{pq}t(\b^{-1}\g)^{\pm1}z_1^{\pm1}z_2^{\pm1})}{\Gamma_e (z_1^{\pm2})\Gamma_e(z_2^{\pm2})}\,.
\ee That is 

\be\label{ortho}
&&\oint_{{\mathbf v}} d\mu_0({\mathbf v};\beta, \gamma, t) \bar \psi^{[0]}_\lambda({\mathbf v},\beta, \gamma, t) \psi^{[0]}_{\mu}({\mathbf v},\beta, \gamma, t) = \oint \frac{dz_1}{2\pi i z_1}\oint \frac{dz_2}{2\pi i z_2} \Delta{(\mathbf z)} \, \psi_\lambda({\mathbf z})\psi_\mu ({\mathbf z}^\dagger)=\delta_{\lambda\mu}\,,\nonumber\\
&&\oint_{{\mathbf v}} d\mu_1({\mathbf v};\beta, \gamma, t) \bar \psi^{[1]}_\lambda({\mathbf v},\beta, \gamma, t) \psi^{[1]}_{\mu}({\mathbf v},\beta, \gamma, t)=\oint \frac{dz_1}{2\pi i z_1}\oint \frac{dz_2}{2\pi i z_2} \widetilde\Delta{(\mathbf z)} \, \widetilde \psi_\lambda({\mathbf z})\widetilde \psi_\mu ({\mathbf z}^\dagger)=\delta_{\lambda\mu}\,.\nonumber\\
\ee

\begin{figure}[htbp]
\begin{center}
\begin{tabular}{c}
\includegraphics[scale=0.21]{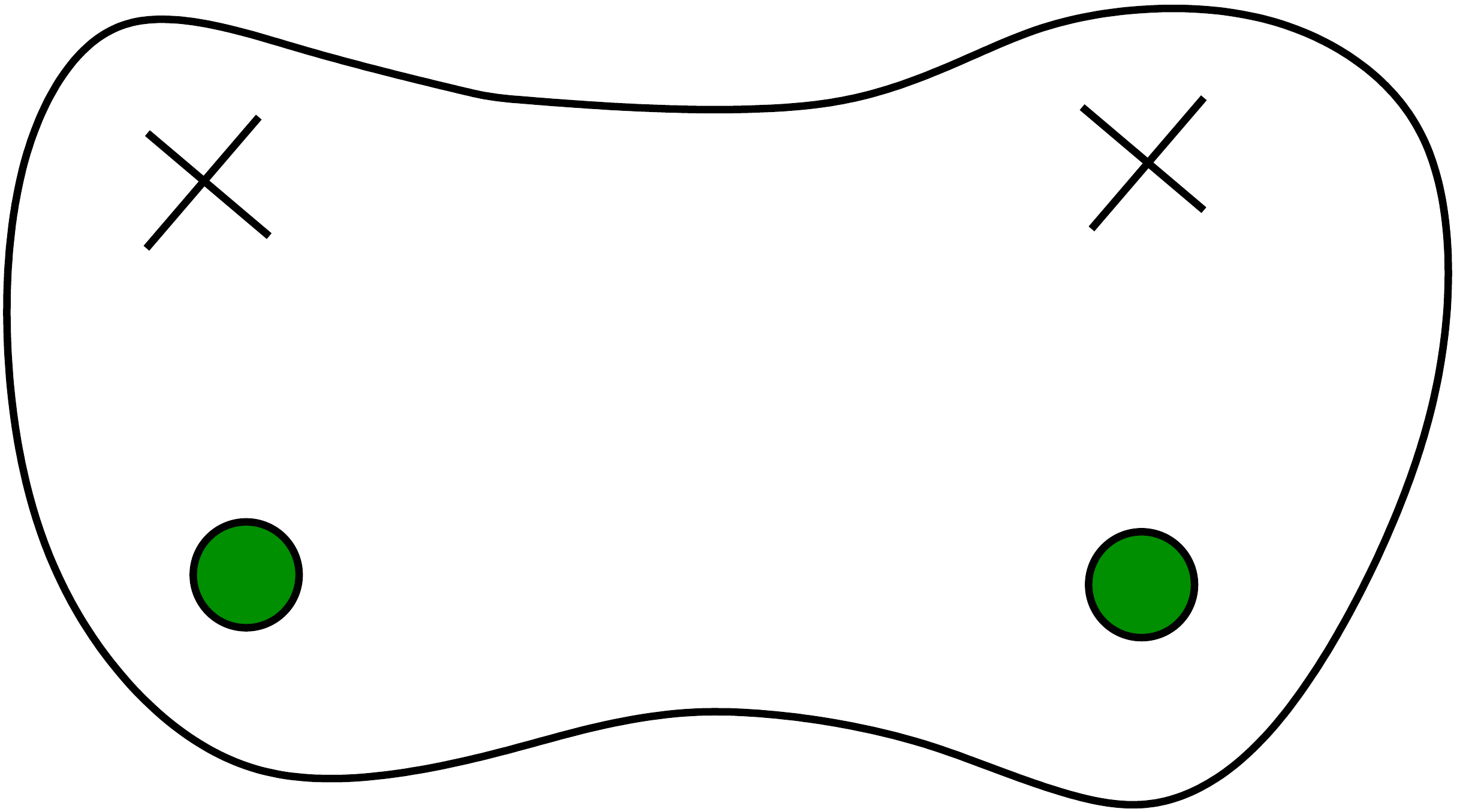}
\end{tabular}
\end{center}
\caption{ The sphere with two maximal punctures of same kind and two minimal punctures.
\label{sphereoneint}}
\end{figure}  
From the free trinions we can obtain the sphere with two maximal punctures, either both upper or both lower, and two minimal punctures.  The two theories are physically isomorphic and differ only by re-labeling of the flavor symmetries. The index of the theory with two lower punctures is,

\be\label{indsimplest}
&&{\cal I}_{{\mathbf u}\delta{\mathbf v}\alpha}(t,\beta,\gamma,p,q)=\left[(q;q)(p;p)\right]^2
\oint\frac{dz_1}{4\pi i z_1}
\oint\frac{dz_2}{4\pi i z_2}
\left[\frac{\Gamma_e(\frac{p\,q}t (\beta\gamma)^{\pm1}\,z_1^{\pm1}\,z_{2}^{\pm1})}{\Gamma_e(z_1^{\pm2})\Gamma_e(z_2^{\pm2})}
\right]\,\times\\
&&\qquad \left[
\Gamma_e(t^{\frac12}z_1^{\pm1}u_1^{\pm1}\beta\delta^{-1})
\Gamma_e(t^{\frac12}z_1^{\pm1}u_2^{\pm1}\gamma^{-1}\delta^{})
\Gamma_e(t^{\frac12}z_2^{\pm1}u_1^{\pm1}\gamma\delta^{})
\Gamma_e(t^{\frac12}z_2^{\pm1}u_2^{\pm1}\beta^{-1}\delta^{-1})
\right]\times\nonumber\\
&&\qquad\qquad
\left[
\Gamma_e(t^{\frac12}z_1^{\pm1}v_2^{\pm1}\beta^{-1}\alpha^{-1})
\Gamma_e(t^{\frac12}z_1^{\pm1}v_1^{\pm1}\gamma\alpha^{})
\Gamma_e(t^{\frac12}z_2^{\pm1}v_2^{\pm1}\gamma^{-1}\alpha^{})
\Gamma_e(t^{\frac12}z_2^{\pm1}v_1^{\pm1}\beta\alpha^{-1})
\right]\,.\nonumber
\ee Here the first line comes from the gauging with the  second and third lines coming from the contributions of the two trinions.
This index has the duality symmetry 

\be\label{symmindex}
{\cal I}_{{\mathbf u}\delta{\mathbf v}\alpha}=
{\cal I}_{{\mathbf u}\alpha{\mathbf v}\delta}=
{\cal I}_{{\mathbf v}\delta{\mathbf u}\alpha}
\,,
\ee which we should expect following our discussion in previous subsections. As discussed before this duality follows from a sequence of Seiberg and S-dualities with the relevant mathematical identities proven in~\cite{rains,Dolan:2008qi,Focco}.

 Using the eigenfunctions this index is given by

\be\label{indquatd}
{\cal I}_{{\mathbf u}\a{\mathbf v}\delta}=\sum_{\lambda} \psi_\lambda(u_1,u_2)\psi_\lambda(v_1,v_2) \phi_\lambda(\alpha)\phi_\lambda(\delta)\,.
\ee  
Gluing many trinions together we can obtain theories with arbitrary number of minimal puncture 
but with only two maximal punctures. Moreover if the number of minimal punctures is even the two maximal punctures are of the same color and if that number is odd they are of different color. To go beyond these constraints we will have to consider RG flows triggered by vacuum expectation values of the theories obtainable from our trinions and we are turning to that task next.

\begin{figure}
\begin{center}
\begin{tabular}{c}
\includegraphics[scale=0.51]{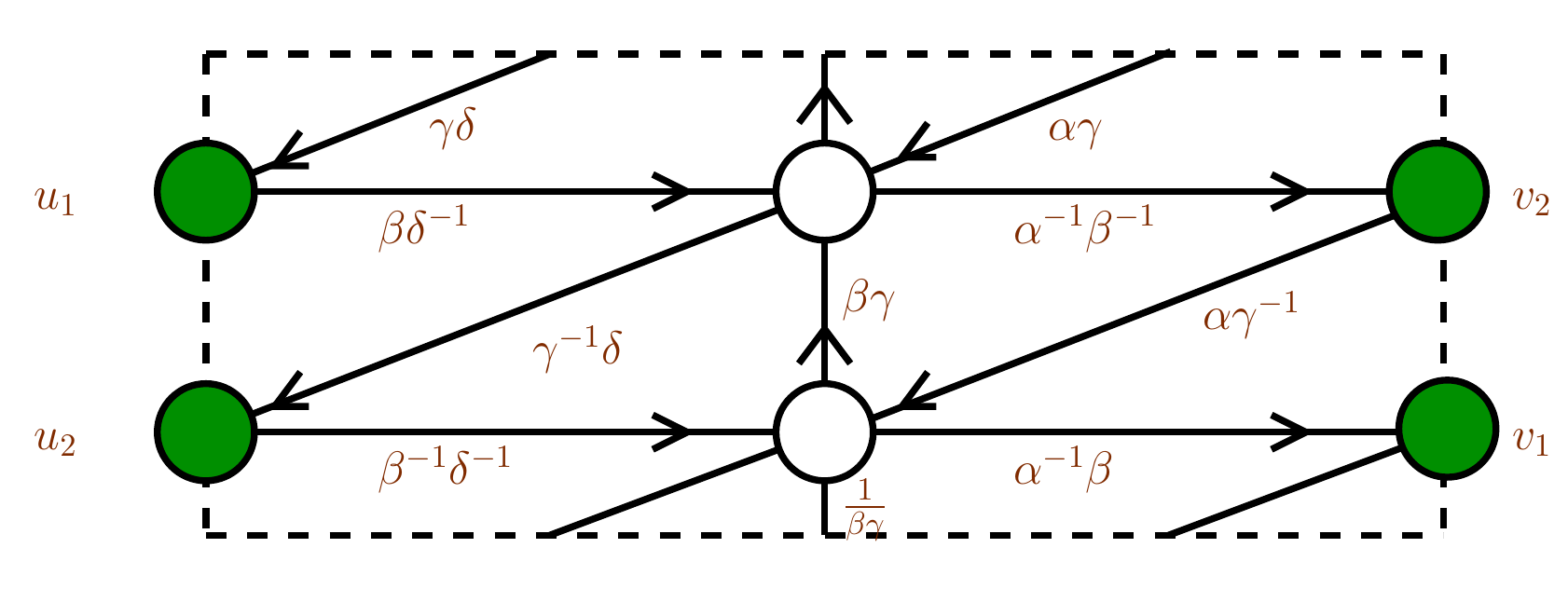}
\end{tabular}
\end{center}
\caption{ The sphere with two maximal and two minimal punctures with the fugacities associated to the 
matter fields. The white nodes correspond to gauged $SU(2)$ groups with the colored one to flavor $SU(2)$ groups. We suppressed the $pq$ and $t$ fugacities. 
\label{interfugs}}
\end{figure}

\

\section{Closing minimal punctures and discrete charges}\label{sect:closemin}

The six-dimensional SCFTs which we conjecture lie behind our story have global symmetries which 
we identify with the four-dimensional intrinsic global symmetries ${\frak G}_k$.  The two copies of $U(1)^k/U(1)$ actually enhance to $SU(k)$ and in four dimensional in general only the Cartan subgroup is not broken.
In the process of compactification from six to four dimensions, 
one has a choice of curvature for the line bundles associated to these global symmetries.
 In the standard class $\CS$ story,   there is a single global symmetry corresponding to $U(1)_t$, and one obtains $\CN=2$ theories for a specific choice of line bundles, and
more general $\CN=1$ theories for other choices. We will later comment on the orbifolded versions of these $\CN=1$ theories,
which should correspond to different choices of $U(1)_t$ curvature. In this section, we would like to assess the four-dimensional meaning of different choices of curvature for the 
$U(1)_\beta$ and $U(1)_\gamma$ bundles.  

In order to do so, we look at the possibility of ``closing'' a minimal puncture, by giving a vev to a chiral operator charged 
under the $U(1)_\alpha$ global symmetry associated to the minimal puncture.
An assumption we make is that such vevs leave us inside the class of theories we are discussing, that is
the theory in the IR can be associated to a certain Riemann surface.  
An obvious choice for the operator to receive a vev is one of  the (anti)baryons
 built from the $N^2$ blocks of the
corresponding free trinion. There are $2k$ such operators, $Q_i^N$ and ${\widetilde Q}_i^N$, and we could be giving a vev to any of them. 

We can do so for any of the minimal punctures. Indeed, the S-duality properties we expect from our theories 
indicate that we should get the same result, up to a relabeling of minimal puncture fugacities, by turning on (anti)baryon vevs in different free trinions.
For our purpose, it is particularly instructive to pick the duality frame where the free trinion we are working on is glued to 
a maximal puncture of some other generic theory $T$ to give a theory $T'$ with an extra minimal puncture and a new maximal puncture.  
The analysis for a free trinion glued to two other theories can be done in a similar manner. 

It is straightforward to see the effect of the vev. 
First of all, the vev Higgses the $SU(N)$ gauge group coupled to the (anti)quarks and identifies it with the $\widetilde{SU}(N)$ flavor group of the (anti)quarks.
The vev converts the cubic superpotential coupling involving that set of (antiquarks)quarks to a mass term for one of the $\Phi$
fields and for one block of (quarks)anti-quarks for the nearby gauge node. We can integrate these away.

When giving a vev to the (anti)quark block, we need to re-define all the symmetry generators, including the R-symmetry one, 
by an appropriate multiple of the generator of $U(1)_\alpha$ in such a way that the (anti)quark block is not charged under any of the assignments
re-defined symmetries. It is easy to verify that the final theory we obtain after Higgsing has the same 't Hooft anomalies for the re-defined symmetries as the initial theory,
as is obvious from anomaly matching. In particular, all the constraints we impose on 't Hooft anomalies of global symmetries associated to the 
punctures we did not close remain true after the Higgsing. 

The resulting theory has three fewer marginal couplings: we lost two superpotential couplings involving the lifted $\Phi$ field 
and one gauge coupling. We also lost three blocks of fields and one global symmetry. Thus we expect to have lost an exactly marginal deformation 
parameter. This is consistent with the intuition that the minimal puncture is gone. 

The resulting theory depends on the choice of which $2k$ (anti)baryons got a vev, and it is not equivalent to the theory $T$.
Indeed, all the $2k$ resulting theories have the same ``punctures'', which differ from the punctures of $T$ 
just by the color of the maximal puncture the original free trinion was glued to. 
Thus we obtained $2k$ potential new class $\CS_k$ theories and discovered that class $\CS_k$ theories need to be labelled by extra data 
besides the choice of Riemann surface and punctures. 

We aim to identify the extra data with a choice of curvature for the $U(1)_\beta \times U(1)_\gamma$ global symmetries 
of the underlying six-dimensional theory. Tentatively, we would say that closing a minimal puncture by a vev of a baryon charged under 
$U(1)_{\beta_i}$ or an anti-baryon charged under $U(1)_{\gamma_i}$ adds a unit of curvature for the corresponding six-dimensional global symmetry. 

If this idea is correct, adding one unit of curvature for each $U(1)_{\beta_\ell}$ should be the same thing as not adding any, as the $U(1)_{\beta_\ell}$ are identified with the Cartan of a 6d $SU(k)$ global symmetry. The same should be true when adding one unit of curvature for each $U(1)_{\gamma_\ell}$. We will see that this is indeed the case, 
as long as we refine slightly our notion of ``closing a minimal puncture'' by adding to the theory some gauge-neutral chiral multiplets 
coupled linearly to the surviving baryons (if we turned on a baryon vev) or anti-baryons (if we turned on an anti-baryon vev) in the free trinion associated to the minimal puncture we closed.  

As we will often have to add gauge-neutral chiral fields with linear couplings to some given chiral operator, it is useful to introduce the notion of 
``flipping'' a chiral operator $O$: an operation which maps a theory with a chiral operator $O$ to a new theory with an extra chiral multiplet $\phi$ coupled to 
$O$ by a superpotential $\phi O$. The new theory lacks the operator $O$, but usually has a new chiral operator $\phi$ with opposite charges to $O$. 
\

Let us now specialize to the $k=2$ case and still hold $N$ general while considering closing two minimal punctures. We consider a duality frame where the two minimal punctures reside in two free trinions glued to each other and also both of them glued to some general models. That is the two minimal punctures reside on a ``tube'' connecting otherwise generic Riemann surfaces. 
If we give vevs to baryons of fugacities  $(\alpha_1^{-1} \sqrt{t} \beta)^N$ and $(\alpha_2^{-1} \sqrt{t} \beta^{-1})^N$, we end up removing most of the chiral fields in the 
two corresponding free trinions. The $SU(N)_{1}$ gauge groups to the left of the two trinions, in the middle, and to the right are all Higgsed to a common $SU(N)_{1}$. 
The other $SU(N)_{2}$ gauge group which glues the two trinions together survives, but most of the fields charged under it are lifted by mass terms:
only bifundamentals connecting it to the left and right $SU(N)_{2}$ gauge groups survive. Thus we have a node with $N_f = N$ flavors. 
The low energy description of such a node is well-known: a set of mesons and baryons constrained by the equation,

\begin{equation}
\det M - B \tilde B = \Lambda^{2N}\,.
\end{equation}

The theory has a vacuum where $M = \Lambda^2$ and $B = \tilde B = 0$, which would precisely Higgs the left and right 
 $SU(N)_{2}$ gauge groups and thus bring us back to the standard theory with two fewer minimal punctures, 
 except for the extra baryonic degrees of freedom $B, \tilde B$. Even at the level of the index, one finds 
 that the $N_f=N$ node effectively produces a delta function \cite{Spiridonov:2014cxa} of the fugacities of the left and right $SU(N)_{z_2}$ gauge groups,
 multiplied by the index of the baryons $B$, $\tilde B$ with fugacities $(\alpha_1^{-1} \sqrt{t} \beta^{-1})^N$  and $(\alpha_2^{-1} \sqrt{t} \beta)^N$ .
 
 We can modify our definition of how to close a minimal puncture, by 
 both turning on a vev for the baryon with, say, $(\alpha_1^{-1} \sqrt{t} \beta)^N$
 and a linear superpotential coupling between the other baryon 
 $B$ with fugacity $(\alpha_1^{-1} \sqrt{t} \beta^{-1})^N$ and a new gauge-neutral chiral field $b$. The superpotential forces us at the origin of the baryonic branch 
 of the $N_f=N$ node and insures the desired Higgsing. It also removes the undesired free baryons after the Higgsing.

\

We can now go back to general $k$. We consider a sequence of $k$ trinions and close the corresponding punctures by giving a vev to 
the baryon charged under $U(1)_{\beta_1}$ in the first trinion, $U(1)_{\beta_2}$ in the second trinion, {\it etc}, and flipping all other baryons charged under the 
$U(1)_{\beta_\ell}$. Because of S-duality, we could have picked any other permutation $\sigma$, turning on at the $\sigma(a)$-th trinion a vev for the baryon charged under 
$U(1)_{\beta_a}$. The order we chose simplify the analysis considerably, though.
 
 After integrating away the chiral fields which receive mass parameters after the vevs, precisely $k-1$ gauge nodes end up with $N_f=N$ flavors. 
 With the help from the superpotential couplings suppressing the corresponding baryons, the mesons for these gauge nodes get vevs, and 
 initiate another set of Higgsing and lifting of pairs of chiral multiplets. This leads to another set of $N_f=N$ nodes, etc. At the end of the RG flow cascade, 
 the $k$ punctures have been completely eliminated.  

\

Notice that using the gauging procedure  discussed in this paper there is a certain degree of correlation between the choice of punctures on a Riemann surface and the possible values for the  
discrete charges. Starting from a theory $T$ with a maximal puncture of some color and orientation, we can glue to it a chain of $k$ 
free trinions and then close the resulting $k$ minimal punctures in several different ways. This produces new theories with the same punctures as $T$ 
but different discrete charges. The discrete charges, though, will necessarily add to a multiple of $k$. 

\

\subsection{A $k=2$ example}
We can give some rather explicit examples of this construction for theories in class $\CS_2$. 
For example, we can start from our basic interacting theory $\tilde T_2$ built from two trinions, with two minimal punctures and maximal punctures of opposite orientation and color 1 and $-1\equiv 1$. 

If we give a diagonal vev to, say, the block of quarks of fugacity $\sqrt{t} \alpha^{-1} \beta$ in the second trinion, 
we are left with an $SU(N)$ gauge theory with $2N$ flavors: two blocks of $N$ quarks with flavour symmetries 
$SU(N)_2$ and $\tilde{SU}(N)_2$ and two blocks of $N$ anti-quarks with flavour symmetries 
$SU(N)_1$ and $\tilde{SU}(N)_1$. We also have neutral chirals in bi-fundamental representations of 
$\tilde{SU}(N)_1 \times \tilde{SU}(N)_2$ and $\tilde SU(N)_1 \times SU(N)_2$ coupled to the (anti)quarks by 
cubic superpotential couplings and an additional  decoupled set $\tilde{SU}(N)_1 \times SU(N)_1$ which 
will enter the definition of the ``mesons'' at the maximal punctures of color 1 and $-1$. Finally, we have the single neutral chiral field 
we use to flip the baryon made out of the quarks of fugacity $\beta^{-2}$, which are charged under $\tilde{SU}(N)_2$.

This is one of four theories which we can obtain by closing the same minimal puncture in $\tilde T_2$ in different ways. 
They all have one minimal puncture and two maximal of opposite orientation and the same color $1$, but different discrete charges. 
They will be important in defining surface defects in section \ref{sect:surface}.

We can also consider theories obtained from $\tilde T_2$ by closing both minimal punctures in such a way that the total discrete charge does not cancel out. 
This produces a variety of ``charged tubes'', {\it i.e.} theories which can be associated to a cylinder with two maximal punctures of the same type 
and some extra discrete charge. Such charged tubes can be glued to a maximal puncture of some other theory $T$ to shift the discrete charges of that theory 
without changing the type of punctures. We will discuss some examples in appendix \ref{app:eig}.

\subsection{The index avatar}\label{subsec:closemin:index}

Let us now translate the discussion above to the language of the index.  To study RG 
flows generated by vacuum expectation values at the level of the index one needs to study its analytical properties. Different poles of the index correspond to operators for which a vev can be turned on with residues being indices of the theories flown to in the IR~\cite{Gaiotto:2012xa}.

\begin{figure}[htbp]
\begin{center}
\begin{tabular}{c}
\includegraphics[scale=0.21]{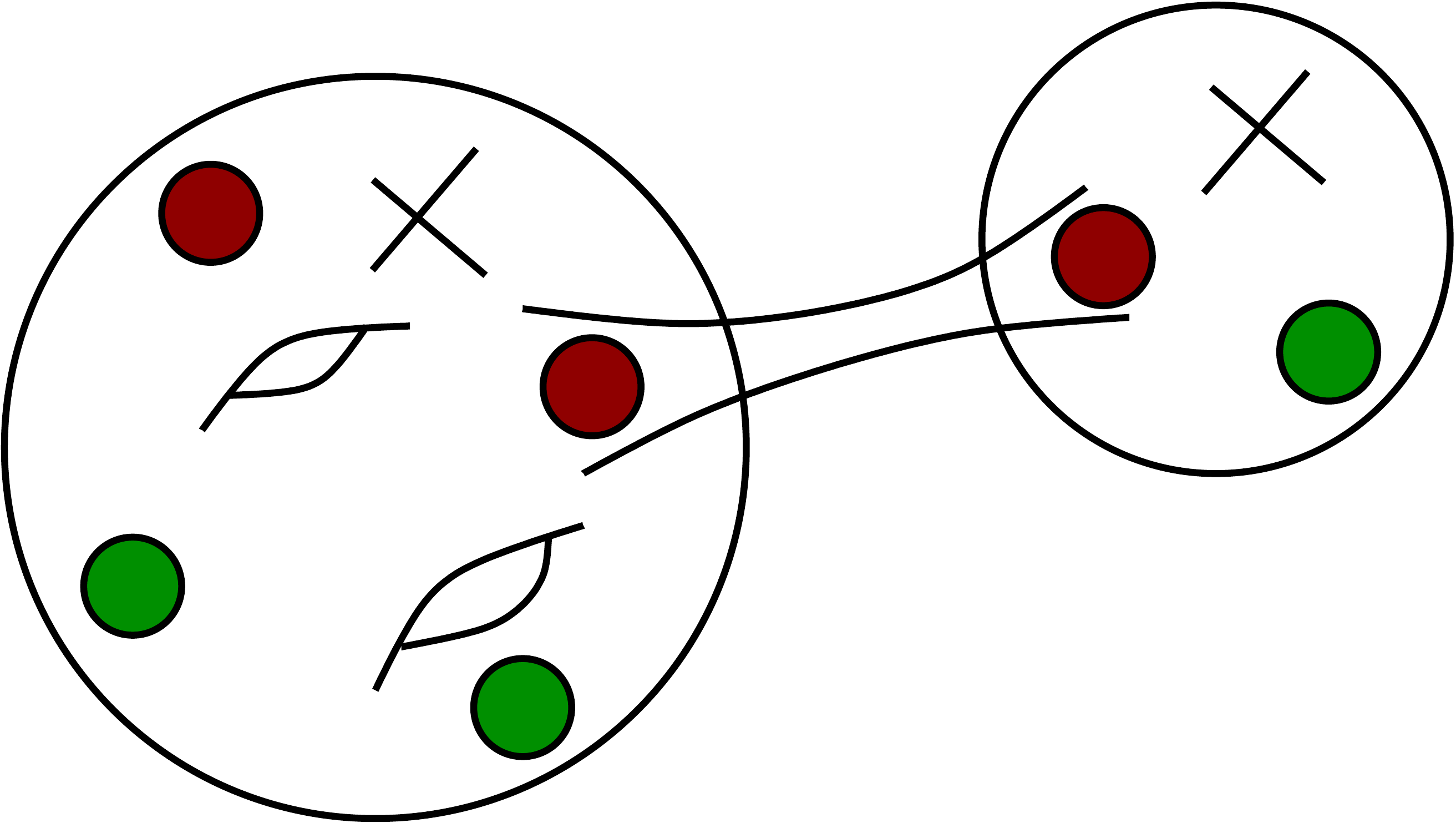}
\end{tabular}
\end{center}
\caption{Gluing the free trinion to a general theory.
\label{polesfiP}}
\end{figure}
\begin{figure}[htbp]
\begin{center}
\begin{tabular}{c}
\includegraphics[scale=0.59]{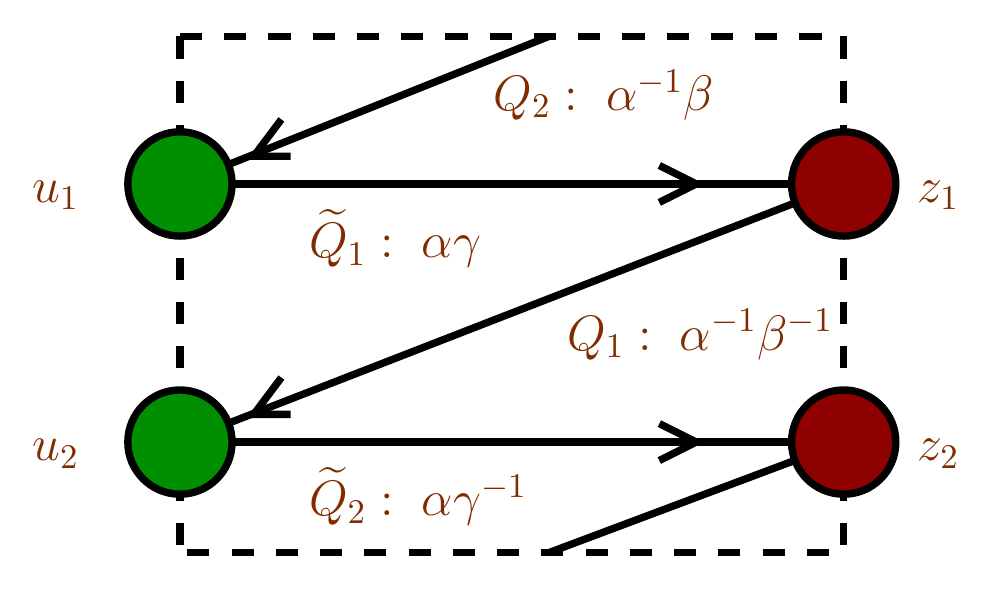}
\end{tabular}
\end{center}
\caption{ The glued sphere with fugacities.
\label{sphertoP}}
\end{figure}

  
To study analytical properties of the index
we first take an index of a generic theory corresponding to a Riemann surface and glue to it the free trinion. See Figure~\ref{polesfiP}. The index of such a theory is given by

\be\label{bubble}
&&{\cal I}_V^2\oint \frac{dz_1}{4\pi i z_1}
\oint \frac{dz_2}{4\pi i z_2}\, {\cal I}({\mathbf z})
\frac{\ge(\frac{pq}{t}(\b\g)^{\pm1}z_1^{\pm1}z_2^{\pm1})}
{\Gamma_e(z_1^{\pm2})\ge(z_2^{\pm2})}
\\
&&\qquad
\ge(t^{\frac12}z_1^{\pm1}u_1^{\pm1}\alpha\gamma )
\ge(t^{\frac12}z_1^{\pm1}u_2^{\pm1}\alpha^{-1} \beta^{-1})
\ge(t^{\frac12}z_2^{\pm1}u_1^{\pm1}\alpha^{-1}\beta )
\ge(t^{\frac12}z_2^{\pm1}u_2^{\pm1}\alpha \gamma^{-1})\,.\nonumber
\ee We depict in Figure~\ref{sphertoP} the fugacities associated to different fields of the free trinion.
 A class of poles in the index above occurs whenever 
the integration contours are pinched while varying the fugacities.
We look thus for pinchings of the integration contours. The poles inside and outside the integration
contour coming from the free trinion are located at

\be\label{poles}
in&:&\;\; z_1=t^{\frac12}u_1^{\pm1}\alpha\gamma q^n p^m\,,\qquad t^{\frac12}u_2^{\pm1}\alpha^{-1}\beta^{-1}q^np^m,\qquad\\
&&z_2=t^{\frac12}u_1^{\pm1}\alpha^{-1}\beta q^np^m,\;t^{\frac12}u_2^{\pm1}\alpha^{}\gamma^{-1}q^np^m, 
\nonumber\\
out&:&\;\; z_1=t^{-\frac12}u_1^{\pm1}\alpha^{-1}\gamma^{-1}q^{-n}p^{-m},\;t^{-\frac12}u_2^{\pm1}\alpha^{}\beta q^{-n}p^{-m},\nonumber\\
&&z_2=t^{-\frac12}u_1^{\pm1}\alpha^{}\beta^{-1}q^{-n}p^{-m},
\;t^{-\frac12}u_2^{\pm1}\alpha^{-1}\gamma q^{-n}p^{-m}\,.\nonumber
\ee 

\

\noindent When some of the {\it in} poles coincide with the {\it out} poles the integration contours are pinched and the index develops poles. Different poles correspond to vevs for some protected operators.
In this section we are interested in the case of the baryonic operators obtaining a vev.

 We consider the pole in $\alpha$, the fugacity associated to the $U(1)_\alpha$ symmetry of a minimal puncture, at $\alpha=t^{\frac12}\beta^{-1}$. 
This pole occurs when an operator with weight $t\beta^{-2}\alpha^{-2}$, the baryon $Q_1^2$, gets a vacuum expectation value. 
Giving a vev to such an operator Higgses the $z_1$ gauge group.
By turning on the vacuum expectation value we break the $U(1)_\alpha$ symmetry, {\it i.e.} we {\it close} the minimal puncture. There are similar poles at $\alpha=t^{\frac12}\beta$ corresponding to vev to baryon $Q_2^2$, and
$\alpha=t^{-\frac12}\gamma^{\pm1}$ corresponding to vevs for baryons $\widetilde Q_i^2$.  
Setting $\alpha=t^{\frac12}\beta^{-1}$ the $z_1$ integral is pinched at $z_1=u_2^{\pm1}$ and the residue of the index becomes

\be\label{bubbleB}
&&
{\cal I}_V\ge(tu_2^{\pm1}u_1^{\pm1}\beta^{-1}\gamma )
\oint \frac{dz_2}{4\pi i z_2}\, {\cal I}(\{u_2,z_2\})\,
\frac{
\ge(\frac{pq}t (\beta\gamma)^{-1}\,u_2^{\pm1}z_{2}^{\pm1})\ge(\beta^2z_2^{\pm1}u_1^{\pm1} )}{\ge(z_2^{\pm2})}
\,.
\ee Here and in what follows by residue we more precisely mean the following operation,

\be
2 {\cal I}_V Res_{{\mathbf u}\to {\mathbf u}^*} \frac1{u_1u_2}{\cal F}({\mathbf u})\qquad \to\qquad \widetilde{ Res}_{{\mathbf u}\to {\mathbf u}^*}{\cal F}({\mathbf u})\,.
\ee This operation is natural as it removes the decoupled free chiral associated to the Goldstone boson
index of which is ${\cal I}_V^{-1}$. The factor of $2$ appears since the $U(1)_\alpha$ charge of the baryonic operator
which is getting a vev is $2$.

 In particular,  if  the general theory of index ${\cal I}(\{u_2,z_2\})$ is just a free trinion, the residue is the index of  an $N_f=4$ $SU(2)$ SQCD
with additional singlet fields and superpotentials.  Let us denote the field corresponding to the prefactor in the integral $M$, which is in bifundamental of $SU(2)_{u_1}\times SU(2)_{u_2}$. The 
fields in the numerator of the integral are a quark $\Phi$ in fundamental of $SU(2)_{u_2}$ and a quark $Q'$ in the fundamental of $SU(2)_{u_1}$. In ${\cal I}(\{u_2,z_2\})$ we have the contribution of 
additional four quarks, ${\cal Q}_1$ and $\widetilde {\cal Q}_1$, and four gauge singlets in a fundamental 
of $SU(2)_{u_2}$ which we denote by ${\cal M}_{1,2}$. The superpotential then is,

\be
\Phi M Q'+\Phi {\cal Q}_1 {\cal M}_1\,,
\ee  with ${\cal M}_2$ being free fields. This theory enjoys an action of large duality group~\cite{Seiberg:1994pq,csakiterningshmaltzskiba,Intriligator:1995ne,Spiridonov:2008zr,Dimofte:2012pd}.

\begin{figure}[htbp]
\begin{center}
\begin{tabular}{c}
\includegraphics[scale=0.21]{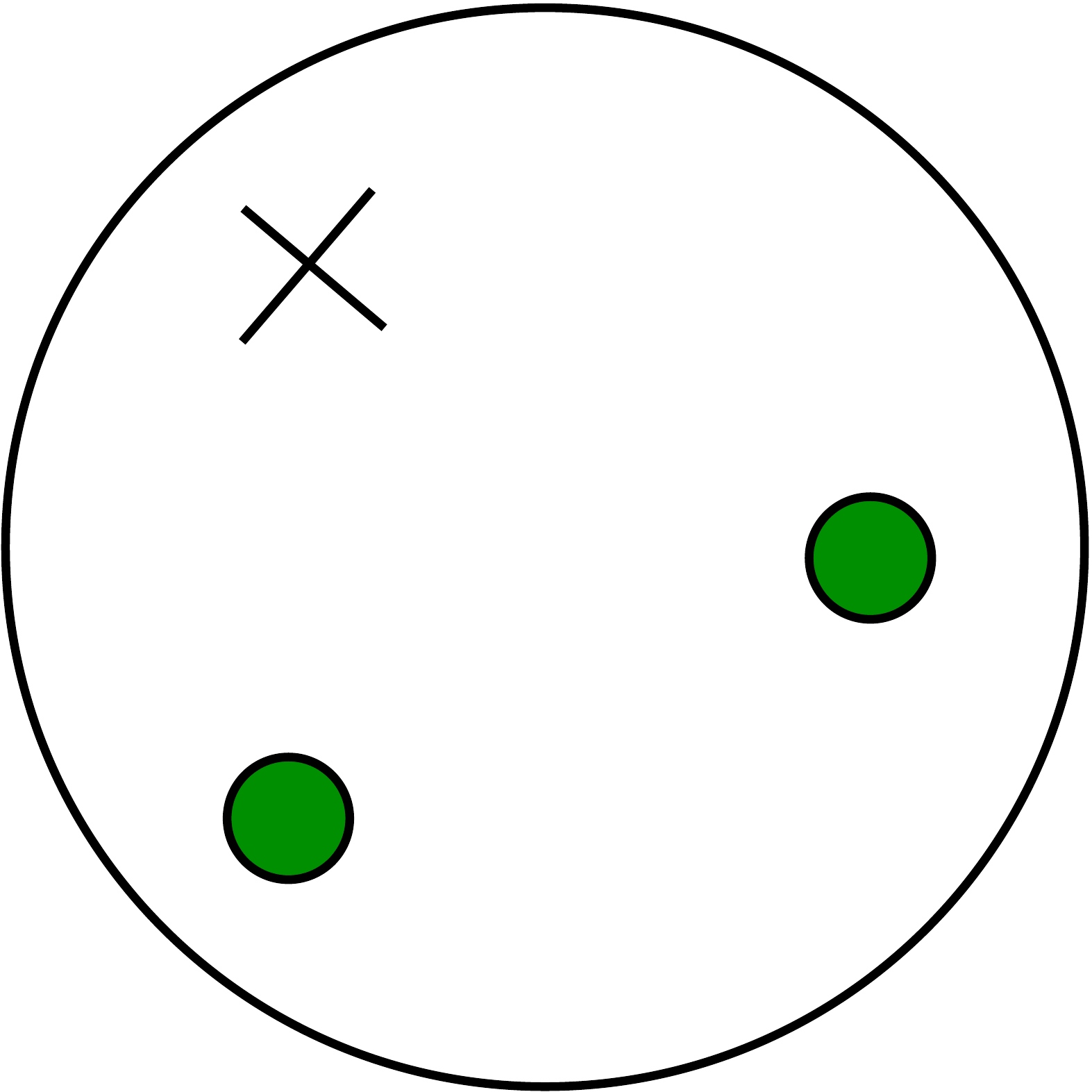}
\end{tabular}
\end{center}
\caption{The sphere with two maximal punctures of the same kind and one minimal puncture.
This corresponds to $N_f=4$ $SU(2)$ SQCD with some singlet fields and a superpotential. 
\label{sphereinP}}
\end{figure}

Under our assumptions that the RG flows generated by vacuum expectation values should leave us in our class of theories, the $N_f=4$ $SU(2)$ gauge theory should be associated to a Riemann surface. In fact it can be only associated to a sphere with one minimal puncture and two maximal punctures of the same type. This is a new, interacting, trinion we discover in our bootstrap procedure. Note that here the two 
maximal punctures are of same color and thus the theory should be invariant under exchanging the two factors of the associated flavor symmetries.  This new trinion is depicted in Figure~\ref{sphereinP}.
For general $k$ such a trinion can be obtained by gluing together $k$ free trinions and closing $k-1$ minimal punctures in certain way.
	
\

Let us study what this residue teaches us about the  functions $\psi_\lambda$, $\widetilde \psi_\lambda$ and $\phi_\lambda$. 
As we take the residue for a minimal puncture fugacity and flip the other (anti)baryon, the wave function  $\phi_{\lambda}(\alpha)$ in the sum
is replaced by the insertion of certain functions of the intrinsic fugacities, {\it i.e. }

\begin{align}\label{defcpm}
C^{(\b,\pm)}_\lambda &\equiv \ge(pq\b^{\pm 4}) \widetilde{Res}_{\alpha\to t^{\frac12}\b^{\pm1}} \phi_{\lambda}(\alpha)\,,\cr
C^{(\g,\pm)}_\lambda &\equiv \ge(pq\g^{\mp 4}) \widetilde{Res}_{\alpha\to t^{-\frac12}\g^{\pm1}} \phi_{\lambda}(\alpha) \,,
\end{align}
which we can interpret as the contribution to the index sum of a unit of positive or negative 
discrete charge for $U(1)_\beta$ or $U(1)_\gamma$. Notice that it must be true that

\be\label{condone}
C^{(\b,+)}_\lambda C^{(\b,-)}_\lambda =1\,, \qquad \qquad C^{(\g,+)}_\lambda C^{(\g,-)}_\lambda =1\,,
\ee since as we discussed closing two minimal punctures by giving vevs to the two different types of (anti)baryons leaves behind no discrete charge.
Applying the residue prescription to a trinion glued to a maximal puncture, we can get a neat 
integral relation between wavefunctions of different color:

\be\label{resphi}
&&C^{(\b,-)}_\lambda \psi_{\lambda}(u_1,u_2)=\\
&&\qquad {\cal I}_V\ge(pq\b^{-4})\ge(tu_2^{\pm1}u_1^{\pm1}\beta^{-1}\gamma)
\oint \frac{dz_2}{4\pi i z_2}\, \widetilde \psi_\lambda(z_2,u_2)\,
\frac{
\ge(\frac{pq}t(\beta\gamma)^{-1}\,u_2^{\pm1}z_{2}^{\pm1})}{\ge(z_2^{\pm2})
}
\ge(\beta^2z_2^{\pm1}u_1^{\pm1} )\,.\nonumber
\ee
 This relation is actually known as an elliptic Fourier transform~\cite{spirinv}, and can be inverted by a second elliptic Fourier transform. The invertibility of the elliptic Fourier transform, though, is precisely the index avatar of the 
Seiberg duality relation for an $SU(N)$ gauge node with $N$ flavors and flipped baryons, which we 
used to show how opposite discrete charges cancel out. Thus the elliptic inversion formula gives us the same result as 
directly taking a residue of a free trinion glued to a $\phi_\lambda$ wave function:

\be\label{resphiInv}
&&C^{(\b,+)}_\lambda \widetilde \psi_{\lambda}(u_1,u_2)=\\
&&\qquad {\cal I}_V\ge(pq\b^{4})\ge(tu_2^{\pm1}u_1^{\pm1}\beta\gamma )
\oint \frac{dz_2}{4\pi i z_2}\, \psi_\lambda(z_2,u_2)\,
\frac{
\ge(\frac{pq}t\beta^{}\gamma^{-1}\,u_2^{\pm1}z_{2}^{\pm1})}{\ge(z_2^{\pm2})}
\ge(\beta^{-2}z_2^{\pm1}u_1^{\pm1})\,.\nonumber
\ee 

Computing residues of the basic four punctured sphere and removing the appropriate 
singlets we thus get new trinions with indices,

\be\label{inttrins}
{\cal I}^{(\b,\pm)}= \sum_\lambda C_\lambda^{(\b,\pm)} \psi_\lambda({\mathbf u}) \psi_\lambda ({\mathbf v}) \phi_\lambda (\a)\,.
\ee
This is the $SU(2)$ $N_f=4$ SQCD with singlets and superpotential we discussed above. It will be very useful when we derive the difference equations satisfied by $\psi_\lambda$. 

We can consider closing the minimal puncture in the new interacting  trinion. The theories obtained in this way would correspond to spheres with two maximal punctures of same color and with non vanishing discrete charges. We will discuss briefly these constructions in appendix \ref{app:eig}.

\

\section{Closing maximal punctures}\label{sect:closemaxi}

Our next aim is to give evidence for existence of theories corresponding to spheres with maximal punctures only. To do so we will study RG flows triggered by the vev of chiral mesonic 
operators which are charged  
only under symmetries associated to a single  maximal puncture and the intrinsic symmetries. 
To have a concrete example, to which we will refer in the discussion below, the generic theory glued to a free trinion in the previous section can be  taken to be a sphere with two maximal and many minimal puncture, but the discussion is completely generic and example independent.

\subsection*{$A_1$ $k=2$}

The analysis of a general case is a bit cumbersome so we choose to start our discussion with $A_1$ and $k=2$ and gradually crank up these parameters. 
In the class $\CS$ theories of type $A_1$, maximal punctures turn out to be equivalent to minimal 
punctures, though they appear different in the brane construction, essentially because mesons and baryons are on the same footing in linear quivers of $SU(2)$ gauge groups. 

For $k=2$ maximal and minimal punctures are clearly different. The simplest choice of chiral operators charged under the global symmetries of a maximal puncture are 
the meson operators $Q_i\widetilde Q_i$ with  fugacities $u_1^{\pm1} u_2^{\pm1} t \left(\frac\beta\gamma\right)^{\pm1}$. 
We refer to fugacities and fields as depicted in Figure \ref{sphertoP}. Turning on vevs for a single meson operator breaks the $SU(2)_{u_1}\times SU(2)_{u_2}$ flavor symmetry down to an $U(1)_\delta$ subgroup. Thus we may hope it will result in a minimal puncture. 

Without loss of generality we can focus at first on giving a vev to the operator $Q_1\widetilde Q_1$, with fugacity $(u_1u_2)^{-1} t \frac\gamma\beta$.
More specifically, we give a vev to the component of $Q_1$ with gauge charge $1$ under the $SU(2)_1$ and $\widetilde Q_1$ with gauge charge $-1$. 
When turning on these vevs, we need to re-define the intrinsic global symmetries and define $U(1)_\delta$ by appropriate combinations of the 
old global and gauge symmetries, in such a way that the fields getting a vev are neutral under the new global symmetries. 

At the level of fugacities, this is accomplished by setting the the $SU(2)_{u_1}\times SU(2)_{u_2}$ fugacities to $(u_1,u_2)=(t^{\frac12}\frac\gamma\delta,t^{\frac12}\frac\delta\beta)$ with $\delta$ being a fugacity for $U(1)_\delta$ and the gauge fugacity for $SU(2)_1$ to $z_1=\a\delta$. 

As we turn on the vev for these chiral fields, some other fields are lifted by the cubic superpotentials, which become mass terms. 
The $SU(2)_{z_1}$ gauge field is Higgsed and only $SU(2)_{z_2}$ is left to glue the other free trinions to the one we triggered the mesonic vev in.  

The crucial observation is that the surviving gauge groups only interact with chiral fields which have the same charge under 
$U(1)_\alpha$ and $U(1)_\delta$. This is obvious for the fields in the general theory we glued to the free trinion, which are only charged under the diagonal combination of the two, 
{\it i.e.} have fugacities depending only on $z_1 = \alpha \delta$.
Among the chiral fields coupled to $SU(2)_{z_2}$, the only ones which have different $U(1)_\alpha$ and $U(1)_\delta$ charges have fugacities 
$\frac\b\g z_2^{\pm1}(\frac{\delta}{\a} )^{\pm1}$ and are exchanged by permuting $\alpha$ and $\delta$. 
Other surviving fields have fugacities $\frac{pq}t(\alpha\delta\beta\gamma)^{\pm1}z_{2}^{\pm1}$.

The surviving chiral fields which are not charged under $SU(2)_{z_2}$ have fugacities 
$t\alpha^2\gamma^2$, $t\delta^{-2}\g^2$, $t\alpha^{-2}\beta^{-2}$, and $t\delta^{2}\beta^{-2}$. In order to find a complete symmetry between $U(1)_\alpha$ and $U(1)_\delta$ 
we need to remove (``flip'') the fields with fugacity $t\delta^{2}\beta^{-2}$ and $ t\delta^{-2}\g^2$,
by adding new fields of fugacity $p q(t\delta^{2}\beta^{-2})^{-1}$ and $p q/( t\delta^{-2}\g^2)$ with quadratic superpotential couplings, 
and add chiral fields with fugacity $t\delta^{-2}\beta^{-2}$, $t \delta^2\gamma^2$ with appropriate superpotential couplings. 

Thus we find that by giving vev to a meson and adding some extra neutral chirals linearly coupled to 
chiral operators of the original theory we arrive to a theory which has an extra explicit symmetry, permuting the 
unbroken $\delta$ fugacity with (any)one of the minimal punctures fugacities. 
In other words, starting say from a linear quiver built by concatenating free trinions  we have produced a quiver gauge theory which can be rightfully labelled by 
a single maximal puncture and several minimal ones. An alternative perspective is that 
the surviving fields in the ``$\alpha$'' free trinion, together with the new chiral fields, 
define a ``quiver tail'' which can be appended to a negative maximal puncture with color $1$ 
in order to convert it to two minimal punctures of fugacities $\alpha$ and $\delta$. 

\

Notice that we could have obtained a similar result starting with the second set of mesons at the original maximal puncture. 
That would have imposed fugacities, say, $z_2 = \alpha \delta'$, $(u_1,u_2)=(t^{\frac12}\b\delta',t^{\frac12}\frac1{\g\delta'})$ pole corresponding to vev for $Q_2\widetilde Q_2$. 
This would have produced an a priori different way to reduce the maximal puncture to a minimal one.

It turns out that we can ascribe the difference between the two possible ways to map maximal to minimal to a difference  
in discrete charges. Indeed, we can probe the difference between these two choices further by closing
the newly-created minimal puncture by giving vev to baryonic operators of fugacity $\delta^2 t \g^{\pm 2}$ 
or $t\frac1{{\delta'}^2} \b^{\pm 2}$. 

At the level of fugacities, if we set, say, $\delta = t^{-\frac12} \g$ in the relations 
$(u_1,u_2)=(t^{\frac12}\frac\gamma\delta,t^{\frac12}\frac\delta\beta)=(t,\frac\g\b)$ for the first type of maximal to minimal reduction
we obtain the same result as if we were setting $\delta' = \sqrt{t}\b^{-1}$ in the same type of maximal to minimal reduction $(u_1,u_2)=(t^{\frac12}\b\delta',t^{\frac12}\frac1{\g\delta'})=(t,\frac\b\g)$ due to Weyl symmetry of $u_2$. Similarly, setting $\delta = t^{-\frac12}\b$ in the first reduction is equivalent to setting $\delta' = \sqrt{t} \g^{-1}$ in the second one. 
A detailed analysis of the reduction procedure shows that these pairs of ways to completely close the maximal puncture are indeed equivalent, 
even when keeping track of the neutral chiral fields we added in the process, as long as we remove an additional singlet chiral field with fugacity $\frac{\b^2}{\g^2}$ in the first way to 
reduce maximal to minimal and $\frac{\g^2}{\b^2}$ in the second way. 

Thus we conclude that reducing a maximal puncture to a minimal puncture by giving a vev to a meson with fugacity proportional to $\frac\gamma\beta$ 
or to a meson with fugacity proportional to $\frac\beta\gamma$ give class $\CS_2$ theories with discrete charges which differ by one unit of $U(1)_\beta$ curvature 
and one unit of $U(1)_\gamma$ curvature. 

As we have identified a ``quiver tail'' which can be attached to a maximal puncture to obtain two minimal punctures, it is natural to 
do the same step which in class $\CS$ leads to the definition of non-trivial trinion theories: we can conjecture the existence of 
SCFTs with one puncture of color $0$ and two of color $1$ (and appropriate choices of discrete charges), 
with the property that attaching a quiver tail to one puncture of color $1$ will produce our basic core theory built from 
two free trinions. As we have two different versions of the quiver tail, we seem to need at least two distinct SCFTs, with discrete charges 
differing by one unit of $U(1)_\beta$ curvature and one unit of $U(1)_\gamma$ curvature.

It is instructive to look a bit further to the combination of vevs which we expect to produce and close a minimal puncture starting from a maximal one. 
It corresponds to giving a vev to mesons with fugacities $t\frac\g\b u_1^{-1}u_2^{-1}$ and $t\frac\b\g u_1^{-1}u_2$. This implies a vev for
both chiral fields with fugacity $u_1^{-1}z_1^{-1} \sqrt{t} \alpha \g$ and $u_2^{-1} z_1 \sqrt{t}\a^{-1}\b^{-1}$ and chiral fields with fugacities
$u_2 z_2^{-1} \sqrt{t} \g^{-1}\a$ and $u_1^{-1}z_2 \sqrt{t}\a^{-1} \b$. If we are working with a standard core theory built from a sequence of trinions, 
these vevs force us to turn on chiral fields in the next free trinion as well, 
because of the cubic superpotential couplings of the second and third fields to the $\Phi$ field of fugacity $\frac{p q}t \b\g z_1^{-1} z_2$: 
the extremum equations for $\Phi$ require us to turn on a vev for the meson of fugacity $t \frac{1}{\b\g} z_1 z_2^{-1}$ in the next free trinion. 
Looking at the theory in detail, we find that the original free trinion has been completely eliminated, while 
the next free trinion is precisely subject to the vev which reduces the maximal puncture to minimal. 
This is just another manifestation of the S-duality relations which 
permute the minimal punctures of the theory. 

\

Although we do not have an intrinsic way to compare discrete charges between theories with different types of punctures, 
we can use some symmetry considerations to set up some useful conventions. Let's declare that the core theories have 
zero discrete charges. We have several different ways to produce a theory with a single maximal puncture and several minimal ones:
a minimal puncture can be produces starting from a maximal puncture of either color, which can be reduced in two ways. 

We saw that the two ways of reducing a puncture of color $0$ differ by one unit of $U(1)_\beta$ curvature and one unit of $U(1)_\gamma$ curvature.
In a similar fashion, the two ways of reducing a puncture of color $0$ differ by one unit of $U(1)_\beta$ curvature and minus one unit of $U(1)_\gamma$ curvature.
Finally, we saw that reducing a puncture of color $0$ and then closing the resulting minimal puncture in a specific way 
is the same as reducing a puncture of color $1$ in a theory with one fewer minimal punctures. 

Thus it is natural to pick the following symmetric convention to compare the discrete charges of theories before and after 
reducing maximal punctures: giving a vev to a meson with fugacity proportional to $\gamma^{\pm1} \beta^{\mp1}$ in a color $0$ 
puncture adds $\mp1/2$ unit of $\beta$ charge and $\pm1/2$ of $\gamma$ charge, while giving a vev to a meson with fugacity proportional to $\gamma^{\pm1} \beta^{\pm1}$ in a color $1$ 
puncture adds $\pm1/2$ unit of $\beta$ charge and $\pm1/2$ of $\gamma$ charge.

\subsection*{$A_1$ general $k$}

The general $k$ case for $N=2$ is not much harder to analyze, except that now both orientation and color matter. 
We will proceed by analogy with our $k=2$ analysis, and give a prescription to reduce a maximal puncture
to minimal.

\begin{figure}[htbp]
\begin{center}
\begin{tabular}{c}
\includegraphics[scale=0.39]{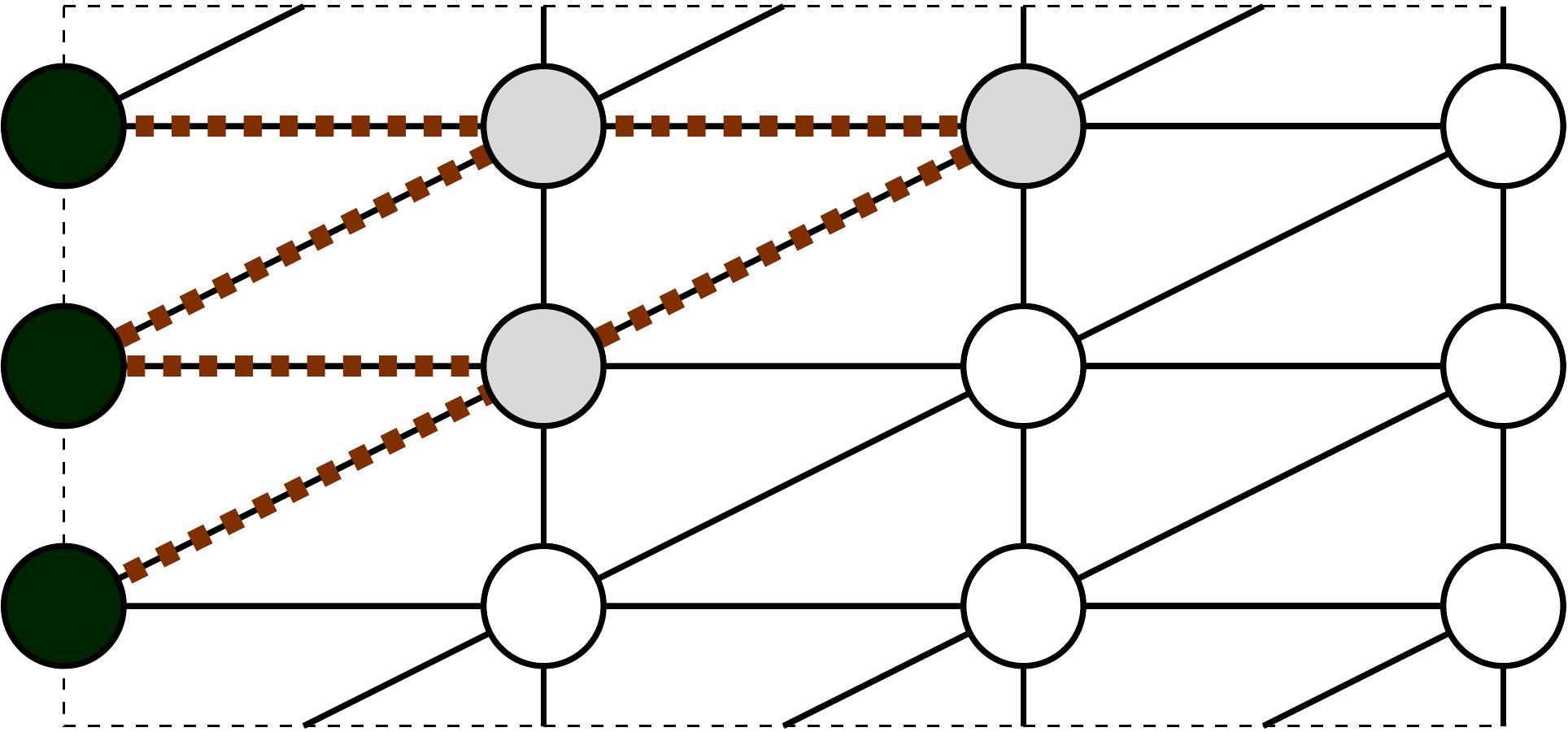}
\end{tabular}
\end{center}
\caption{Example of closing a maximal puncture in a linear quiver with $k=3$. Giving a vev to the two mesons denoted by dashed brown lines in the left column the $SU(2)^3$ group denoted by filled green dots is broken down to $U(1)$.  These vevs also Higgs two of the three gauge groups in the second column. The vevs for the mesons through superpotential interactions generate vevs for a meson in the second column, again denoted by dashed brown lines. This meson Higgses another $SU(2)$ gauge group. We end up with a ``triangle'' of unbroken gauge groups denoted by white dots. 
\label{closemanymax}}
\end{figure}

Our prescription will be to give a vev to mesons of fugacity 
$u_i u_{i+1}^{-1} t \beta_i \gamma_i^{-1}$. We can give a vev to up to $k-1$ of them, say for $i=1, \cdots, k-1$. 
Thus we give a vev to chiral fields of fugacity $u_i z_i^{-1} \sqrt{t} \beta_i \alpha^{-1}$ and 
$z_i u_{i+1}^{-1} \sqrt{t}\gamma_i^{-1}\a$ and Higgs $k-1$ $SU(2)$ groups, leaving only $SU(2)_k$. 
The chiral field vevs enter the cubic superpotential couplings of the $\Phi$ fields between $SU(2)_i$ and 
$SU(2)_{i+1}$ for $i=1, \cdots, k-2$. They force us to also give a vev to the mesons in the nearby trinion,  with fugacities $z_i z_{i+1}^{-1} t\beta_{i+1}\gamma_i^{-1}$ for $i=1, \cdots, k-2$.
In turn, these fugacities Higgs $k-2$ $SU(2)_i$ gauge fields in the next column, for $i=1, \cdots, k-2$,
but also force us to turn on $k-3$ mesons at the next trinion, {\it etcetera}.
The result is that one end of our quiver of $(k-1)k$ $SU(2)$ gauge groups is modified to take a triangular shape,  with columns of $k-1$, $k-2$, $\cdots$, $1$ $SU(2)$ gauge groups.  See Figure \ref{closemanymax} for illustration.

We can parameterize the $z_i$ and $u_i$ in terms of a parameter $\delta$, so that in general the 
$z_i$ are proportional to $(\alpha \delta)^{-1}$ and the $u_i$ proportional to $\delta^{-1}$. 
Thus in order to identify a symmetry exchanging $\alpha$ and $\delta$ we need first to make sure that the 
fields charged under $SU(2)_k$ satisfy such a symmetry. Then we can try to impose the symmetry on neutral fields  by adding extra neutral chirals. 
The relevant fields have fugacities $u_k^{\pm1} z_k^{\pm1} \sqrt{t} \beta_k \alpha^{-1}$ and 
$z_k^{\pm1} u_1^{\pm1} \sqrt{t}\gamma_k^{-1}\alpha$.
Half of these receive masses by the meson vevs, the other half are
$u_k^{-1} z_k^{\pm1} \sqrt{t} \beta_k \alpha^{-1}$ and $z_k^{\pm1} u_1 \sqrt{t}\gamma_k^{-1}\alpha$. They are symmetric if we set 
$\delta^{-2} = u_1 u_k \gamma_k^{-1} \beta_k^{-1}$. Notice that we have set $u_2 = u_1 t \beta_1\gamma_1^{-1}$, {\it etcetera}. Thus $u_1 u_k = u_1^2 t^{k-1} \prod_{i=1}^{k-1} \beta_i\gamma_i^{-1}$
and thus $\delta^{-1} = u_1 t^{(k-1)/2}\beta_k^{-1}$. 

Next we can focus on the lack of symmetry exchanging the $\a$ and $\delta$ symmetries of neutral fields. The trinion fields have fugacities $u_i^{\pm1} z_i^{\pm1} \sqrt{t} \beta_i \alpha^{-1}$ and 
$z_i^{\pm1} u_{i+1}^{\pm1} \sqrt{t}\gamma_i^{-1}\alpha$, but the ones which do not get vevs nor masses and are not eaten by the Higgs mechanism are
the ones with fugacities $u_i^{-1} z_i^{\pm1} \sqrt{t} \beta_i \alpha^{-1}$ and $z_i^{\pm1} u_{i+1} \sqrt{t}\gamma_i^{-1}\alpha$.
Half of the fields have fugacities proportional to $\alpha^2$, {\it i.e. }
$t \beta^2_i \alpha^{-2}$ and $t\gamma_i^{-2}\alpha^{2}$, and are remnants of the baryons. 
The other half has $\alpha$-independent fugacities $u_i^{-2}$ and $u_{i+1}^2$
proportional to $\delta^{\pm 2}$. 
If we remove them through linear couplings to new neutral chirals, 
and replace them with chirals of fugacities $t \beta^2_i \delta^{-2}$ and $t\gamma_i^{-2}\delta^{2}$ and appropriate 
superpotential couplings, we arrive at a theory symmetric in $\alpha$ and $\delta$. 
For the consistency of the picture with further closing the minimal puncture we might need to decouple additional singlet fields charged only under the intrinsic symmetry; we will not analyze this here.

\

\begin{figure}[htbp]
\begin{center}
\begin{tabular}{c}
\includegraphics[scale=0.39]{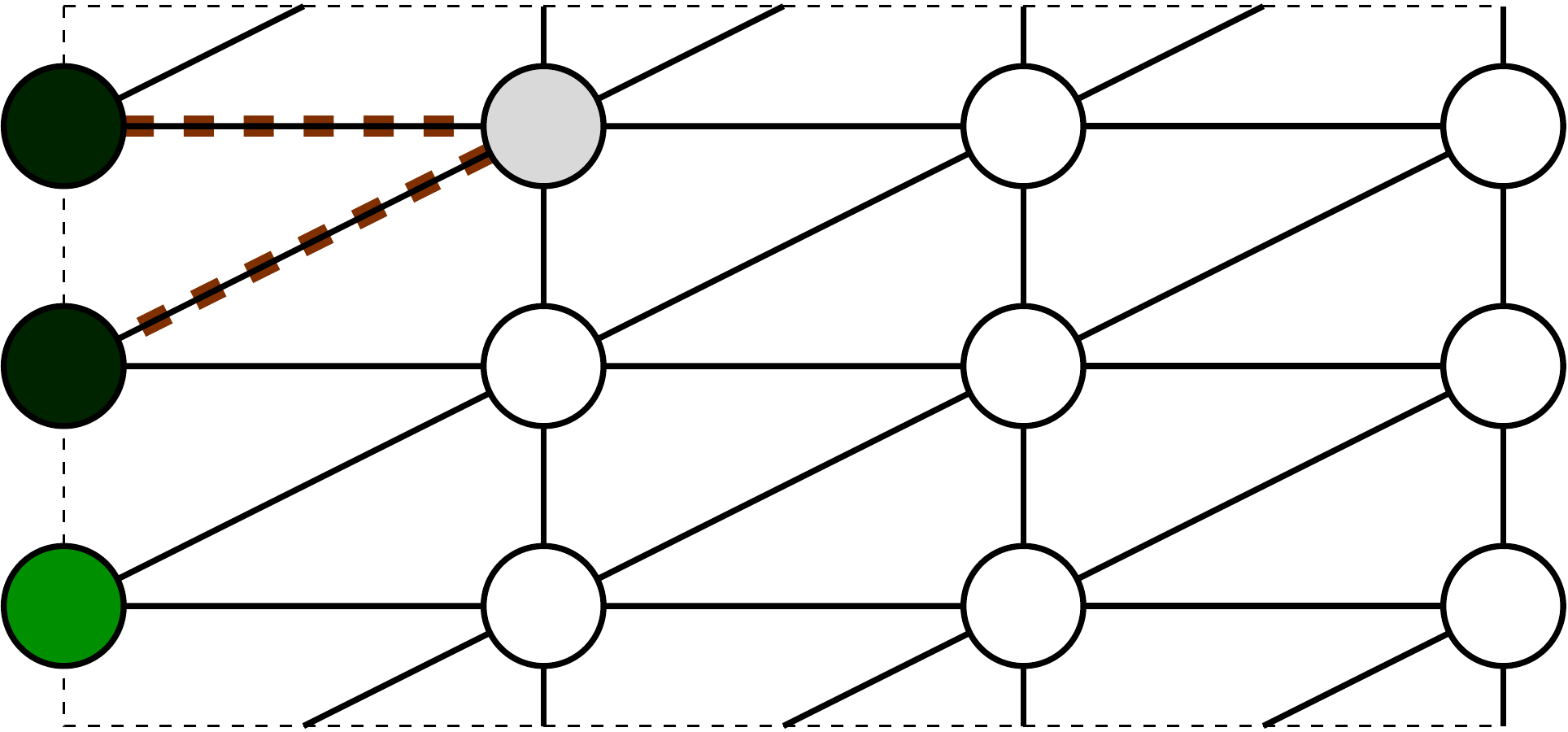}
\end{tabular}
\end{center}
\caption{Example of closing a maximal puncture in a linear quiver with $k=3$ down to a non-minimal one. We give a vacuum expectation value to a single meson. This breaks the flavor group to $SU(2)^2 \times U(1)$ and Higgses one of the gauge groups in the next column. This, next to maximal, puncture together with the maximal and minimal ones are the only ``regular'' punctures of the $A_1$ class ${\cal S}_3$. This puncture comes in several varieties depending on the color of the maximal puncture we start with and the particular meson we choose to trigger the flow.
\label{nexttomaxim}}
\end{figure}

Thus we learned how to convert a maximal puncture to a minimal puncture, in $k$ different ways. As for $k=2$, these different procedures 
leave one with different amounts of discrete charges on the surface.  For general $k$ we can consider giving vevs to different combinations of mesons closing a maximal puncture down to a puncture $\Lambda$ with symmetry $U(1)\subset G_\Lambda\subset SU(2)^k$. 
On the quiver such choices are classified by carving out multiple triangular wedges from the tail.  We will not embark on tail classification here though it is a very interesting problem to discuss. See Figure 
\ref{nexttomaxim} for illustration. We can call punctures accessible from maximal punctures by RG flows triggered by vevs of mesons ``regular'' punctures in analogy with class ${\cal S}$. There might be other types of punctures one would want to consider but that goes beyond the analysis of this paper. 

\

\subsection*{$A_{N-1}$ and general $k$}

Let us now analyze the reduction of maximal punctures in $A_{N-1}$ class ${\cal S}_{k}$ theories.
The basic idea is the same as in $A_1$ case but now we have many more mesons and thus the details are more involved. 
For general $A_{N-1}$ even in class $\CS$  there is a variety of possible punctures labelled by Young diagram \cite{Gaiotto:2009we} 
and to get from a maximal one to minimal one has to turn on vevs for many mesons.

We begin our most general $A_{N-1}$ class ${\cal S}_k$ discussion by prescribing a vev for mesons of fugacity 
$u^{(a)}_{i+1} (u^{(a)})_i^{-1} t \beta_i \gamma_i^{-1}$. We can give a vev to up to $(N-1)(k-1)$ of them, say for $i=1, \cdots, k-1$ and $a = 1, \cdots, N-1$. 
Thus we give a vev to chiral fields of fugacity $z^{(a)}_i (u^{(a)}_i)^{-1} \sqrt{t} \beta_i \alpha^{-1}$ and 
$u^{(a)}_{i+1} (z^{(a)}_i)^{-1} \sqrt{t} \gamma_i^{-1} \alpha$ and Higgs $k-1$ $SU(N)$ groups, leaving only $SU(N)_k$.
These vevs are not yet sufficient for our purposes, as they leave $N-1$ $U(1)$ symmetries unbroken. We will thus also 
turn on an extra set of $N-2$ mesons, with fugacities $u^{(a+1)}_{1}(u^{(a)}_k)^{-1} t \beta_k\gamma_k^{-1}$, with $a = 1, \cdots, N-2$.
Thus we give a vev to chiral fields of fugacity $z^{(a)}_k (u^{(a)}_k)^{-1} \sqrt{t} \beta_k \alpha^{-1}$ and 
$u^{(a+1)}_{1} (z^{(a)}_k)^{-1} \sqrt{t}\gamma_k^{-1} \alpha$ with $a = 1, \cdots, N-2$. These vevs Higgs $SU(N)_k$ to $SU(2)_k$. 

The vevs enter the cubic superpotentials and force us to turn on also vevs for certain mesons in the next free trinion.  Namely, we need mesons with fugacities $z^{(a)}_{i+1} (z^{(a)}_i)^{-1} t \beta_{i+1}\gamma_i^{-1}$ for $i=1, \cdots, k-2$ and $a = 1, \cdots, N-1$,
$z^{(a+1)}_1(z^{(a)}_k)^{-1} t \beta_1\gamma_k^{-1}$ and $z^{(a)}_k(z^{(a)}_{k-1})^{-1} t \beta_k\gamma_{k-1}^{-1}$ with $a = 1, \cdots, N-2$.
These vevs will Higgs the next column of $SU(N)_i$ gauge groups to nothing, except for the last two, Higgsed again to $SU(2)$.  

These meson vacuum expectation values are implemented by vacuum expectation values for chiral fields in the next trinion of fugacities 
$z^{(a)}_{i+1}(y^{(a)}_i)^{-1} \sqrt{t}\gamma_i^{-1}\a$ and $y^{(a)}_i(z^{(a)}_i)^{-1} \sqrt{t} \beta_{i+1} \a^{-1}$ for $i=1, \cdots, k-2$ and $a = 1, \cdots, N-1$,  as well as
$z^{(a+1)}_1(y^{(a)}_k)^{-1} \sqrt{t}\gamma_k^{-1}\a$, $y^{(a)}_k (z^{(a)}_k)^{-1} \sqrt{t} \beta_1 \a^{-1}$, 
$z^{(a)}_k (y^{(a)}_{k-1})^{-1} \sqrt{t} \gamma_{k-1}^{-1}\a$ and $y^{(a)}_{k-1} (z^{(a)}_{k-1})^{-1} \sqrt{t} \beta_k \a^{-1}$ with $a = 1, \cdots, N-2$.
Here the $y^{(a)}_i$ are the fugacities of the next row of gauge groups. 
These enforce vevs of mesons at the next trinion with fugacities 
$y^{(a)}_{i+1}(y^{(a)}_i)^{-1} t \beta_{i+2}\gamma_i^{-1}$ for $i=1, \cdots, k-3$ and $a = 1, \cdots, N-1$, 
$y^{(a+1)}_1 (y^{(a)}_k)^{-1} t \beta_{2}\gamma_k^{-1}$ and $y^{(a)}_k (y^{(a)})_{k-1}^{-1} t \beta_1\gamma_{k-1}^{-1}$ and $y^{(a)}_{k-1}(y^{(a)}_{k-2})^{-1} t \beta_k\gamma_{k-2}^{-1}$
for $a = 1, \cdots, N-2$, thus Higgsing all but three $SU(N)$s to $SU(2)$. As the vevs propagate along the quiver, the number of $SU(2)$ groups increases linearly. At some point $SU(3)$ groups appear, etc. 

We can parameterize the fixed $z^{(a)}_i$ and $u^{(a)}_i$ in terms of a parameter $\delta$, in such a way that 
$z^{(a)}_i$ are proportional to $\alpha \delta$ and $u^{(a)}_i$ are proportional to $\delta$. 
Thus in order to check for a symmetry between $\alpha$ and $\delta$ we only need to focus on the 
initial trinion. 

Few fields charged under the surviving $SU(2)$ survive Higgsing and do not get a mass:
$z^{(a)}_k(u^{(N)}_k)^{-1} \sqrt{t} \beta_k \alpha^{-1}$ and $u^{(N)}_{1}(z^{(a)}_k)^{-1} \sqrt{t}/\gamma_k\alpha$ for $a = N-1, N$. 
In terms of the $SU(2)$ fugacity $\tilde z$, such that $\tilde z^2 = z^{(N-1)}_k (z^{(N)}_k)^{-1}$, 
we can write $z^{(N-1)}_k = \tilde z \sqrt{z^{(N-1)}_k z^{(N)}_k}$ and $z^{(N)}_k = \tilde z^{-1} \sqrt{z^{(N-1)}_k z^{(N)}_k}$, 
and the flavor fugacities as $$(z^{(N-1)}_k z^{(N)}_k)^{\frac12}(u^{(N)}_k)^{-1} \sqrt{t} \beta_k \alpha^{-1} \quad \text{and} \quad \;u^{(N)}_{1}(z^{(N-1)}_k z^{(N)}_k)^{-\frac12} \sqrt{t}\gamma_k^{-1}\alpha.$$
The flavor fugacities have a ratio $ u^{(N)}_{1}u^{(N)}_k(z^{(N-1)}_k z^{(N)}_k\gamma_k\beta_k\a^{-2})^{-1}$ which we want to identify with $\alpha^N/\delta^N$.
Thus we set $\delta^{-N} =  u^{(N)}_{1}u^{(N)}_k(z^{(N-1)}_k z^{(N)}_k  \gamma_k\beta_k)^{-1} \alpha^{2-N}$. As $u^{(a)}_{i+1}= u^{(a)}_it^{-1}\beta_i^{-1} \gamma_i$, 
we have $u^{(a)}_{k}= u^{(a)}_1t^{1-k}\beta_k\gamma_k^{-1}$ and $u^{(N)}_{k}= u^{(N)}_1 t^{(N-1)(k-1)} (\beta_k\gamma_k^{-1})^{1-N}$.
Also, 
$$(u^{(N-1)}_ku^{(N)}_k)^{-1} = \prod z^{(a)}_k = (u^{(1)}_{1}u^{(N)}_{1})^{-1} (t^{-1}\g_k\a^{-1})^{2-N},$$
and thus $\delta^{-N} = u^{(N)}_1(u^{(1)}_{1})^{-1} t^{(N-1)(k-1)+\frac12(N-2)} \beta^{-N}_k$. 

With this choice of $\delta$, we have defined our global symmetries in such a way that $U(1)_\alpha$ and $U(1)_\delta$ act in the same way on chiral fields which carry gauge charge. 
We will have a bunch of gauge-neutral fields charged under both symmetries and by appropriate flips
the resulting theory can be made to be fully symmetric under exchanging the $U(1)_\a$ and $U(1)_\delta$ symmetries.

We have learned how to convert a maximal puncture into a minimal one. 

\

\

\subsection{The index avatar}\label{sect:closemaxi:index}

\

Let us discuss the above at the level of the index. As usual we specialize to the $A_1$ $k=2$ case.

The discussion above makes it clear that one should be able to produce $\phi_\lambda$, up to a normalization factor, by taking a residue of 
either $\psi_\lambda$ or $\widetilde \psi_\lambda$ at appropriate values of the fugacities. When we defined $\phi_\lambda$, we have normalized in such a way 
that the free trinion would have a simple expansion of the  form \eqref{indtrind}.
In a 2d TFT language, it would be more natural to introduce structure constants $C_\lambda$ and write the index associated to a Riemann surface of genus ${\frak g}$ with $n$ punctures $p_a$ and charges $q_i$ in the schematic form

\be
{\cal I}^{(p_a),(q_i)} = \sum_\lambda C_\lambda^{2{\frak g} -2 + n} \prod_i \left( C^i_\lambda \right)^{q_i} \prod_a \psi^{p_a}_\lambda\,.
\ee 
Thus if we use a convention where core theories have charge $0$, we should write $\phi_\lambda = C_\lambda \psi^m_\lambda$, with 
$\psi^m_\lambda$ being a properly normalized minimal puncture wave-function. 
Then with the symmetric charge assignments discussed above, we can write 

\be\label{phipsiRel}
&&\sqrt{C^{(\b,-)}_\lambda C^{(\g,+)}_\lambda} \psi^m_\lambda(\a)=\\
&&\qquad\quad\; \,\ge(pq\b^{-2}\g^{2})
\frac
{\ge(t\a^{-2}\b^{-2})\ge(t\a^2\g^{2})}
{\ge(t\a^{2}\b^{-2})\ge(t\a^{-2}\g^{2})}
\widetilde  {Res}_{(u_1,u_2)\to (t^{\frac12}\frac\gamma\a,\,t^{\frac12}\frac\a\b)}  \psi_\lambda(u_1,u_2)\,.\nonumber
\ee 
We can obtain three other similar relations involving the other way to reduce $\psi_\lambda$ and the two other ways 
to reduce $\widetilde \psi_\lambda$, with other prefectures of the form $\sqrt{C^{(\b,\pm)}_\lambda C^{(\g,\pm)}_\lambda}$. Note that \eqref{phipsiRel}  together with \eqref{defcpm} and \eqref{condone} 
completely determine $C_\lambda$ and $C_\lambda^{(\b/\g,\pm)}$. We will give explicit expressions 
for these in appendix \ref{app:eig}.

We can also describe the quiver tail which converts a maximal puncture to two minimal punctures. To obtain this tail we glue to a maximal puncture a free trinion and subsequently partially close the remaining maximal puncture of the free trinion to a minimal one.
At the level of the index we start from the general expression for the index of a theory glued to a free trinion, 
set $(u_1,u_2)=(t^{\frac12}\frac\gamma\delta,t^{\frac12}\frac\delta\beta)$ and pinch the contour at $z_1=\a\delta$. 
The residue of the index is given  by 

\be\label{bubbleA}
&&{\cal I}={\cal I}_V
\ge(t\alpha^2\gamma^2 )\ge(t\delta^{-2}\g^2)\ge(t\alpha^{-2}\beta^{-2}) \ge(t\delta^{2}\beta^{-2})\\
&&\qquad\oint \frac{dz_2}{4\pi i z_2}\, {\cal I}(\{\alpha\delta,z_2\}^\dagger)
\frac{
\ge(\frac{pq}t(\alpha\delta\beta\gamma)^{\pm1}z_{2}^{\pm1})}{\ge(z_2^{\pm2})
}
\ge(\frac\b\g z_2^{\pm1}(\frac{\delta}{\a} )^{\pm1})\,.\nonumber
\ee  
The above manipulations imply a concrete relation between the wavefunctions corresponding to maximal punctures and to minimal ones.
For the functions $\psi_\lambda$, $\widetilde \psi_\lambda$, and $\psi^m_\lambda$~\eqref{bubbleA} implies that,

\be\label{respsi}
&&C_\lambda \sqrt{C^{(\b,-)}_\lambda C^{(\g,+)}_\lambda} \psi^m_\lambda(\alpha) \psi^m_\lambda(\delta) =\\
&&\qquad\ge(pq\b^{-2}\g^{2})
\ge(t\alpha^2\gamma^2 )\ge(t\delta^{2}\g^2)\ge(t\alpha^{-2}\beta^{-2})\ge(t\delta^{-2}\beta^{-2})\nonumber\\
&&
\qquad{\cal I}_V\oint \frac{dz_2}{4\pi i z_2}\, \widetilde \psi_\lambda(z_2,\alpha\delta)
\frac{
\ge(\frac{pq}t(\alpha\delta\beta\gamma)^{\pm1}z_{2}^{\pm1})}{\ge(z_2^{\pm2})
}
\ge(\frac\b\gamma z_2^{\pm1}(\frac{\delta}{\a})^{\pm1})
\,.\nonumber
\ee  
\

\begin{figure}[htbp]
\begin{center}
\begin{tabular}{c}
\includegraphics[scale=0.21]{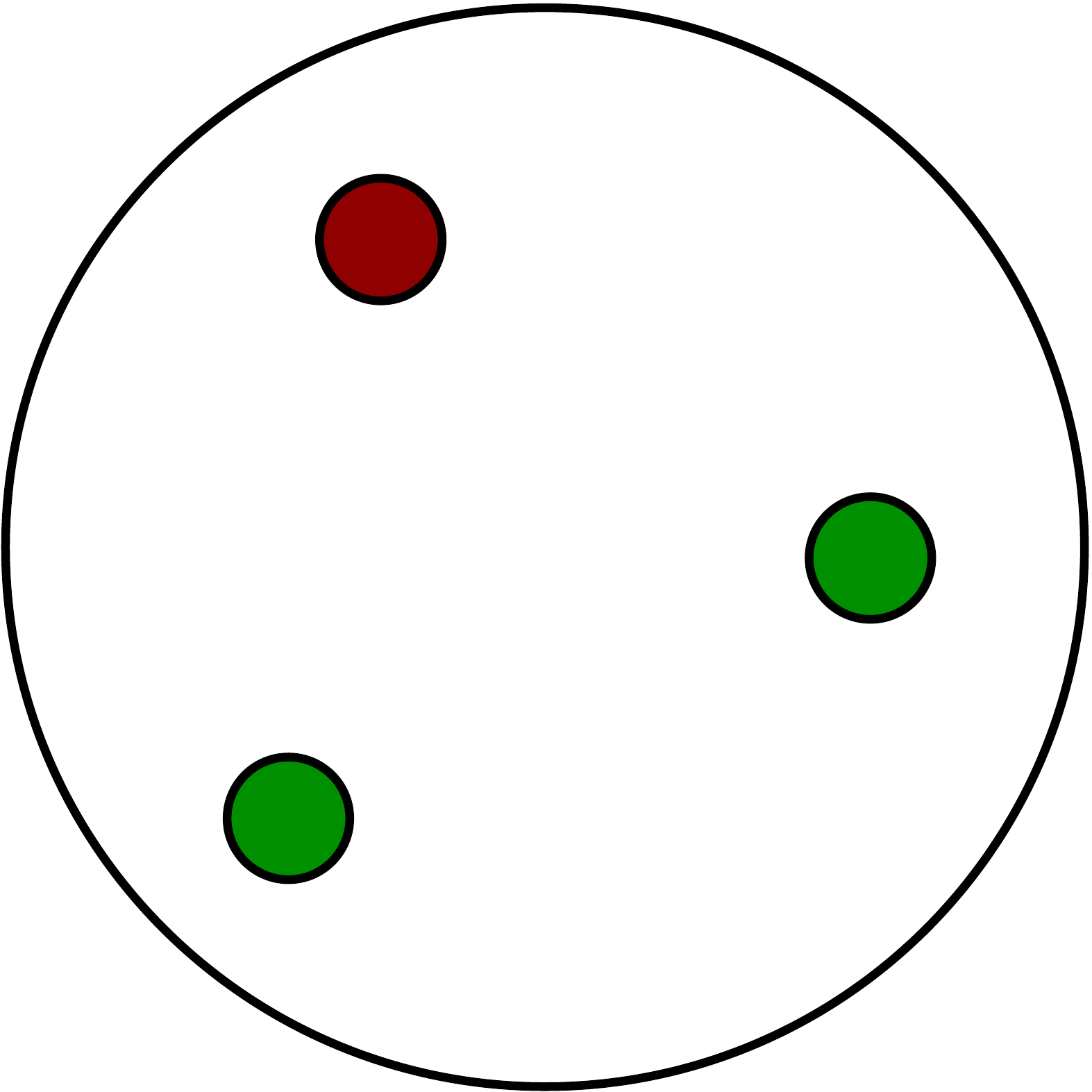}
\end{tabular}
\end{center}
\caption{The sphere with two lower maximal punctures and one upper maximal  puncture. 
\label{sphereinbP}}
\end{figure}  

\

The relation~\eqref{respsi} can be used to write the index of our core example of the four-punctured sphere in a very suggestive form. We note that

\be\label{asframe}
&&{\cal I}_{{\mathbf u}\alpha{\mathbf v}\delta}=\sum_\lambda C^2_\lambda \psi_\lambda({\mathbf u})\psi_\lambda({\mathbf v})\psi^m_\lambda(\a)\psi^m_\lambda(\delta)=\\
&&\qquad\qquad 
\ge(pq\frac{\g^2}{\b^2})\ge(t\alpha^2\gamma^2 )\ge(t\delta^2\gamma^2)\ge(t\beta^{-2}\alpha^{-2} )\ge(t\beta^{-2}\delta^{-2})\times\nonumber \\
&&\qquad
{\cal I}_V\oint \frac{dz_2}{4\pi i z_2}\, {{{\cal I}^{\{z_2,\alpha\delta\}}}}_{{\mathbf u} {\mathbf v}}\,
\frac{
\ge(\frac{pq}t(\alpha\delta\beta\gamma)^{\pm1}z_{2}^{\pm1})}{\ge(z_2^{\pm2})}
\ge(\frac\beta\gamma z_2^{\pm1}(\frac{\delta}{\alpha})^{\pm1})\,,
\ee where
\be
{{{\cal I}^{\mathbf h}}}_{{\mathbf u} {\mathbf v}}=\sum_\lambda C_\lambda \sqrt{C^{(\b,+)}_\lambda C^{(\g,-)}_\lambda} \psi_\lambda({\mathbf u})\psi_\lambda({\mathbf v})\widetilde \psi_\lambda({\mathbf h})\,.
\ee
This index can be interpreted as the index of a new strongly interacting trinion with two lower maximal punctures and one upper puncture and discrete charges $(1/2,-1/2)$. 

The second equality in~\eqref{asframe} can be interpreted 
as the computation of the index in a dual description of the core theory, involving a quiver tail attached to that strongly-interacting SCFT 
associated to a sphere with three maximal punctures. 
We  can invert the integral in ~\eqref{asframe} using the Spiridonov-Warnaar inversion formula, {\it aka} elliptic Fourier transform~\cite{spirinv}, to obtain explicitly the index of the strongly coupled theory without using the eigenfunction. This is the same inversion procedure used in \cite{Gadde:2010te} to obtain the index of the $T_3$ theory. 

We will propose an additional, related, duality which connects this 
interacting trinion to a different Lagrangian theory in appendix~\ref{app:tensant}: 
$SU(4)$ SQCD with four fundamental flavors and two flavors in antisymmetric representation supplemented with gauge singlets and a superpotential.
\

This conjectural, strongly interacting trinion theory can be used as a building block together with the free trinion (with or without closed minimal punctures)
to assemble class $\CS_2$ theories labelled by a generic Riemann surfaces of genus ${\frak g}$ with arbitrary numbers of maximal and minimal punctures and arbitrary discrete charges.

\

\subsection{Discrete charges  for $U(1)_t$}\label{sect:chargeuonet}

\begin{figure}
\begin{center}
\begin{tabular}{c}
\includegraphics[scale=0.51]{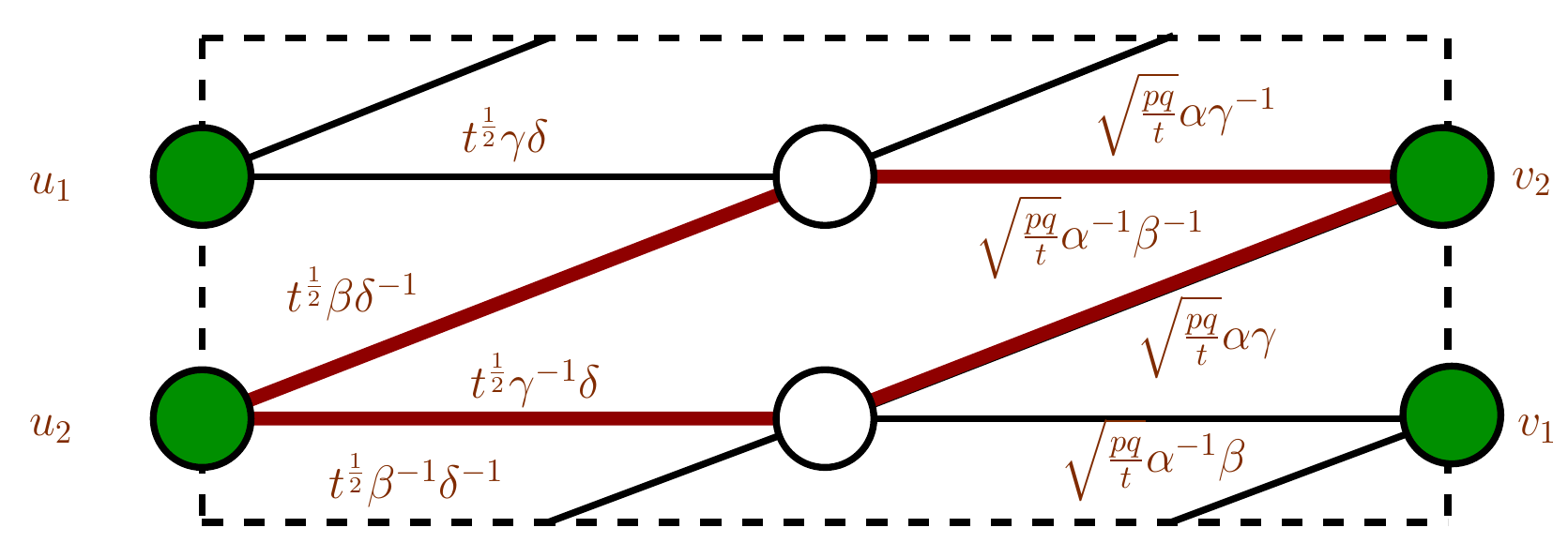}
\end{tabular}
\end{center}
\caption{ An example of a theory with flux for $U(1)_t$.
 The $A_1$ $k=2$ sphere with two maximal and two minimal punctures built from two trinions of opposite type with the fugacities associated to the 
matter fields. The white nodes correspond to gauged $SU(2)$ groups with the colored one to flavor $SU(2)$ groups. Note that in our conventions the quarks in the trinion on the left have R-charge zero and on the right R-charge one. Each gauge node has four flavors. We turn on superpotentials for the two ``diamond'' paths in the quiver.
\label{interfugst}}
\end{figure}

In our investigations we found that for the space of theories to be closed under gluings and RG flows
we have to consider the discrete curvatures for $U(1)_\beta$ and $U(1)_\gamma$. However we did not have to incorporate such curvatures for $U(1)_t$. Nevertheless, we can consider turning on these curvatures too as was done for $k=1$, class ${\cal S}$, in~\cite{Bah:2012dg}. Let us here briefly outline how to apply this generalization.  To generalize our story we first should allow two types of theories. These are two copies of theories we discussed till now but with  R-symmetry of the two types of theories related as

\be
R_+= R_-+2q_{t^-}\,,\qquad R_-= R_++2q_{t^+}\,.
\ee 
The intrinsic and puncture symmetries of the two theories are related as

\be\label{chmapt}
U(1)_t\times SU(k)_\b\times SU(k)_\g\qquad\quad&\to& \qquad\quad\; U(1)_{t^{-1}}\times SU(k)_{\b^{-1}}\times SU(k)_{\g^{-1}}\,,\\
SU(N)^k_{\mathbf u} \qquad\quad&\to&\qquad\quad\; SU(N)^k_{{\mathbf u}^{\dagger}}\,.\nonumber
\ee  
 To  trinions of type $+$ we associate a new $\Z$ valued charge $+1$ and to trinions of type $-$ we associate charge $-1$. We glue two trinions of the same type with the appropriate measure we discussed till now, {\it i.e.} introducing bi-fundamental fields $\Phi$ and coupling them to the mesons associated to the gauged maximal puncture through superpotential. We glue two trinions of different types along a maximal puncture without introducing any extra fields but turning on a quartic superpotential which is a product of the mesons associated to the gauged puncture from the two glued theories. At each gauge node we have here $N_f=2N$. See Figure~\ref{interfugst} for an example.
A way to derive the map of symmetries \eqref{chmapt} is to study anomaly free gluings of two free 
trinions of different type as depicted in Figure~\ref{interfugst}.

If we consider only one type of theories as we did till now the new $\Z$ valued charge will have a fixed value, which we can take to be $\pm(\frak g-2+s)$ with $s$ being the total number of punctures and the sign 
determined by which class of trinions we use. 
However when we glue theories of the two types together and start triggering RG flows closing minimal punctures arbitrary values of the new $\Z$ valued charge can be achieved. It would be interesting to develop this generalization in more detail. We make some comments on the index of theories with 
$U(1)_t$ discrete charges in appendix~\ref{app:indext}.

\

\section{Surface defects}\label{sect:surface}

In this section we will use the strategy of \cite{Gaiotto:2012xa} in order to produce difference operators which act diagonally on wavefunctions, 
associated to BPS surface defects produced by a ``vortex construction'', {\it i.e.} RG flows initiated by position-dependent vevs of 
chiral operators. Consider two theories $T_{UV}$ and $T_{IR}$ connected by an RG flow initiated by a constant vev of a chiral operator
of charge $1$ under some $U(1)_\alpha$ global symmetry. If we couple the theory $T_{UV}$ to a ``vortex'', {\it i.e.} background connection for the
$U(1)_\alpha$ global symmetry, with $n$ units of flux concentrated near the origin of the plane, we can give the chiral operator a vev which is constant at infinity, 
but has a zero of order $n$ near the origin. An RG flow far to the IR will leave us with a surface defect in $T_{IR}$. 

At the level of the index, the vortex construction of surface defect is very simple. The index of $T_{UV}$ has a pole at some value $\alpha_0$ of the fugacity of $U(1)_\alpha$
whose residue is the index of $T_{IR}$. The pole is accompanied to an infinite family of poles located at $\alpha_0 q^n$, whose residues give the index of 
$T_{IR}$ in the presence of the vortex defect of order $n$.

Let us take a general theory in class ${\cal S}_k$ and attach to it a sequence of $k$ free trinions to produce a new theory 
with $k$ extra minimal punctures and the same discrete charges. We know that we can recover the original theory by 
turning on a baryonic vev for each new minimal puncture, as long as we pick $k$ baryons with distinct $U(1)_{\beta_i}$ charges or 
anti-baryons with distinct $U(1)_{\gamma_i}$ charges.

Thus we can produce $2k$ classes of interesting surface defects $\CS_{\b_i,n}$ and $\CS_{\g_i,n}$ 
by making the vev of one of the $k$ baryons position dependent, or one of the $k$ anti-baryons.

Of course, we could also state the construction in term of the interacting trinions built from a sequence of $k$ free trinions by closing 
$k-1$ minimal punctures. This will save us a bit of work below. 
\
 
\subsection{The index avatar}\label{sect:surface:index}

The surfaces defects are very useful in the index considerations since they allow to fix the functions $\psi_\lambda({\mathbf u})$ by specifying them as eigenfunctions of certain difference operators.
As usual, we specialize here to $k=2$ and $N=2$. 

The index of the interacting trinion is given by,

\be\label{trinindint}
&&{\cal I}^{(\b,-)} = \Gamma_e(pq\b^{-4})\Gamma_e(t\g\b^{-1}v_1^{\pm1}v_2^{\pm1})
\Gamma_e(t^{\frac12}\b\a^{-1}u_1^{\pm1}v_2^{\pm1})
\Gamma_e(t^{\frac12}\g^{-1}\a u_2^{\pm1}v_2^{\pm1})\times\nonumber\\
&&\quad  {\cal I}_V\oint \frac{dz}{4\pi i z} \frac{\Gamma_e(\frac{pq}{t\b\g}v_2^{\pm1}z^{\pm1})}{\Gamma_e(z^{\pm2})}\Gamma_e(\b^2v_1^{\pm1}z^{\pm1})
\Gamma_e(\frac{t^{\frac12}}{\b\a}u_2^{\pm1}z^{\pm1})
\Gamma_e(t^{\frac12} \g\a u_1^{\pm1} z^{\pm1})\,.\nonumber\\
\ee We remind the reader that this theory was obtained by closing a minimal puncture of a sphere with two minimal and two maximal punctures by giving a vev to a baryon which set $\delta=t^{\frac12}\beta^{-1}$.
The symmetry between ${\mathbf u}$ and ${\mathbf v}$ is not manifest here, it follows from the duality 
property of the basic four punctured sphere. We glue this trinion to a generic theory  by gauging one of the maximal punctures. The resulting index is

\be\label{indbig}
{\cal I}={\cal I}_V^2\oint\frac{du_1}{4\pi i u_1}\oint \frac{du_2}{4\pi i u_2}
\frac{\Gamma_e(\frac{pq}{t}(\frac\b\g)^{\pm1}u_1^{\pm1}u_2^{\pm1})}{\Gamma_e(u_1^{\pm2})\Gamma_e(u_2^{\pm2})}
{\cal I}_0({\mathbf u}^\dagger) {\cal I}^{(\b,-)}({\mathbf u},{\mathbf v},\a)\,.
\ee  
\begin{figure}[htbp]
\begin{center}
\begin{tabular}{c}
\includegraphics[scale=0.21]{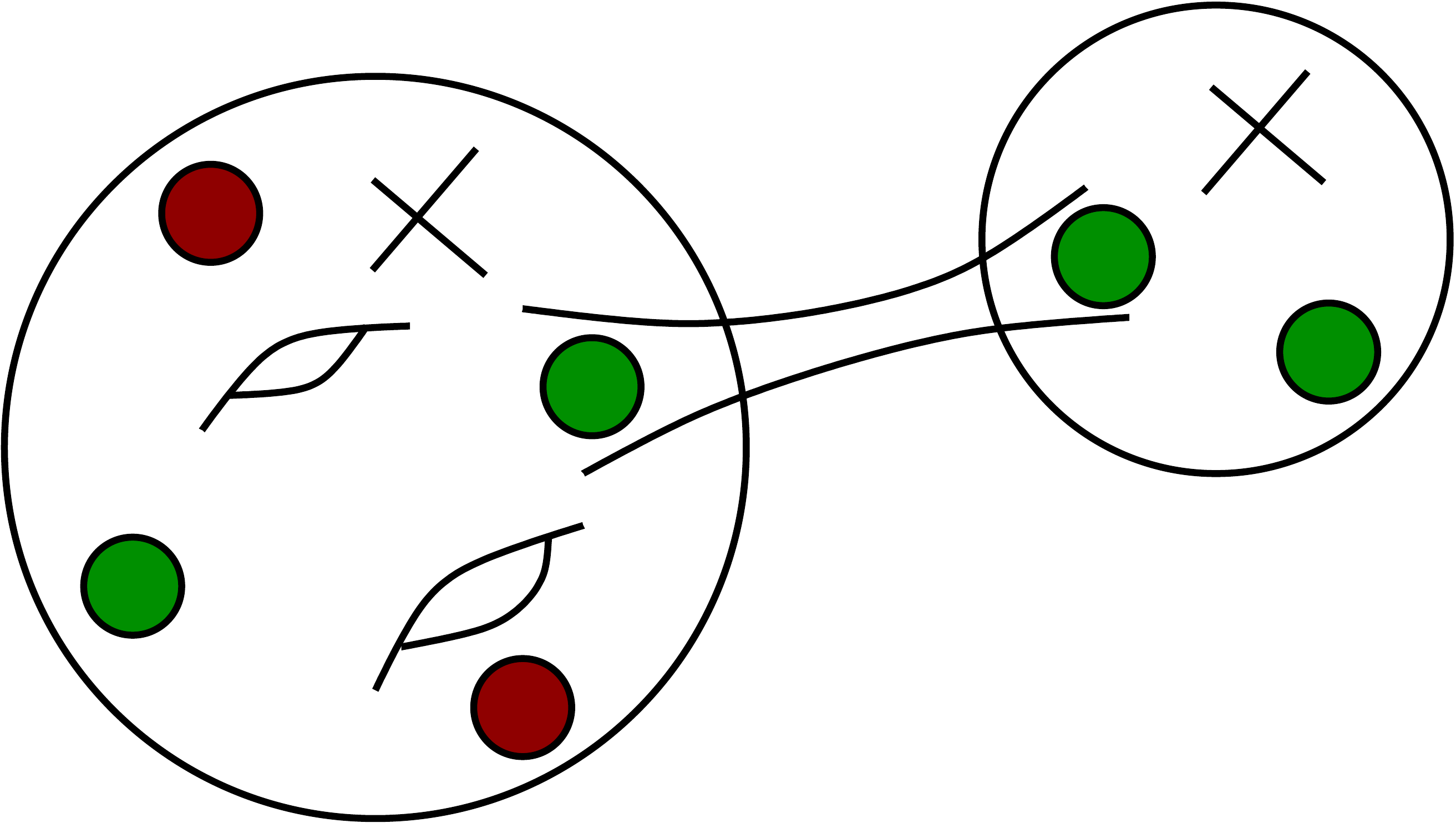}
\end{tabular}
\end{center}
\caption{Gluing the intercting trinion to a general theory.
\label{polesPi}}
\end{figure}  
This index has many interestng poles. If one computes the residue at
$\a=t^{\frac12}\beta$ we will erase the minimal puncture and obtain precisely the index  of the generic theory we glued the trinion to.
As an operatorial statement we say that computing the residue at $\a=t^{\frac12}\beta$ amounts to acting with identity operator on ${\cal I}_0$. 

Next we can consider poles in $\alpha$ which have additional powers of $q$.
We claim that this index has,
among many others, a pole when $\a=t^{\frac12} \beta q^{\frac12}$. This corresponds to 
the simplest position-dependent vev for the baryon operator. 

The pole is produced by pinching the integration contours between poles of the integrand. The poles arise from the 
$\Gamma_e$ functions associated to the chiral multiplets which receive a vev. In particular, we have a collision of poles of 
$\Gamma_e(t^{\frac12}\b\a^{-1}u_1^{\pm1}v_2^{\pm1})$ and obtain a finite residue  at\footnote{
Setting $\a=t^{\frac12}\b q^{\frac12}$ and $u_1=q^{\pm\frac12}v_2^{\pm1}$ in \eqref{trinindint} one flavor becomes massive and we can evaluate the integral in terms of the Seiberg dual mesons~\cite{Spiridonov3}. With the particular charges and generic fugacities for global symmetries appearing in \eqref{trinindint} one of the mesons contributes zero. However tuning $u_2=q^{\pm\frac12}v_1^{\pm1}$ another meson contributes a pole which cancels the zero.} 

\be\label{pinchings}
u_1=q^{\pm\frac12}v_2^{\pm1}\,,\qquad 
u_2=q^{\pm\frac12}v_1^{\pm1}\,.
\ee
Computing the residue is tedious, but straightforward.
 If we define 

\be\label{difoppp}
{\cal T}(v_1,v_2;\b,\g,t)=\frac{\theta(\frac{t v_1^{-1} v_2^{-1}}{q}(\frac\g\b)^{\pm1};p)\theta(\frac{t\b v_1}{\g  v_2};p)\theta(\frac{t\b^3\g v_2}{v_1};p)}{\theta(v_1^{2};p)\theta(v_2^{2};p)}\,,
\ee
the residue is computed by acting with the following operator on  ${\cal I}_0$,

\be\label{fop}
{\frak S}^{(\b,-)}_{(0,1)}\cdot f(v_1,v_2)=\sum_{a,b=\pm1} {\cal T}(v_1^a,v_2^b;\b,\g,t) f(q^{\frac{a}2} v_1,q^{\frac{b}2} v_2)\,.
\ee 
Since the theories we consider enjoy S-duality we can act with the difference operator on any of the maximal punctures with the same outcome~\cite{Gaiotto:2012xa}. The functions $\psi_\lambda$ are eigenfunctions of ${\frak S}^{(\b,-)}_{(0,1)}$. This operator introduces a surface defect into the ${\cal N}=1$ theories of $A_1$ class ${\cal S}_2$.

We can start from the trinion with opposite $\beta$ discrete charge and close a minimal puncture with $t^{\frac12}\b^{-1}$, or start with one of the trinions with $\gamma$ discrete charge and close the minimal punctures appropriately. The difference operators one obtains  are all related to the above,

\be\label{difopsm}
&&{\frak S}^{(\g,-)}_{(0,1)}\cdot f(v_1,v_2)=\sum_{a,b=\pm1} {\cal T}(v_1^a,v_2^b;\g,\b,t) f(q^{\frac{a}2} v_1,q^{\frac{b}2} v_2)\,,\\
&&{\frak S}^{(\b,+)}_{(0,1)}\cdot f(v_1,v_2)=\sum_{a,b=\pm1} {\cal T}(v_2^a,v_1^b;\b^{-1},\g^{-1},t) f(q^{\frac{a}2} v_1,q^{\frac{b}2} v_2)\,,\nonumber\\
&&{\frak S}^{(\g,+)}_{(0,1)}\cdot f(v_1,v_2)=\sum_{a,b=\pm1} {\cal T}(v_2^a,v_1^b;\g^{-1},\b^{-1},t) f(q^{\frac{a}2} v_1,q^{\frac{b}2} v_2)\,.\nonumber
\ee The functions $\psi_\lambda$ should be simultaneous eigenfunctions of all these operators and indeed it can be checked that these operators do commute. 

We can now in principle compute more difference operators introducing more general surface defects 
by computing other residues of the index above, with higher $n$. It is a priori straighforward but tedious exercise and we refrain from  doing it here.

\

We leave a systematic study of the difference operators and wavefunctions with generic fugacities to future work. 
Here we will take some degeneration limits which simplify the analysis considerably and allow us to write down some simple, explicit 
formulae. Although for a general $\CN=1$ theory the limit $p\to 0$ of the index may not be well defined, or useful, 
the indices of class $\CS_k$ theories with the choice of fugacities used in this paper appear to have a reasonable $p \to 0$ limit, akin to the Macdonald limit of the index of 
$\CN=2$ gauge theories. In this Macdonald-like limit, the eigenfunctions of this difference operator orthonormal under the vector multiplet measure become (experimentally) polynomials up to an universal pre-factor. 

The measure under which 
the polynomial part of the eigenfunction is orthogonal is 
\be\label{measureKtwo}
\Delta_{k,N}({\mathbf z})=
\prod_{\ell=1}^k\frac{\prod_{a\neq b}^N((z^a_\ell)(z^b_\ell)^{-1};q)}{\prod_{a,b=1}^N(t\,\b_\ell^{-1}\gamma_\ell\,z^a_\ell (z^b_{\ell+1})^{-1};q)}\,,
\ee with $k=1$, $N=2$, and $\b_1=\b$, $\g_1=\g$. This is a generalization of the $A_{N-1}$ Macdonald measure.  We will discuss a straightforward algorithm to compute the eigenfucntions by diagonalizing the difference operators in Macdonald limit in appendix \ref{app:eig}.

Let us here quote the results if we further set $\b,\g=1$ and take the Hall-Littlewood-like limit $p,q=0$. For the first several wavefunctions we find 
\be\label{eigensHL}
&&\hat \psi_{(0)}=\frac{1}{(1-t\,z_1^{\pm1}z_2^{\pm1})^2},\\
&&\hat \psi_{(1)_\pm}=\frac{1}{(1-t\,z_1^{\pm1}z_2^{\pm1})^2}((z_1+z_1^{-1})\pm(z_2+z_2^{-1})),\nonumber\\
&&\hat \psi_{(2)_0}=\frac{1}{(1-t\,z_1^{\pm1}z_2^{\pm1})^2}((z_1^2+z_1^{-2})-(z_2^2+z_2^{-2})),\nonumber\\
&&\hat \psi_{(2)_\pm}=\frac{1}{(1-t\,z_1^{\pm1}z_2^{\pm1})^2}\times\left(
-\frac{\left(\pm\sqrt{2-t^2}+t\right) \left(z_1^4+1\right)}{2 \left(t^2-1\right) {z_1}^2}-\frac{\left(\pm\sqrt{2-t^2}+t\right) \left({z_2}^4+1\right)}{2 \left(t^2-1\right) {z_2}^2}\right.\nonumber\\
&&\qquad\qquad\left.\pm\sqrt{2-t^2}-t+\left({z_1}+\frac{1}{{z_1}}\right) \left({z_2}+\frac{1}{{z_2}}\right)\right)\,.\nonumber
\ee The hat on the functions reminds us that these functions are not normalized to be orthonormal. Note that the coefficients of the polynomials here are algebraic expressions, roots of polynomial equations,  as opposed to rational expressions in the ${\cal N}=2$ case. 
{\it The ${\cal N}=1$ theories are in this sense irrational though algebraic}. 
The indices are single valued sums over roots of algebraic equations.
We denote the normalized functions

\be\label{deffunc}
\psi_\lambda({\mathbf z})=\frac{\chi_\lambda({\mathbf z})}{(1-t\,z_1^{\pm1}z_2^{\pm1})^2}\,,
\ee where $\chi_\lambda({\mathbf z})$ is a polynomial.  
The HL index of the free trinion  is then given by

\be\label{trinionIndex}
{\cal I}_{{\mathbf z}{\mathbf y}\alpha}=\left[\frac{1}{1-t z_1^{\pm1}z_2^{\pm1}}
\frac{1}{1-t y_1^{\pm1}y_2^{\pm1}}\frac{1-t^2}{1-t \alpha^{\pm2}}\right]^2\sum_\mu
\frac{
\chi_\mu(z_1,z_2)\,\chi_\mu(y_1,y_2)\,\chi_\mu(t^{\frac12}\alpha,t^{\frac12}\alpha^{-1})}
{ \chi_\mu(1,t)}\,.\nonumber\\
\ee 
As we will see in appendix \ref{app:eig} the index of the interacting trinion with three maximal punctures becomes

\be\label{trinionIndexI}
{\cal I}_{{\mathbf z}{\mathbf y}{\mathbf x}}=\left[\frac{1}{1-t z_1^{\pm1}z_2^{\pm1}}
\frac{(1-t^2)^2}{1-t y_1^{\pm1}y_2^{\pm1}}\frac{1}{1-t x_1^{\pm1}x_2^{\pm1}}\right]^2\sum_\mu
\frac{
 \chi_\mu(z_1,z_2)\,\chi_\mu(y_1,y_2)\, \chi_\mu(x_1,x_2)}
{ \chi_\mu(1,t)}\,.\nonumber\\
\ee When $\b,\g=1$ there is no difference between the colors of punctures and we are blind to the 
different discrete charges.

In appendix \ref{app:eig} we will write the general expressions for the index 
of any theory in class ${\cal S}_2$ when $p=0$. 

\

\section{Five dimensional interpretation}\label{sect:fivedimi}

In this section we will re-examine our four-dimensional discussion in the language of boundary conditions and interfaces for a $\CN=1$ 
five-dimensional gauge theory $\CN_{N,k}$, the necklace quiver formed by $k$ $SU(N)$ gauge groups. This model is the world-volume theory of 
 D4 branes sitting at an $A_{k-1}$ singularity and conjecturally arising from the compactification on a circle of the six-dimensional SCFTs which inspire our work. Our intuitive picture is that a Riemann surface with $s$ semi-infinite tubes labels interfaces between 
$s$ copies of $\CN_{N,k}$ defined on half-spaces. 

Our first task is to review the properties and definitions of boundary conditions and interfaces for five-dimensional $\CN=1$ gauge theories

\

\subsection{Generalities}

A five-dimensional $\CN=1$ gauge theory is labelled by a gauge group $G$, a flavor group $F$ and a quaternionic representation of $G \times F$ 
which specifies the hypermultiplet content. The hypermultiplets can be given real masses associated to the Cartan sub-algebra of $F$. 
The gauge couplings of $G$ can be identified as real masses associated to ``instanton''
global symmetries $U(1)_I$ whose conserved currents are of the schematic form $*\Tr F \wedge F$.
 We can denote the full global symmetry group as $\hat F = F \times U(1)_I$.
The theory is further labelled by a choice five-dimensional Chern-Simons couplings $\kappa$, which may be integral or half-integral depending on the amount of matter fields. 

Although the gauge theories may have strongly-coupled UV completions, in the IR they are free. That implies that the gauge groups 
can be treated as very weakly coupled when describing a boundary condition. A simple class of boundary conditions is labelled by 
two pieces of data: the subgroup $G_\partial$ of the gauge symmetry preserved at the boundary and 
a choice of boundary condition for the hypermultiplets. Somewhat more general boundary conditions are possible, which involve a
generalization of the Nahm pole which occurs in the maximally symmetric case. We will come back to these later in the section. 

The boundary condition for the hypermultiplets can be strongly coupled. If the hypermultiplets pseudoreal representation is the sum of two conjugate representations, 
a general construction is available, which starts from some free boundary conditions and adds superpotential couplings to extra four-dimensional $\CN=1$ degrees of freedom living at the boundary. 
If we denote the two halves of the hypermultiplets as $X$ and $Y$, we can start from a $Y=0$, $\partial_\perp X=0$ boundary condition 
and add a linear superpotential coupling 
\begin{equation}
W = X O_X\,,
\end{equation}
to an operator $O_X$ in a boundary theory $B_X$. Alternatively, we can start from a $X=0$, $\partial_\perp Y=0$ boundary condition 
and add a linear superpotential coupling 
\begin{equation}
W = Y O_Y\,,
\end{equation}
to an operator $O_Y$ in a boundary theory $B_Y$. As long as we focus on F-terms only, the two constructions are 
essentially equivalent: $B_Y$ can be obtained from $B_X$ by ``flipping'' $O_X$, i.e. by introducing 
a new chiral field $\phi$ with superpotential coupling $\phi O_X$. Then $O_Y \equiv \phi$. 
A particular case is that the $X=0$, $\partial_\perp Y=0$ can be obtained from $Y=0$, $\partial_\perp X=0$ by adding a chiral with $X \phi$ coupling, and vice versa. 

Boundary conditions for a five-dimensional gauge theory may have various anomalies. 
The boundary cubic gauge anomaly receives three contributions: 
\begin{itemize}
\item The bulk Chern-Simons coupling
\item The boundary theory 
\item The boundary condition for the bulk hypermultiplet: $Y=0$, $\partial_\perp X=0$ contributes half of the anomaly of a boundary chiral field with the same charge as $X$. \footnote{That follows from the symmetry between 
$Y=0$, $\partial_\perp X=0$ and $X=0$, $\partial_\perp Y=0$ and the fact that we can switch from one to the other by a flip}
This half is the reason the bulk CS coupling may sometimes be half-integral 
\end{itemize}
The total gauge anomaly for $G_\partial$ must cancel out. 

We also have various sources of 't Hooft anomalies. This includes cubic and mixed anomalies arising from the boundary theory, the boundary condition for the hypermultiplets
and the bulk CS couplings. The R-symmetry anomaly also receives contributions from the boundary conditions for the gauge fields, which is half of what a four-dimensional 
$G_\partial$ multiplet would give. 

\

\subsection{A review of $\CN_{N,k}$}
\begin{figure}[htbp]
\begin{center}
\begin{tabular}{c}
\includegraphics[scale=0.45]{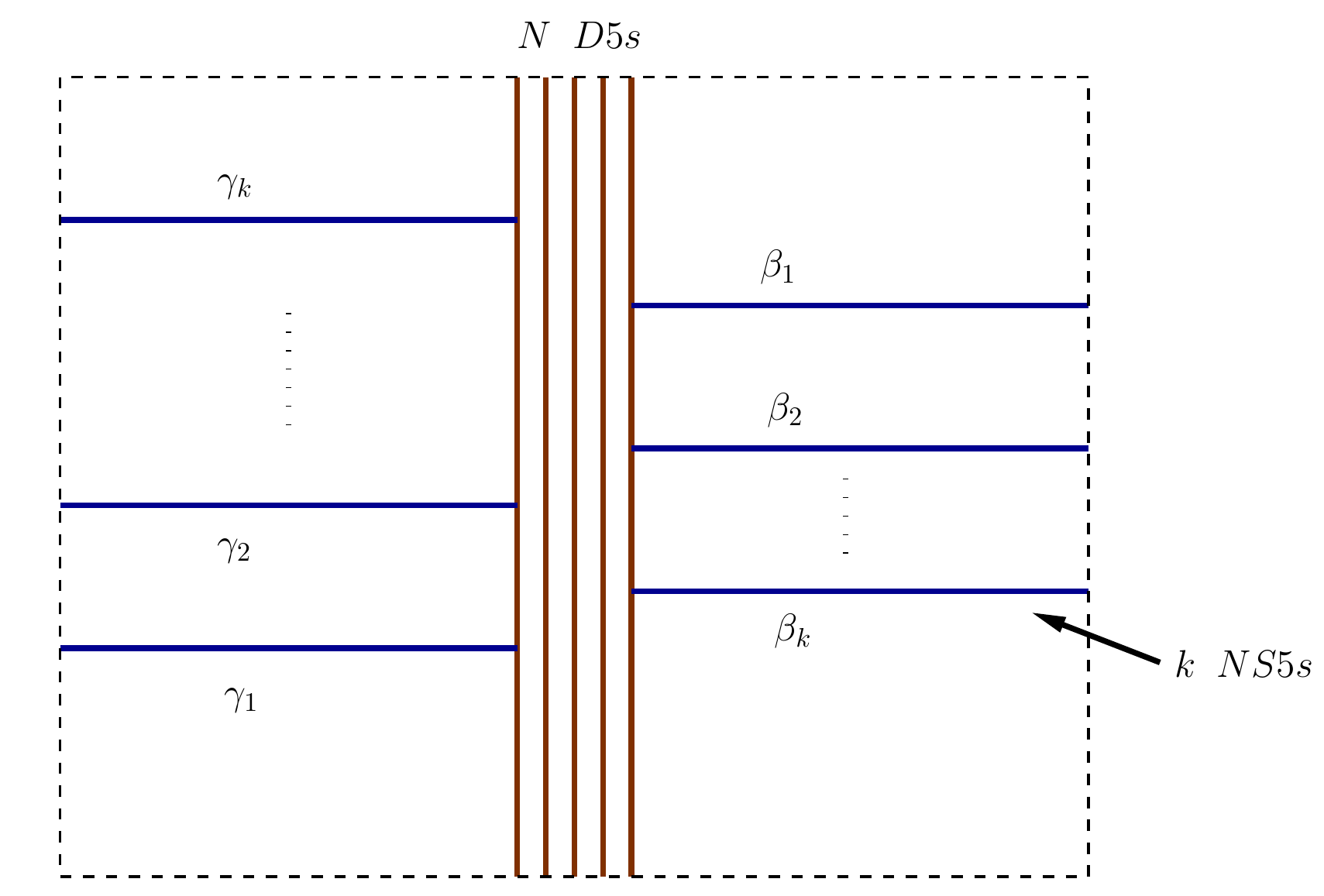}
\end{tabular}
\end{center}
\caption{The brane picture for the necklace quiver ${\cal N}_{N,k}$. The top and bottom edges are identified forming a cylinder.
\label{neckcylindr}}
\end{figure}

The necklace quiver theory $\CN_{N,k}$ has a $U(1)^{2k}$  global symmetry: rotations of the bi-fundamental hypermultiplets and instanton symmetries. See figure \ref{neck}.
A natural way to parameterize these symmetries follows from the UV realization of the gauge theory as a web of fivebranes drawn 
on a cylinder, with $k$ infinite NS5 branes and $N$ circular D5 branes, see Figure \ref{neckcylindr}. 

The transverse position of the $i$-th top half-infinite NS5 brane is the mass parameter for a $U(1)_{\hat \beta_i}$ symmetry which acts with charge $1$ on the $X^{a_{i+1}}_{a_{i}}$ fields of 
the $i$-th hypermultiplet, $-1$ on the $Y_{a_{i+1}}^{a_{i}}$ fields, $1/2$ on the instanton charge at the $(i+1)$-th node and $-1/2$ on the instanton charge at the $i$-th node.
We define the $k-1$ symmetry generators $\left(\frac{U(1)^k}{U(1)} \right)_{\beta}$ by quotienting by the diagonal symmetry $U(1)_t$. 

Similarly, the transverse position of the $i$-th bottom half-infinite NS5 brane is the mass parameter for a $U(1)_{\hat \gamma_i}$ symmetry which acts with charge $-1$ on the $X^{a_{i+1}}_{a_{i}}$ fields of 
the $i$-th hypermultiplet, $1$ on the $Y_{a_{i+1}}^{a_{i}}$ fields, $1/2$ on the instanton charge at the $(i+1)$-th node and $-1/2$ on the instanton charge at the $i$-th node.
We define the $k-1$ symmetry generators $\left(\frac{U(1)^k}{U(1)} \right)_{\gamma}$ by quotienting by the diagonal symmetry $U(1)_t$. 

Finally, we need to pick a $U(1)_p$ symmetry generator whose mass parameter is associated to the sum of all gauge couplings, {\it i.e.} the size of the cylinder. We pick it to act on the instanton charge at the first node. 

In terms of fugacities, the hypermultiplet fields $X^{a_{i+1}}_{a_{i}}$ have fugacities $t \beta_i \gamma_i^{-1}$, and the CS coupling at the $i$-th node is $\sqrt{\frac{\beta_{i-1}\gamma_{i-1}}{\beta_i \gamma_i}}$
for $i\neq 1$, $p \sqrt{\frac{\beta_{k}\gamma_{k}}{\beta_1 \gamma_1}}$ otherwise. 

In the following we will encounter variants of this symmetry labeling, where the $U(1)_{\beta_i}$ and $U(1)_{\gamma_i}$ are permuted among themselves
by permutations $\sigma$ and $\tau$ respectively, and the $U(1)_p$ symmetry generator act on the $i$-th node. We can denote that as 
$\CN_{N,k}^{\sigma, \tau,i}$. We will also often consider boundary conditions which break $U(1)_p$. In that case, we can refer to 
$\CN_{N,k}^{\sigma, \tau, *}$.

In the low energy quiver gauge theory, the mass parameters for these global symmetries are constrained by 
the requirement that the gauge couplings should be positive. The UV description in terms of fivebranes suggest that symmetry enhancements should occur 
when some semi-infinite branes are brought together. A maximal case is when all semi-infinite branes are brought together, giving rise to a 
$SU(k)_\beta \times SU(k)_\gamma$ symmetry enhancement. Indeed, as the branes live on a cylinder there are $k$ distinct ways to 
reach such a symmetry enhancement, as one bring the branes together in the order $i, i+1, \cdots, i-1$. 
The parametrization of $\CN_{N,k}^{\sigma, \tau,i}$ is adapted to that order. 

We should also remember the six-dimensional UV completion of the five-dimensional quiver gauge theory, 
in terms of a circle compactification of the $(1,0)$ SCFT corresponding to $N$ M5 branes in an $A_{k-1}$ singularity. 
Then $U(1)_p$ becomes the KK momentum and $SU(k)_\beta \times SU(k)_\gamma \times U(1)_t$ become six-dimensional 
global symmetries. Then the five-dimensional mass parameters are lifted to the inverse radius of the compactification circle and to Wilson lines for the six-dimensional global symmetries.

Intuitively, we may hope to find ``duality walls'' in the five-dimensional gauge theory, which express the invariance of the UV theory under 
permutation of two consecutive symmetries $U(1)_{\beta_{i}}$ and $U(1)_{\beta_{i+1}}$ or $U(1)_{\gamma_{i}}$ and $U(1)_{\gamma_{i+1}}$.
More precisely, such a wall would arise as the low energy limit of a Janus configuration in the UV, where the mass parameters for these symmetries 
are brought across each other as we move along the fifth direction. 

The notion of duality walls for five-dimensional gauge theories is explored in depth in a separate publication \cite{HEE}. 
Here we mention them because the domain walls which are associated to spheres with two punctures and non-trivial discrete charges
will turn out to coincide, amazingly, with the duality walls for the $\CN_{N,k}^{\sigma, \tau,i}$ theories associated to permutations of the 
$U(1)_{\beta_i}$ and $U(1)_{\gamma_i}$ 5d mass parameters, i.e. of the 6d flavor Wilson lines. This provides a powerful 
check of our conjectural interpretation of discrete charges as curvature charges for the six-dimensional $U(1)_{\beta_i}$ and $U(1)_{\gamma_i}$
global symmetries.

\

\subsection{Boundary conditions from maximal punctures}

Consider any of our four-dimensional gauge theories, with a positive puncture of color $0$. 
We will build from it a $U(1)_p$-breaking right boundary condition for $\CN_{N,k}$, i.e. a boundary condition for $\CN_{N,k}^*$

The anomalies and superpotential couplings match nicely. We can deform the $X=0$ boundary conditions by coupling the $Y_{a_{i+1}}^{a_{i}}$ to the mesons $M^{a_{i+1}}_{a_{i}}$.
The gauge anomaly cancels out because of the balance between fundamental and anti-fundamental fields. 
The $SU(N)_i$ gauge group has a mixed anomaly associated to the fugacity $t \beta_{i-1} \gamma_i^{-1}$
from the boundary fields, $\sqrt{(t \beta_{i-1} \gamma^{-1}_{i-1})^{-1}(t \beta_{i} \gamma^{-1}_{i})^{-1}}$ from the hypermultiplet boundary conditions,
for a total of $\sqrt{\frac{\beta_{i-1} \gamma_{i-1}}{\beta_i \gamma_i} }$, which precisely cancels against the contribution from the CS coupling 
for a right boundary condition. 
Similarly, a positive puncture of color $n$ can be coupled from the right to $\CN_{N,k}^{(n+1,n+2,\cdots), (1,2,\cdots), *}$.

We can use negatively oriented punctures to define left boundary conditions as well. For example, if we have a negatively oriented puncture of color $0$, 
shifting the $i$ indices by one so that mesons have fugacity $t \beta_{i} \gamma^{-1}_{i}$ and anomalies $t \beta_{i} \gamma_{i-1}^{-1}$, 
the total anomaly $\sqrt{\frac{\beta_i \gamma_i}{\beta_{i-1} \gamma_{i-1}} }$ precisely cancels against the contribution from the CS coupling 
for a left boundary condition. Thus a negative puncture of color $n$ can be coupled from the left to $\CN_{N,k}^{(n+2,n+3,\cdots), (2,3,\cdots), *}$.

A crucial example of interface is a free trinion coupled on the left to  
$\CN_{N,k}^{(2,3,\cdots), (1,2,\cdots), *}$ and on the right to $\CN_{N,k}^{*}$. The fact that the interface breaks $U(1)_p$ is
consistent with the six-dimensional interpretation of the system as an infinite tube with a minimal puncture with a specific location on the cylinder. 
Upon closing the minimal puncture, though, we will find that $U(1)_p$ is restored.

\

\subsection{Closing minimal punctures and duality walls}

Next, we close the minimal puncture in the interface between 
$\CN_{N,k}^{(2,3,\cdots), (1,2,\cdots), *}$ and $\CN_{N,k}^{*}$
by giving a vev to the bifundamental chirals of fugacity $\sqrt{t} \beta_1 \alpha$ 
which are coupled to the five-dimensional gauge groups $SU(N)_1^L$ and $SU(N)_1^R$. 

The vev Higgses the two gauge groups together, so that they form a single $SU(N)_1$ gauge group 
stretching across the interface. The vev also couples the bifundamentals of 
fugacities $\sqrt{t} \gamma_1^{-1} \alpha^{-1} = t \beta_1/\gamma_1$ linearly to the five-dimensional 
hypermultiplets on the right hand side of the interface, and $\sqrt{t} \gamma_k^{-1} \alpha^{-1} = t \beta_1/\gamma_k$
on the left hand side, converting the corresponding $X=0$ to $Y=0$ boundary conditions. 

The match between the global symmetries across the new interface works almost exactly as before, 
except for a crucial difference: the $U(1)_p$ symmetry acting on the $SU(N)_1$  instanton charge 
both on the left and on the right of the interface is now unbroken! This is consistent with the picture that the minimal puncture 
has been erased. We have obtained an interface between $\CN_{N,k}^{(2,3,\cdots), (1,2,\cdots), 1}$ and $\CN_{N,k}^{1}$.

It turns out that such an interface can be actually decomposed into $k-1$ simpler interfaces, 
each corresponding to a simple duality operation. The rightmost interface is between 
$\CN_{N,k}^{(2,1,3,4,\cdots), (1,2,\cdots), 1}$ and $\CN_{N,k}^{1}$ and we can associate it to the permutation of 
$\beta_1$ and $\beta_2$. The next interface is between $\CN_{N,k}^{(2,3,1,4,\cdots), (1,2,\cdots), 1}$ and $\CN_{N,k}^{(2,1,3,4,\cdots), (1,2,\cdots), 1}$ and we can associate it to permutation of $\beta_1$ and $\beta_3$, etc. 

The simple interface associated to a permutation of $\beta_1$ and $\beta_2$ 
lets all five-dimensional gauge groups go through the interface, except for $SU(N)_2^L$ and 
$SU(N)_2^R$, which are coupled by a bi-fundamental chiral field $Q$ of fugacity 
$\beta_2/\beta_1$. In turn, $\det Q$ is coupled to a gauge-neutral chiral operator 
$b$ by a linear superpotential $W = b \det Q$. 

The hypermultiplets also just go through the interface, except the 
ones charged under $SU(N)_2^{L,R}$. On the right, we set to zero the $X_2^R$ field of fugacity $t \beta_2/\gamma_2$
and the $Y_1^R$ field of fugacity $t^{-1} \beta_1^{-1} \gamma_1$. On the left, we set to zero the 
$Y_2^L$ field of fugacity $t^{-1} \beta_1^{-1} \gamma_2$ 
and the $X_1^L$ field of fugacity $t \beta_2 \gamma_1^{-1}$.
We introduce superpotential couplings $Q Y_1^L X_1^R + Q X_2^L Y_2^R$.

This gives an interface between $\CN_{N,k}^{(2,1,3,4,\cdots), (1,2,\cdots), i}$ and $\CN_{N,k}^{i}$ for every $i$ 
except when $i=2$. We would like to interpret it as the duality interface for permuting $\beta_1$ and $\beta_2$. 

In a similar manner, we could consider an interface between theories $\CN_{N,k}^{(1,2,\cdots), (2,1,3,4,\cdots), i}$ 
and $\CN_{N,k}^{i}$ for every $i$ except when $i=2$, involving a bi-fundamental chiral field $\tilde Q$ of fugacity 
$\gamma_2/\gamma_1$ going in the opposite direction. 
We would need superpotential couplings $\tilde Q X_1^L Y_1^R + \tilde Q Y_2^L X_2^R$.
We would like to interpret it as the duality interface for permuting $\gamma_1$ and $\gamma_2$.

In order to test our interpretation, we should check that these interfaces fuse in a manner consistent with the expected 
properties of the permutation group of the $\beta_i$ and the $\gamma_i$. 

The first test is obvious: permuting $\beta_1$ and $\beta_2$ twice should give the identity, and thus 
concatenating two copies of the corresponding duality interface should give a trivial interface. 
This works indeed as expected: the $SU(N)_2$ gauge theory in between the interfaces 
gives at low energy a four-dimensional $SU(N)$ gauge group with $N_f=N$, as 
the boundary conditions at the interfaces eliminate completely the bulk hypermultiplets 
in the interval, and only the two sets of $Q$ chiral multiplets remain, together with the 
baryonic couplings. We know that such a theory in the IR reduces to the $QQ$ mesons 
with a non-zero vev, which Higgses the five-dimensional gauge theories on the two sides 
of the fused interface together. 

A second test is that permuting $\beta_1$ and $\beta_2$ should commute with 
permuting $\gamma_1$ and $\gamma_2$. It turns out that this follows from a neat application of Seiberg duality. 
The $SU(N)_2$ gauge theory in between the interfaces now has $N_f = 2 N$: the $Q$ and $\tilde Q$ 
give $2 N$ ``quarks'', while the bulk hypers in the interval give $2N$ ``antiquarks''. 
Seiberg duality at that node exchanges the setups which arise from fusing the interfaces in two different orders. 

It is even easier to check that the interface for permuting of $\beta_1$ and $\beta_2$ commutes with, say, the 
interface for commuting $\gamma_2$ and $\gamma_3$. 

The $\sigma_{12} \sigma_{23} \sigma_{12} = \sigma_{23} \sigma_{12} \sigma_{23}$ relations in the permutation group of the 
$\beta_i$ can also be checked. If we concatenate the corresponding triplets of interfaces, we get interfaces which support 
four-dimensional gauge theories which are related by the Seiberg duality of a $SU(N)$ theory with $N_f = 2N$,
up to a slight mismatch in the superpotential couplings which may perhaps explained away by operator mixing. 
We leave a more detailed analysis to \cite{HEE}.

Finally, we can comment on the relation between different ways of closing a minimal puncture and 
``curvature charges'' in the six-dimensional puncture. By closing a minimal puncture 
with chirals of fugacities proportional to $\beta_i$, we obtain an interface between 
$\CN_{N,k}^{(2,3,\cdots,1), (1,2,\cdots), i}$ and $\CN_{N,k}^{i}$. The final order of the $\beta_j$ 
is independent of our choice, but the interfaces glue the $U(1)_p$ action on the two sides in different ways. 

We can try to match the interface 
with the behaviour of an Abelian $SU(k)$ connection on an infinite tube, asymptotically 
flat at the two ends, but with curvature $(1,0,0, \cdots)$ in between. 
The curvature will force the Wilson line parameter in the first eigenline 
to vary as $m_1 \to m_1 + 1/R$ as we go from right to left, thus passing 
across all other Wilson line parameters and back to the initial position on the circle. 
This seems to roughly match what we see in the interface. 

\

\subsection{Punctures and orbifold Nahm poles}

It is natural to identify a maximal puncture with a $D_X$ Dirichlet boundary condition for the five-dimensional 
gauge theory, setting to zero the $Y_{a_{i+1}}^{a_{i}}$ half of the hypermultiplets: If we map a 4d theory $T$
with a maximal puncture to a boundary condition $B$ by coupling it to the 5d theory, we can recover the 4d theory $T$
from the 5d theory on a segment, with boundary condition $B$ at one end and Dirichlet at the other end. 
The surviving half $X^{a_{i+1}}_{a_{i}}$ of the hypermultiplets provide the expected ``mesons'' 
and the boundary condition produces the desired 't Hooft anomalies at the boundary. 

The five-dimensional perspective is useful in describing other types of punctures as well. 
For example, in class $\CS$ general punctures are associated to a variant of Dirichlet boundary conditions, the Nahm pole 
boundary conditions. In this section we generalize that notion to the necklace quiver theories $\CN_{N,k}$.

The boundary conditions we are after are boundary conditions for BPS equations which describe field
configurations of the 5d theory which preserve four-dimensional super-Poincare invariance. 

There are D-term and F-term equations. The F-term equations set the complex moment maps to be zero and require the hypermultiplet vevs to be 
covariantly constant under a complexified gauge connection of the form $\CD_5 = D_5 - \Phi$,
where $\Phi$ is the scalar super partner of the gauge bosons. The D-term equations 
set $D_5 \Phi$ to be equal to the real moment map for the hypermultiplets. For the necklace quiver, we can write schematically

\be
&&X_i Y_i = Y_{i+1} X_{i+1}\,,\qquad
D_5 X_i= \Phi_{i+1} X_i - X_i \Phi_i \,,\qquad
D_5 Y_i= \Phi_{i} Y_i - Y_i \Phi_{i+1} \,,\\
&&D_5 \Phi_{i+1} = X_i X^\dagger_i - Y^\dagger_i Y_i - X^\dagger_{i+1} X_{i+1}+ Y_{i+1} Y^\dagger_{i+1}\,.\nonumber
\ee

We are interested in deformations of the $D_X$ boundary conditions, which still set the $Y_i$ fields to zero but 
enforce a singular behavior of the $X_i$, $\Phi_i$ fields,

\begin{equation}
X_i \sim -\frac{B_i}{x^5}\,, \qquad \qquad \Phi_i \sim -\frac{A_i}{x^5}\,,
\end{equation} 
with 

\begin{equation}
B_i = A_{i+1} B_i - B_i A_i\,,  \qquad \qquad A_{i+1} = B_i B^\dagger_i - B^\dagger_{i+1} B_{i+1}\,.
\end{equation}
We can call these boundary conditions orbifold Nahm poles. 

Notice that we can organize the $A_i$ and $B_i$ matrixes into two $k N \times k N$ matrices $t^3$ and $t^+$, 
with $t_3$ block diagonal of blocks $A_i$ and $t^+$ with blocks $B_i$ under the diagonal. The $t^+$ and $t^3$ 
give an $SU(2)$ embedding $\rho$ into $SU(kN)$, with the property that $t^3$ commutes with 
a diagonal matrix $\Omega$ with $N$ eigenvalues equal to $1$, $N$ equal to $e^{2 \pi i/k}$, etc. 
while $t^+$ has charge $1$ under the action of $\Omega$. 

If we start from $D_X$ and we give a vev to the ``mesons'', i.e. the boundary values of the $X_i$, 
such that the $k N \times kN$ matrix $M$ with blocks $X_i$ under the diagonal is nilpotent, 
it is natural to expect the boundary condition to flow in the IR to the orbifold Nahm pole 
labelled by the $su(2)$ embedding $\rho$ in $SU(kN)$ associated to $M$. 

The vevs we used to fully close a maximal puncture are a perfect example of this setup: the ``chain'' of $k (N-1)$ meson vevs engineers a nilpotent matrix $M$ with a single, large Jordan block.

\

\section{Discussion}\label{sect:summa}

In this paper we have given a basic description of some the properties of class $\CS_k$ theories. 
Starting from core theories which we associate to spheres with two maximal and a bunch of minimal punctures we discovered that in order to build classes of theories closed under gaugings and RG flows 
triggered by vevs for  a very particular set of chiral operators we should consider theories 
which are naturally associated to spheres with more general combinations of punctures. Moreover, 
the theories should be labelled not just by a punctured Riemann surface, but also by a collection of discrete charges. Some of the theories which followed from our considerations are strongly coupled SCFTs. Some
of these can be thought of as IR fixed points of a Lagrangian theory but some lack such a description.
Using these SCFTs we in principle now can construct theories corresponding to Riemann surfaces 
of arbitrary genus.

\

Our analysis is rather incomplete in many respects and leaves room for many new insights to be uncovered. We can supplement our analysis with a very partial list of interesting questions and open problems which we would like to see addressed in the near future. 

\

\subsection*{Comparison with bipartite theories}
Although we have focussed on quivers drawn on a cylinder and compactifications of the six-dimensional theories 
on surfaces of genus $0$, we could have also readily defined core theories associated to tori with $k n$
minimal punctures, by gluing together the two maximal punctures of our standard core theories. 
It would be interesting to explore in full the relation between such genus $1$ class $\CS_k$ 
theories and the standard bipartite quiver gauge theories associated to toric Calabi Yau singularities. 
See appendix \ref{app:tor} for some examples.

\

\subsection*{Classification of punctures}

We have discussed at first two types of punctures, maximal and minimal, the former coming in $k$ varieties. We have argued that more general ``regular'' punctures may be defined by turning vevs of collections of mesons at a maximal puncture, and proposed a five-dimensional classification 
in terms of $su(2)$ embeddings in $SU(kN)$ commuting with a $\Z_k$ subgroup. The duality walls we encountered 
in five dimensions also suggest that one may define maximal or other regular punctures labelled by general permutations of the $\beta_i$ and $\gamma_i$ as well. A natural open problem is to complete a systematic classification of ``regular'' punctures and the corresponding quiver tails.

\

\subsection*{Precision study of the spectrum}

We have given a prescription to compute, at least in principle, the index for any of the theories in $A_{N-1}$ class ${\cal S}_k$. 
It would be interesting to actually extract from the indices information about the operators of
the putatively new strongly coupled SCFTs. For example, their marginal and relevant deformations~\cite{Beem:2012yn}. It would be also interesting to find a systematic way to determine their conformal anomalies. 

\

\subsection*{Extension to other $(1,0)$ SCFTs}

In this paper we discussed theories obtained, conjecturally, by reducing the $\CT^N_k$ $(1,0)$ SCFTs associated to N $M5$ branes on $A_k$ singularity.  It would be interesting to extend the discussion to other types of $(1,0)$ SCFTs.

Some $(1,0)$ SCFTs can be obtained from RG flows initiated by Higgs branch vevs in the $\CT^N_k$ theories. 
It may be possible to track the four-dimensional image of these RG flows, perhaps by giving vevs to chiral operators which are 
charged under intrinsic symmetries only. 

It should be possible to extend our work to other 4d gauge groups and 6d SCFTs by considering ``core'' theories defined by
brane systems enriched by orientifold planes.

\

\subsection*{Reduction to three dimensions}

We can consider reducing theories of class ${\cal S}_k$ to three dimensions. 
When we have a description of the theory in four dimensions in terms of a conformal Lagrangian, in three dimensions the description will not in general be conformal. Moreover, the reduction on a circle 
will produce additional superpotentials involving monopole operators~\cite{Aharony:2013dha}.
 Such superpotentials break explicitly symmetries in three dimensions which are anomalous in four dimensions.
Keeping track of such superpotentials is crucial to have dualities work for the compactified theories. 

One interesting aspect of the dimensional reduction in class ${\cal S}_1$ was that the reduced theories
possess a mirror description which was always Lagrangian, a star-shaped quiver 
with arms associated to the punctures of the original theory~\cite{Benini:2010uu}. The difference operators and wave-functions~\cite{Nishioka:2011dq} had a simple interpretation in the language of domain walls interpolating between two S-duality frames of ${\cal N}=4$ SYM
with $SU(N)$ gauge groups~\cite{Gaiotto:2008ak}. The eigenvalue equation was associated to the S-duality relation between a Wilson line a 't Hooft line on the two sides of the duality wall~\cite{Gaiotto:2012xa,Bullimore:2014nla,Razamat:2014pta}.
 
We do not know how much of this will generalize to class ${\cal S}_k$. The difference operators do appear to be related to 
't Hooft line operators in a necklace quiver. It may be that the 3d limit of the wave-functions is still related to 
domain walls interpolating between different duality frames of $\CN=2$ four-dimensional necklace quivers. 
It would be interesting to explore this idea and perhaps build universal mirrors for the compactification of class 
 ${\cal S}_k$ theories. 

\

\subsection*{Surface defects}

We have discussed RG flows introducing surface defects into our theories. It would be interesting to study properties of these defects in more detail. 
They appear to give a broad generalization of the elliptic RS difference operators encountered in class $\CS$ theories. 

\

\subsection*{Holography}

Class ${\cal S}_k$ theories can be in principle studied in large $N$ limit. For example one can consider 
theories corresponding to Riemann surfaces without punctures to avoid proliferation of flavor symmetries. Such $AdS_5$ backgrounds were recently considered in \cite{Apruzzi:2015wna,Apruzzi:2015zna} generalizing some of the  $k=1$ results discussed in~\cite{Gaiotto:2009gz}.  It would be interesting to study the relations between holography and our results in more detail.

\

\subsection*{Geometrization of Seiberg dualities}

Starting from quiver theories with bifundamental matter and employing different types of dualities 
one can in principle generate gauge theory with matter in more intricate, tensor, representations. See appendix \ref{app:tensant} for example. It would be interesting to understand whether 
class ${\cal S}_k$ can serve as a natural setup to systematize the diverse variety of
Seiberg dualities.  

\

\subsection*{Quantum mechanical models}

It will be interesting to study in more detail the quantum mechanical models for which our wavefunctions are eigenfunctions. Here we have discussed in some detail the case of $A_1$ and $k=2$ and the generalization to higher $k$ and $N$ though rather straightforward might be interesting. One can also try and compute  other partition functions, such as lens index~\cite{Benini:2011nc}, for class ${\cal S}_k$ theories. In terms of eigenfunctions while the supersymmetric index is related to symmetric functions (and polynomials) the lens index in general is a natural generalization to non-symmetric functions~\cite{Alday:2013rs,Razamat:2013jxa}. The lens index should also provide a window into 
subtleties with global properties of the gauge and flavor groups \cite{Razamat:2013opa}.

\

\

\section*{Acknowledgments}

We would like to thank
Chris Beem and 
Brian Willett
for useful discussions.
The research of SSR was  partially supported by
"Research in Theoretical High Energy Physics" grant DOE-SC00010008.  
SSR gratefully acknowledges support from the Martin~A.~Chooljian and Helen Chooljian 
membership during his stay at the Institute for Advanced Study, and
would like to thank KITP, Santa Barbara, and the Simons Center, Stony Brook, for hospitality and support during different stages of this work. 
The research of
DG was supported by the Perimeter Institute for Theoretical Physics. Research
at Perimeter Institute is supported by the Government of Canada through Industry
Canada and by the Province of Ontario through the Ministry of Economic Development
and Innovation.

\

\

\appendix

\section{The supersymmetric index}\label{app:index}

The supersymmetric index~\cite{Romelsberger:2005eg,Kinney:2005ej} counts with signs and weights the protected operators of the theory. It can be defined either as an $\S^3\times \S^1$ partition function with supersymmetric boundary conditions on the $\S^1$ or as a trace over the Hilbert space on $\S^3$. The latter definition takes the following form,

\be
{\cal I}(p,q; {\mathbf u})=\Tr_{{\cal H}_{\S_3}} (-1)^F p^{j_1+j_2-\frac12 r} q^{j_1-j_2-\frac12 r} \prod_{a\in {\frak F}} u_a^{q_a}\,.
\ee   Here $j_i$ are the Cartans of the $SO(4)\sim SU(2)_1\times SU(2)_2$ isometry of $\S^3$ and $r$ is the $U(1)_r$ R-symmetry. The charges $q_a$ correspond to $U(1)$ global flavor symmetries with the set 
of these symmetries denoted by ${\frak F}$. The supersymmetric index of a free chiral field of R-charge $R$ is given by

\be
{\cal I}_\chi = \prod_{i,j=0}^\infty\frac{1-p^{1-\frac{R}2+i}q^{1-\frac{R}2+j} u^{-1}}{1-p^{\frac{R}2+i}q^{\frac{R}2+j}u} = \Gamma((pq)^{\frac{R}2}u;p,q)\equiv \Gamma_e ((pq)^{\frac{R}2}u)\,.
\ee
For the sake of brevity In this paper $\Gamma_e(z)$ will stand for the elliptic Gamma function $\Gamma(z;p,q)$ implicitly defined above. When an $SU(N)_{\mathbf z}$ flvor symmetry of a theory with index ${\cal I}({\mathbf z})$ is gauged, the index of the gauge theory is given by,

\be
{\cal I}= \frac{(q;q)^{N-1}(p;p)^{N-1}}{N!} \oint\prod_{i=1}^{N-1}\frac{dz_i}{2\pi i z_i} \prod_{i\neq j}^N \Gamma_e(z_i/z_j) \, {\cal I}({\mathbf z})\,,
\ee with 
\be
(z;q)\equiv \prod_{i=0}^\infty (1-z q^i)\,,
\ee being the q-Pochhammer symbol. The index of a free vector field is given by,

\be
{\cal I}_V = (p;p)(q;q)\,.
\ee We will also encounter theta functions which we will define as

\be
\theta(z;q)=(z;q)(q z^{-1};q)\,.
\ee
An important property of the supersymmetric index is that it is independent of the continuous parameters 
of the theory, such as marginal couplings, and also is independent  of the RG scale. Thus it can be computed for example in the UV using a non-conformal description and be the same as the superconformal index of the IR CFT. 

We will adopt the standard convention that ``ambiguous'' powers in the arguments of a function 
denote product over all the possibilities. For example the index of the ${\cal N}=2$ hypermultiplet is given by,

\be
\ge(t^{\frac12}z^{\pm1}) = \ge(t^{\frac12} z) \ge(t^{\frac12} z^{-1})\,.
\ee

\

\section{Eigenfunctions for $A_1$ class ${\cal S}_2$}\label{app:eig}

We will discuss here explicit algorithm to construct eigenfunctions for $A_1$ theories of class ${\cal S}_2$ in the $p=0$ limit. The simplifying factor here is that the eigenfunctions turn out to be proportional to polynomials. For $p\neq 0$, as also is the case for class ${\cal S}_1$, the computation is much more involved. However, in principle a perturbative computation in $p$ around $p=0$ solution can be set up. See for example ~\cite{Razamat:2013qfa}.   

\

For $p=0$ the computation is rather straightforward. 
The equation we have to solve is 

\be
{\frak S}^{(\b,-)}_{(0,1)}\cdot \psi({\mathbf u}) = {\cal E} \,\psi({\mathbf u})\,.
\ee Here ${\frak S}^{(\b,-)}_{(0,1)}$ was explicitly given in section \ref{sect:surface:index}.
A way to proceed is to make an anzats,

\be\label{anzg}
\psi({\mathbf u})=K_{max}({\mathbf u}) \chi({\mathbf u})\,,
\ee with 

\be
K_{max}({\mathbf u};\b,\g)&& = \frac1{(t(\frac\b\g)^{\pm1}u_1^{\pm1}u_2^{\pm1};q)}\,.
\ee Then we assume $\chi({\mathbf u})$ to be a polynomial in $u_i$ with general coefficients of maximal degree $n$ and symmetric in $u_i\to u_i^{-1}$. Since our indices are invariant under ${\mathbf u}\to -{\mathbf u}$ we also can assume that $\chi({\mathbf u})$ has well defined parity.  We will have then  for polynomials of order $n$ order $\frac14 n^2$ free parameters, and we also have the eigenvalue as a parameter. Expanding  the eigenvalue equation in $u_i$ and demanding it to hold for any monomial term will give us a system of nonlinear equations for these coefficients. The non-linearity comes because the eigenvalue can multiply other coefficients. We then should solve these system of algebraic equations which in general has a finite set of solutions. 

\

Let us illustrate this explicitly for the lowest orders. Assuming the polynomial is of degree zero we get 
that 

\be
{\frak S}^{(\b,-)}_{(0,1)}\cdot 1 -{\cal E}\propto {\cal E}-1-\beta ^4 t^4+\beta ^2 \gamma ^2 t^2+\frac{\beta ^2 t^2}{\gamma ^2}\,,
\ee and thus the order zero polynomial is an eigenfunction given that,

\be
{\cal E}=1+\beta ^4 t^4-\beta ^2 \gamma ^2 t^2-\frac{\beta ^2 t^2}{\gamma ^2}\,.
\ee
The first eigenfunction is then given by,

\be
\psi_0({\mathbf u}) = K_{max}({\mathbf u})\sqrt{(t^2;q)(t^2\biggl(\frac{\b}{\g}\biggr)^{\pm2};q)}\,.
\ee We have normalized it to have unit norm under the gluing measure.

At next order we make an ansatz,

\be
\chi({\mathbf u})=u_1+\frac1{u_1}+H(u_2+\frac1{u_2})\,.
\ee Then we compute taking $q\to 0$ to avoid horrendous expressions

\be
&&\left.\lim_{q\to 0}q^{\frac12}({\frak S}^{(\b,-)}_{(0,1)}-q^{-\frac12}{\cal E}')\cdot \chi({\mathbf u})\right|_{u_1^0}=\\
&&\quad\;-\frac{\left(u_2^4-1\right) \left(H \left(t^2 \left(\beta ^4 \gamma ^2-\beta ^2-\gamma ^2\right)-\gamma ^2 ({\cal E}'-1)\right)+\beta  \gamma  t \left(\beta ^2 \gamma ^2 t^2-1\right)\right)}{\gamma ^2 u_2} =0\,,\nonumber\\
&&\left.\lim_{q\to 0}q^{\frac12}({\frak S}^{(\b,-)}_{(0,1)}-q^{-\frac12}{\cal E}')\cdot \chi({\mathbf u})\right|_{u_1^3}=\\
&&\quad\;\frac{\left(u_2^2-1\right) \left(\beta  H t \left(t^2-\beta ^2 \gamma ^2\right)-\gamma  \left({\cal E}'+\beta ^2 \gamma ^2 t^2-1\right)\right)}{\gamma }=0\,.\nonumber
\ee Here ${\cal E}=q^{-\frac12}{\cal E}'$. This system of equations can be reduced to a quadratic equation in one of the two variables $H$ or ${\cal E}'$ and thus has two solutions. Ultimately we obtain,

\be
&&H =\\&&\;\frac{\left(\mp\sqrt{4 \b^4 \g^4 (1-t^2\b^{-2}\g^{-2})  \left(1-t^2\beta ^2 \gamma ^2 \right)+t^2 (1-\b^2  \g^2)^2 \left(\b^2+\g^2\right)^2}+t  (1-\b^2  \g^2) \left(\beta ^2+\gamma ^2\right)\right)}{2 \b^3  \g^3  (1-t^2\b^{-2}  \g^{-2} )}\,.\nonumber
\ee We should further normalize $\chi({\mathbf u})$ to have norm one. We note that this expression is algebraic in fugacities. Taking for simplicity $\b,\g=1$ we obtain that here $H=\pm1$ recovering the result advertised in section \ref{sect:surface:index}.  Note also that with this value for $H$ all the symmetry properties \eqref{relpsitildepse} are satisfied.

We can continue to derive eigenfunctions in this manner. Going to higher orders we get higher order polynomial equations which do not have closed form solutions, but we can solve them perturbatively in the fugacities.

\subsection*{The self-adjointness of the difference operator}

Let us check that the basic difference operator we computed is self adjoint under the gauging measure.
That is for two sufficiently nice behaving functions $f({\mathbf u})$ and $g({\mathbf u})$ we have, 

\be\label{adj}
&&\oint \frac{du_1}{u_1}\oint \frac{du_2}{u_2}\frac{ \Gamma_e(\frac{pq}t(\frac{\beta}\gamma)^{\pm1} u_1^{\pm1}u_2^{\pm1})}{\Gamma_e(u_1^{\pm2})\Gamma_e(u_2^{\pm2})} f({\mathbf u}) {\frak S}^{(\b,-)}_{(0,1)}({\mathbf u}^\dagger) g({\mathbf u}^\dagger)=\\
&&\qquad\oint \frac{du_1}{u_1}\oint \frac{du_2}{u_2} \frac{\Gamma_e(\frac{pq}t(\frac{\beta}\gamma)^{\pm1} u_1^{\pm1}u_2^{\pm1})}{\Gamma_e(u_1^{\pm2})\Gamma_e(u_2^{\pm2})} ( {\frak S}^{(\b,-)}_{(0,1)}({\mathbf u})f({\mathbf u})) g({\mathbf u}^\dagger)\,.\nonumber
\ee
We note that,

\be
\left.\frac{ \Gamma_e(\frac{pq}t(\frac{\beta}\gamma)^{\pm1} u_1^{\pm1}u_2^{\pm1})}{\Gamma_e(u_1^{\pm2})\Gamma_e(u_2^{\pm2})}\right|_{u_i\to q^{\frac12}u_i} =
\frac{\theta(q^{-1}u_1^{-2};p)}{\theta(u_1^2;p)}
\frac{\theta(q^{-1}u_2^{-2};p)}{\theta(u_2^2;p)}
\frac{\theta(\frac{t}q(\frac{\beta}\gamma)^{\pm1} \frac1{u_1 u_2};p)}{\theta(t(\frac{\beta}\gamma)^{\pm1} u_1 u_2;p)}
\frac{ \Gamma_e(\frac{pq}t(\frac{\beta}\gamma)^{\pm1} u_1^{\pm1}u_2^{\pm1})}{\Gamma_e(u_1^{\pm2})\Gamma_e(u_2^{\pm2})}\,.\nonumber\\
\ee Then we have that 

\be
&&\left.\frac{ \Gamma_e(\frac{pq}t(\frac{\beta}\gamma)^{\pm1} u_1^{\pm1}u_2^{\pm1})}{\Gamma_e(u_1^{\pm2})\Gamma_e(u_2^{\pm2})}{\cal T}(u_2^{-1},u_1^{-1})
g(q^{-\frac12}u_2,q^{-\frac12}u_1)f(u_1,u_2)
\right|_{u_i\to q^{\frac12}u_i}=\\
&&\;\;\frac{\Gamma_e(\frac{pq}t(\frac{\beta}\gamma)^{\pm1} u_1^{\pm1}u_2^{\pm1})}{\Gamma_e(u_1^{\pm2})\Gamma_e(u_2^{\pm2})}{\cal T}(u_1,u_2)
g(u_2,u_1)f(q^{\frac12}u_1,q^{\frac12}u_2)\,.\nonumber
\ee This implies that  under change of coordinates $\{u_i\}\to \{q^{\frac12}u_i\}$ the  first term, out of four, on the left hand side of~\eqref{adj} in the expansion of the difference operator maps exactly to the fourth term on the right hand side. This can be repeated for the other three terms and thus implies that the difference operator is self-adjoint.

\

\subsection*{The index of generic class ${\cal S}_2$ $A_1$ theory}

Let us here write down the index of a generic theory residing in $A_1$ class ${\cal S}_2$. We will specialize to the case $p=0$ where we can write very explicitly expressions for the index. 

The index is given in terms of the following building blocks. The eigenfunctions $\psi_\lambda({\mathbf u})\equiv K_{max}({\mathbf u}) \chi_\lambda({\mathbf u})$ are associated to the maximal punctures.
From \eqref{condone},\eqref{defcpm}, and \eqref{phipsiRel} we deduce that 

\be
C_\lambda^{(\g,+)}\equiv \Phi_\lambda(\b,\g)=\left(\frac{(t^2\g^2\b^{\pm2};q)}{(t^2\g^{-2}\b^{\pm2};q)}\right)^{\frac12}
\left(\frac{\chi_\lambda(t,\frac\g\b;\b,\g)}{\chi_\lambda(t\g^2,\frac1{\b\g};\b,\g)}\right)^{\frac12}\,.
\ee 
We find that the eigenfunctions satisfy the symmetry properties~\eqref{relpsitildepse}, from which and
from \eqref{condone} we can deduce that

\be
C_\lambda^{(\b,+)}(\b,\g)=(C_\lambda^{(\b,-)}(\b,\g))^{-1}=C_\lambda^{(\g,-)}(\g,\b)=(C_\lambda^{(\g,+)}(\g,\b))^{-1}\,.
\ee
For minimal punctures we use the above and the relation \eqref{phipsiRel} to write

\be
&&\psi^m_\lambda(\a)\equiv K_{min}(\delta)\chi^m_\lambda (\delta)=(C^{(\b,-)}_\lambda C^{(\g,+)}_\lambda)^{-\frac12} \\
&&\qquad\quad\; \,(\b^{2}\g^{-2};q)
\frac{(t\a^{2}\b^{-2};q)(t\a^{-2}\g^{2};q)}
{(t\a^{-2}\b^{-2};q)(t\a^2\g^{2};q)}
\chi_\lambda(t^{\frac12}\frac\g\delta,t^{\frac12} \frac\delta\b;\b,\g)\,.\nonumber
\ee Thus we deduce that

\be
\chi_\lambda^m(\delta)=\left(\frac{
\sqrt{\chi_\lambda(t\g^2,\frac1{\b\g};\b,\g)
\chi_\lambda(t\b^2,\frac1{\b\g};\b,\g)
}}{\chi_\lambda(t,\frac\g\b;\b,\g)}\right)^{\frac12}
\chi_\lambda(t^{\frac12}\frac\g\delta,t^{\frac12} \frac\delta\b;\b,\g)\,,
\ee
with the $K$-factors  given by,

\be
K_{max}({\mathbf u};\b,\g)&& = \frac1{(t(\frac\b\g)^{\pm1}u_1^{\pm1}u_2^{\pm1};q)}\,, \qquad\quad\\
K_{min}(\delta;\b,\g) &&= \frac{(t^2\b^{-2}\g^{-2};q)}{(t^2\b^{2}\g^{2};q)(t^2\frac{\g^2}{\b^2};q)}\frac{1}{(t^2;q)(t\b^{\pm2}\delta^{-2};q)(t\g^{\pm2}\delta^2;q)}\,.\nonumber
\ee 
Finally the structure constant is set from \eqref{condone} and \eqref{defcpm} to be

\be
C_\lambda=(t^2;q)\sqrt{\frac{(t^2\g^2\b^{\pm2};q)^3(t^2\b^2\g^{-2};q)}{(t^2\g^{-2}\b^{-2};q)}}
\frac1{\sqrt{\chi_\lambda^m(t^{\frac12}\b)\chi_\lambda^m(t^{\frac12}\b^{-1})}}\,.
\ee 
The index of a theory corresponding to genus ${\frak g}$ surface with $m_u$ upper maximal punctures, $m_d$ maximal lower punctures, $m_m$ minimal punctures, charge $\ell_\b$ under $U(1)_\b$ discrete symmetry, and charge $\ell_\g$ under $U(1)_\g$ discrerte symmetry is given by,

\be
&&\prod_{i=1}^{m_d}K_{max}({\mathbf u}_i;\b,\g)\prod_{j=1}^{m_u}K_{max}({\mathbf u}_j;\b^{-1},\g)
\prod_{k=1}^{m_m}K_{min}(\delta_k;\b,\g)\times\\
&&\sum_\lambda \biggl(C_\lambda\biggr)^{2\frak g-2+m_u+m_d+m_m} 
\Phi(\b,\g)^{\ell_\g}\Phi(\g,\b)^{-\ell_\b} 
\prod_{i=1}^{m_d}\chi_\lambda({\mathbf u}_i;\b,\g)\prod_{j=1}^{m_u}\chi_\lambda({\mathbf u}_j;\b^{-1},\g)
\prod_{k=1}^{m_m}\chi^m_\lambda(\delta_k)\,.\nonumber
\ee For example the index of the free trinion is given by

\be
&&{{\cal I}_{{\mathbf u}\alpha}}^{{\mathbf v}}= \frac{(t^2\b^{\pm2}\g^{\pm2};q)^{\frac12}}
{(t(\frac\b\g)^{\pm1}u_1^{\pm1}u_2^{\pm1};q)(t(\b\g)^{\pm1}v_1^{\pm1}v_2^{\pm1};q)(t\b^{\pm2}\a^{-2};q)(t\g^{\pm2}\a^2;q)}\times\nonumber\\
&&\qquad\qquad\sum_\lambda\frac{\chi_\lambda({\mathbf u};\b,\g)\chi_\lambda({\mathbf v};\b^{-1},\g)\chi_\lambda(t^{\frac12}\frac\g\a,t^{\frac12}\frac\a\b;\b,\g)}{\sqrt{\chi_\lambda(\frac\g\b,t;\b,\g)\chi_\lambda(\g\b,t\b^{-2};\b,\g)}}\,.
\ee

Note that if we specialize fugacities for different symmetries, {\it i.e.} ignore/break corresponding symmetries the indices simplify. For example, taking $\gamma=\beta$, and thus identifying the two corresponding $U(1)$ symmetries, the difference between $\ell_1$ and $\ell_2$ is gone and we label the theories by Riemann surface and one integer.  Taking $\beta=1$ and thus neglecting the $U(1)_\beta$ symmetry there is no difference between upper and lower punctures and also $\ell_1$ is not a meaningful  number any more.  Let us mention here that setting both $\b=\g=1$ one can derive from \eqref{respsi} the following factorization property,

\be\label{factor}
\frac{\chi_\lambda(t^{\frac12}\a,t^{\frac12}\alpha^{-1})\chi_\lambda(t^{\frac12}\delta,t^{\frac12}\delta^{-1})}{\chi_\lambda(1,t)}=
\chi_\lambda(\a\delta^{-1},\a\delta)\,.
\ee Which in particular implies that switching off $\b$ and $\g$ we cannot distinguish in the index
maximal punctures from pairs of minimal ones.

The eigenfunctions satisfy many properties which guarantee them to be consistent with our considerations. For instance we can check that
\be
\left(\frac{\chi_\lambda(\frac\g\b,t;\b,\g)}{\chi_\lambda(\g\b,t\b^{-2};\b,\g)}\right)^{\frac12}
\chi_\lambda(t^{\frac12}\frac\g\delta,t^{\frac12} \frac\delta\b;\b,\g)=
\left(\frac{\chi_\lambda(t\g^2,\frac1{\b\g};\b,\g)}{\chi_\lambda(t,\frac\g\b;\b,\g)}\right)^{\frac12}
\chi_\lambda(t^{\frac12}\b\delta,t^{\frac12} \frac1{\delta\g};\b,\g)\,,\nonumber\\
\ee which implies that reducing maximal puncture in two different ways to minimal punctures discussed in section \ref{sect:closemaxi:index} gives the same result.

\

\subsection*{Sphere with two maximal punctures}

We can start from the $(\b,-)$ interacting trinion of section \ref{subsec:closemin:index} and close the minimal puncture. There are here only two bayons available for which we can turn on vevs. These correspond to either $\delta=t^{\frac12}\b^{-1}$ or
$\delta=t^{-\frac12}\g^{-1}$ as can be seen from \eqref{trinindint}. In both cases the gauge group is Higgsed and we get a collection of chiral fields.
In the former case the index becomes (in Macdonald limit for simplicity),

\be
{\cal I}^{(\b,-2)}({\mathbf u},{\mathbf v}) = 
\frac{(t\b\g v_2^{\pm1}u_2^{\pm1};q)}{
(\frac{t}{\b\g} v_2^{\pm1}u_2^{\pm1};q)}
\frac{(\b^{4};q)^2}{(\b^2 v_1^{\pm1}u_2^{\pm1};q)(\b^2 v_2^{\pm1} u_1^{\pm1};q)(t\frac\g\b u_1^{\pm1}u_2^{\pm1};q)( t\frac\g\b v_1^{\pm1}v_2^{\pm1};q)}\,,\nonumber\\
\ee
and in the latter case we obtain,

\be
{\cal I}^{(\b,-1)(\g,-1)}({\mathbf u},{\mathbf v}) = 
\frac{(\g^{-4};q)(\b^4;q)}{(\b^2 v_1^{\pm1}u_1^{\pm1};q)(\g^{-2} v_2^{\pm1} u_2^{\pm1};q)(t\frac\g\b u_1^{\pm1}u_2^{\pm1};q)( t\frac\g\b v_1^{\pm1}v_2^{\pm1};q)}\,.\nonumber\\
\ee
Again, we can close the maximal punctures. Gluing these  two-punctured spheres to a general
theory does not change the numbers of punctures but shifts the discrete charges. For eigenfunctions this implies that they are ``eigenfunctions'' also of the following integral operators with well prescribed eigenvalues,

\be
&&(q;q)^2\oint \frac{du_1}{4\pi i u_1}\oint \frac{du_2}{4\pi i u_2}(u_1^{\pm2};q)(u_2^{\pm2};q)
(t(\frac\b\g)^{\pm1} u_1^{\pm1}u_2^{\pm1};q){\cal I}^{(\b,-1)(\g,-1)}({\mathbf u}^\dagger,{\mathbf v}) \psi_\lambda ({\mathbf u}) \nonumber\\
&&\qquad\qquad\qquad\;  =\Phi_\lambda(\b,\g)^{-1}\Phi_\lambda(\g,\b)  \; \psi_\lambda({\mathbf v})\,,\\
&&\nonumber\\
&&(q;q)^2\oint \frac{du_1}{4\pi i u_1}\oint \frac{du_2}{4\pi i u_2}(u_1^{\pm2};q)(u_2^{\pm2};q)
(t(\frac\b\g)^{\pm1} u_1^{\pm1}u_2^{\pm1};q){\cal I}^{(\b,-2)}({\mathbf u}^\dagger,{\mathbf v}) \psi_\lambda ({\mathbf u})\nonumber\\
 &&\qquad\qquad\qquad\; =\Phi_\lambda(\g,\b)^{2}  \; \psi_\lambda({\mathbf v})\,.\nonumber
\ee
One can check that these equations hold for the eigenfunctions that we derived.
Thus the functions $\psi_\lambda$ are eigenfunctions of difference operators with eigenvalues related to surface defects, and are eigenfunctions of integral operators with eigenvalues being related to discrete charges for intrinsic symmetries.

\

\subsection*{Sphere with one maximal and two minimal punctures}

We can start from the $(\b,-)$ interacting trinion of section \ref{subsec:closemin:index} and partially close one of the maximal punctures to obtain a trinion with one maximal and
two minimal punctures. As we discussed we can do this in general in two different ways by giving vevs to two different mesons. However, in this case since the theory is not generic and has rather small flavor symmetry only one meson exists, and it corresponds to setting $(u_1,u_2)\to (t^{\frac12}\frac\g\e,t^{\frac12}\frac\e\b)$. This can be clearly seen from the index \eqref{trinindint}. 
After turning on the vev the gauge part of the theory is $SU(2)$ with three flavors as one flavor acquires mass. Thus, in the Seiberg dual frame it is given by a collection of chiral fields coupled through a superpotential. This theory has $-\frac32$ units of $U(1)_\b$ discrete charge and $\frac12$ unit of $U(1)_\g$ discrete charge.
The index of this theory in Macdonald limit is 

\be
&&{\cal I}^{(\b,-\frac32),(\g,+\frac12)}({\mathbf u},\delta,\epsilon)=\\
&&\qquad \frac{(\b^4;q)(\frac{\beta^2}{\gamma^2};q)(q;q)}{(\frac{t\epsilon}{\delta}u_1^{\pm1};q)(\frac{t\delta\epsilon}{\beta\gamma}u_2^{\pm1};q)(\frac{\beta^2}{\epsilon\delta}u_1^{\pm1};q)(\frac{\delta\beta}{\gamma\epsilon}u_2^{\pm1};q)(\frac{t}{\beta^2\epsilon^2};q)(t\gamma^2\epsilon^2;q)(\frac{t^2\gamma^2}{\beta^2};q)}\times\nonumber\\
&&\qquad\quad\oint\frac{dz}{4\pi i z}
\frac{(z^{\pm2};q)(t^{\frac32}\gamma\epsilon z^{\pm1};q)}{(t^{\frac12}\frac{1}{\beta\delta} u_2^{\pm1}z^{\pm1};q)(t^{\frac12}\gamma\delta u_1^{\pm1} z^{\pm1};q)(\frac{\beta^2\epsilon}{t^{\frac12}\gamma} z^{\pm1};q)}=\nonumber\\
&&
\frac{(\b^4;q)(\frac{\beta^2}{\gamma^2};q)(t\beta\gamma\delta\epsilon u_2^{\pm1};q)}{
(t\beta^{-1}\gamma^{-1}\delta\epsilon u_2^{\pm1};q)
(\frac{\beta}\gamma(\frac\delta\epsilon)^{\pm1}u_2^{\pm1};q)
(\beta^2(\delta\epsilon)^{\pm1} u_1^{\pm1};q)
(t\frac\gamma\beta u_1^{\pm1}u_2^{\pm1};q)
(\frac{t}{\beta^2\delta^2};q)(\frac{t}{\beta^2\epsilon^2};q)
(t\gamma^2\delta^2;q)(t\gamma^2\epsilon^2;q)}\,.
\nonumber
\ee Note that the result is explicitly invariant under exchanging the two minimal punctures.
We can attach this theory to a trinion with opposite discrete charges  and three maximal punctures of the same color to obtain yet another duality frame for the basic four punctured sphere.
For the eigenfunctions the above implies the following relation,

\be
&&C_\lambda \Phi_\lambda(\b,\g)^{\frac32}\Phi_\lambda(\g,\b) \psi^m_\lambda(\delta)\psi^m_\lambda(\epsilon)=\\
&&\qquad\;\; (q;q)^2\oint \frac{du_1}{4\pi i u_1}\oint \frac{du_2}{4\pi i u_2} {\cal I}^{(\b,-\frac32),(\g,+\frac12)}({\mathbf u}^\dagger,\epsilon,\delta)\, (u_1^{\pm2};q)(u_2^{\pm2};q)(t(\frac{\gamma}{\beta})^{\pm1} u_1^{\pm1}u_2^{\pm1};q) 
\psi_\lambda({\mathbf u})\,.\nonumber
\ee 
Thus we can glue our new trinion to an interacting trinion with appropriate discrete charges to obtain yet another Argyres-Seiberg like frame for our basic interacting four-punctured sphere.

One might consider closing punctures in the new trinion.
We can further close the maximal puncture in two different ways to obtain a theory corresponding to sphere with three maximal punctures. We also can close a minimal puncture in four different ways and obtain a sphere with one maximal and one minimal punctures.

\

\section{Fun with tori}\label{app:tor}

\begin{figure}[htbp]
\begin{center}
\begin{tabular}{c}
\includegraphics[scale=0.31]{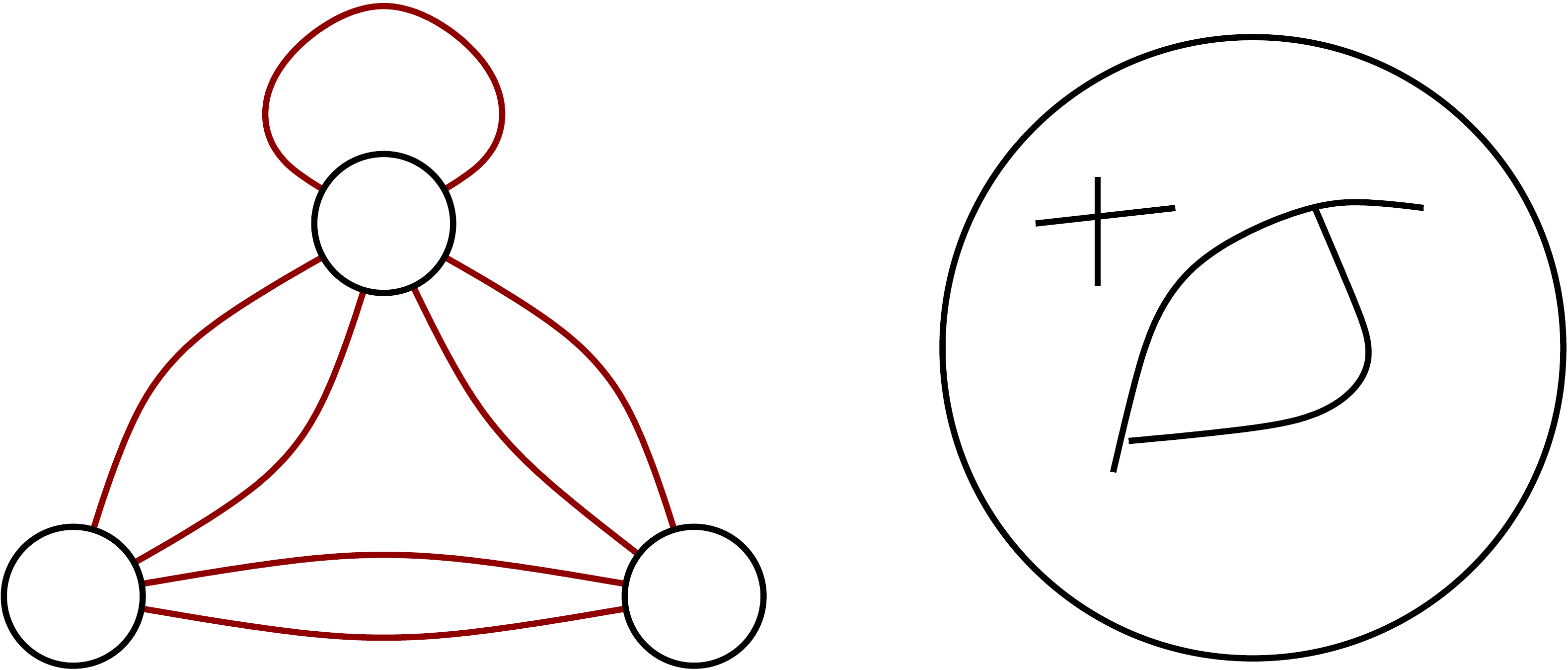}
\end{tabular}
\end{center}
\caption{Torus with one puncture and one of the $\ell_i=\pm1$. The nodes are $SU(2)$ gauge groups.
We have an additional  gauge singlet field coupled to one of the bifundamental chirals connecting two nodes without an adjoint field. 
\label{figtor}}
\end{figure}

\begin{figure}[htbp]
\begin{center}
\begin{tabular}{c}
\includegraphics[scale=0.71]{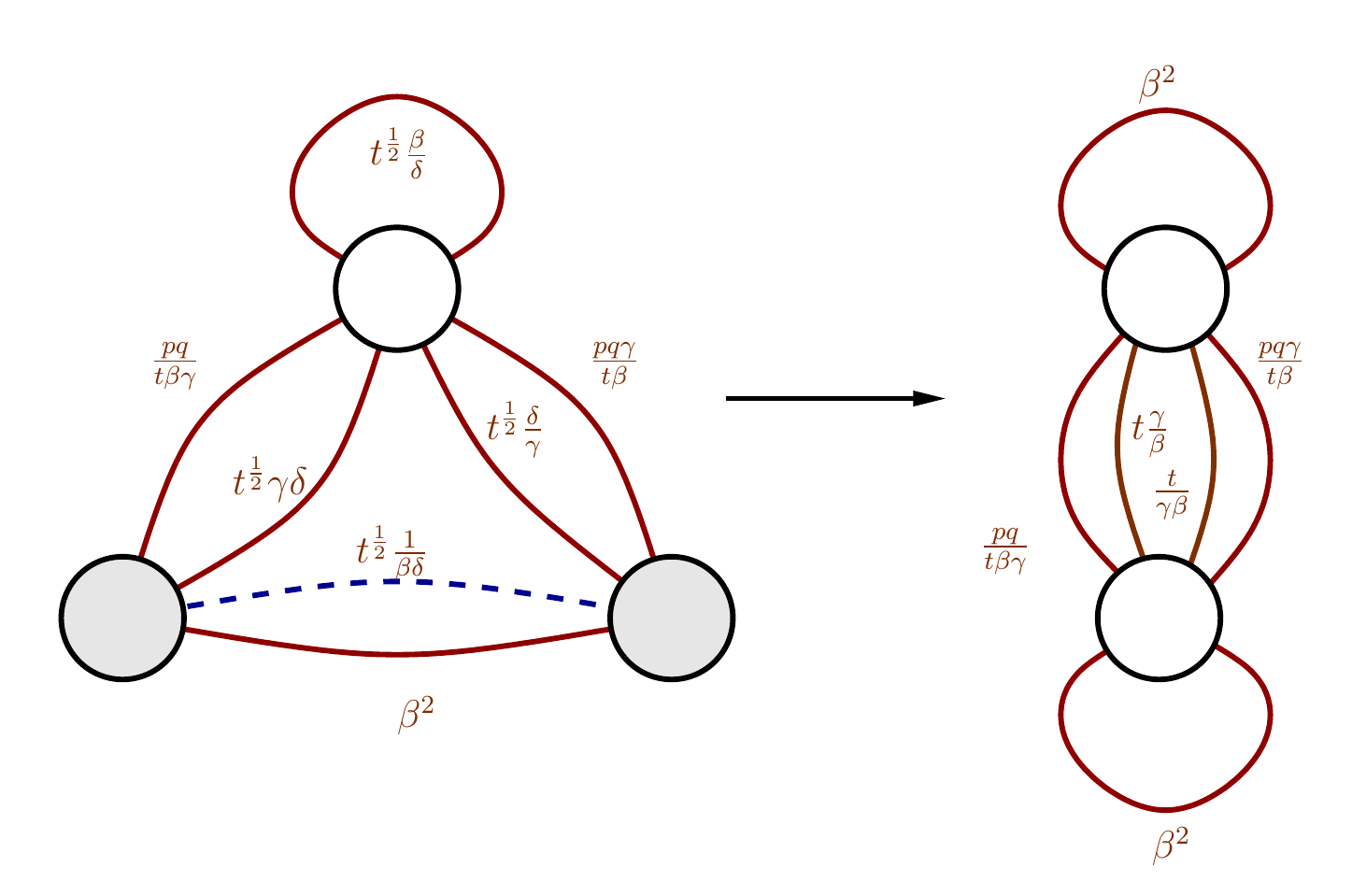}
\end{tabular}
\end{center}
\caption{We turn on a vacuum expectation value for the bifundamental denoted by dashed blue line, {\it i.e.} in the index computing the pole at $\delta=t^{\frac12}\b^{-1}$. The two gauge groups connected by these quarks are Higgsed to a diagonal combination.  The quarks with the vacuum expectation value do not couple through superpotential and thus do not generate mass terms. The resulting theory in the IR is depicted on the right. It is ${\cal S}_1$ theory of type $A_1$ corresponding to torus with two punctures. The three class ${\cal S}_2$ $U(1)_\b\times U(1)_\g\times U(1)_t$ symmetries map to the $U(1)_t$ symmetry of class ${\cal S}_1$ as well to the two $U(1)$ symmetries corresponding to class ${\cal S}_1$  punctures. 
\label{figtortosone}}
\end{figure}

Let us discuss here several simple examples of theories corresponding to torus with one minimal puncture or no puncture in the $k=2$ $A_1$ case.
Torus with one minimal puncture  can be obtained for example by gluing together the two maximal punctures of the trinion with two maximal punctures of same color and a minimal puncture 
with one of the discrete charges, $\ell_i$, being $\pm1$; the theory we obtained by closing a minimal puncture in our core example of an interacting theory. Since here we
have a Lagrangian, depicted in Figure~\ref{figtor}, we can observe that the theory has one-dimensional conformal manifold.  Giving a vev to one of the chiral fields as depicted in Figure ~\ref{figtortosone} we Higgs two of the gauge groups and obtain a theory which we can associate to torus with no punctures but with two units of one of the discrete charges in class ${\cal S}_2$. This theory has an alternative interpretation as  $A_1$ theory of class ${\cal S}_1$ corresponding to torus with two  punctures (with  additional singlet fields flipping the quadratic operators built from the two adjoints). This theory is also know as $Y_{p=1,q=1}$ (modulo the extra superpotential terms) in the $Y_{p,q}$  nomenclature \cite{Benvenuti:2004dy}. 

\begin{figure}[htbp]
\begin{center}
\begin{tabular}{c}
\includegraphics[scale=0.71]{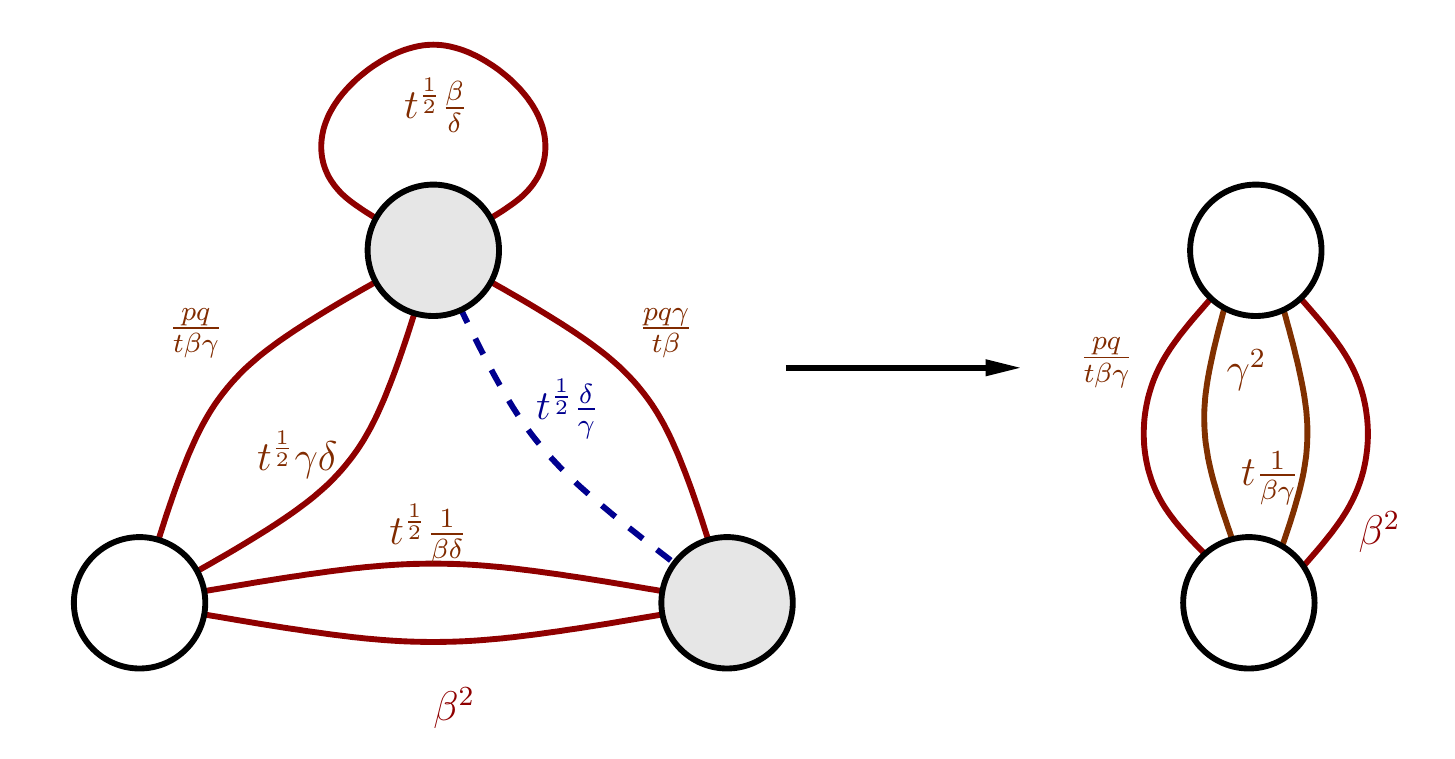}
\end{tabular}
\end{center}
\caption{Giving a vacuum expectation value to bifundamental  denoted by dashed blue line, {\it i.e.} in the index computing the pole at $\delta=t^{-\frac12}\gamma$, the two gauge groups connected by these quarks are Higgsed to a diagonal combination.  The quarks with the vacuum expectation value do couple through superpotential and thus generate mass terms for two other quarks. The resulting theory in the IR is depicted on the right. It is the conifold, $T_{1,1}$, theory. We have a quartic superpotential involving all the bifundamentals. Moreover there are two additional superpotential terms flipping the mesons built from $\b^2$ and $\g^2$ fields. 
\label{figtoroneone}}
\end{figure}

Another two equivalent ways to obtain a torus with no punctures come from giving vacuum expectation
value to bifundamental built from quarks connecting one of the other pairs of gauge groups.
This theory has a unit of both $\b$ and $\g$ discrete charges.
The theory so obtained is the $T_{1,1}$ theory or $Y_{p=1,q=0}$ theory (again with extra singlets and superpotential), $SU(2)^2$ gauge theory with four bifundamental chirals and a superpotential. See figure~\ref{figtoroneone}. 

\begin{figure}[htbp]
\begin{center}
\begin{tabular}{c}
\includegraphics[scale=0.71]{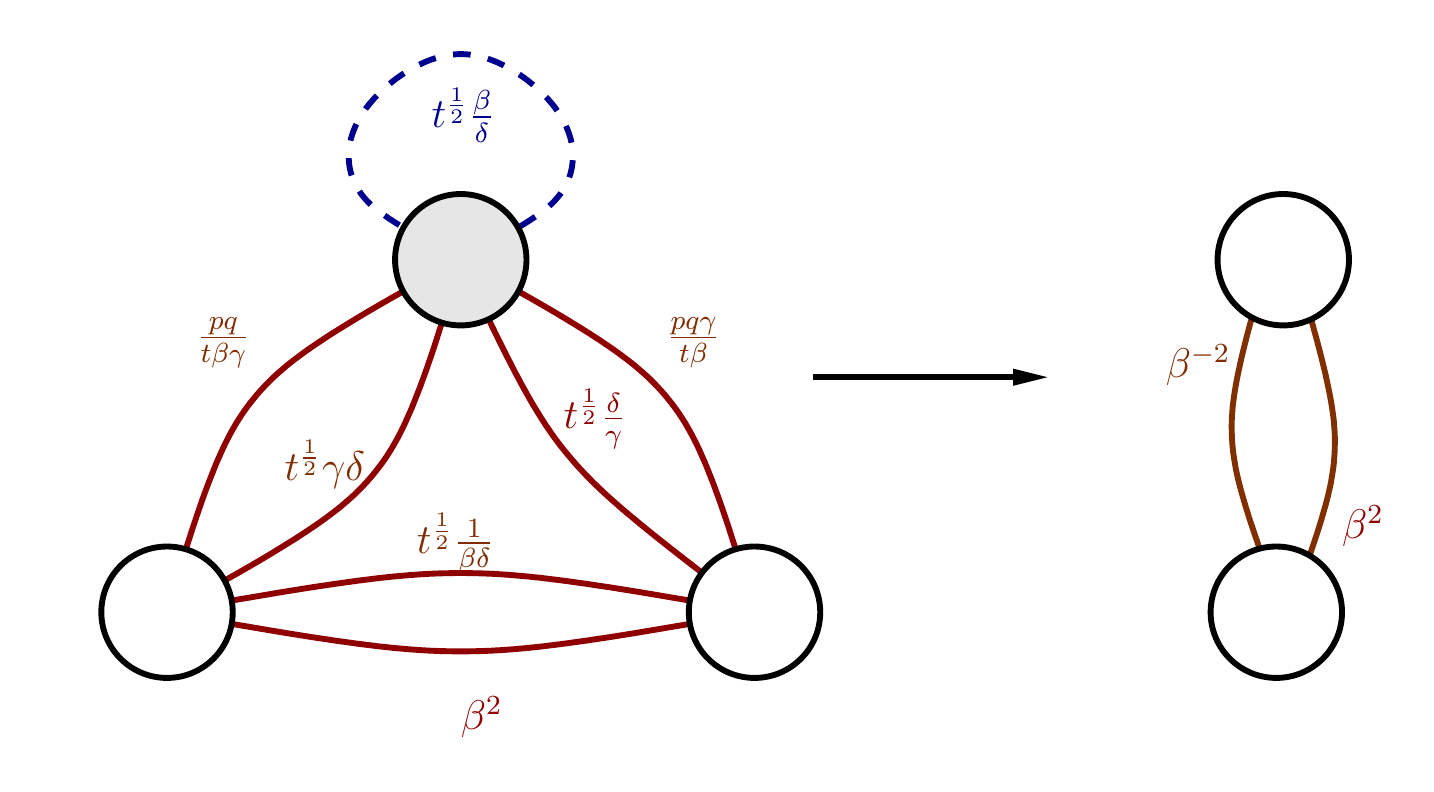}
\end{tabular}
\end{center}
\caption{Giving a vacuum expectation value to the singlet built from the adjoint chiral denoted by dashed blue line, {\it i.e.} in the index computing the pole at $\delta=t^{\frac12}\beta$, the corresponding gauge group is Higgsed and most of the chiral fields acquire mass. 
\label{figsing}}
\end{figure}

Finally we can give a vacuum expectation value to the quadratic singlet built from the adjoint field. This amounts to setting $\delta=t^{\frac12}\beta$.  This vacuum expectation value Higgses the corresponding group and gives masses to four of the quarks. The remaining theory is the $SU(2)^2$ with two bi-fundamental flavors and two additional singlets and is associated to torus with no punctures and no discrete charges. See figure~\ref{figsing}.

\

\section{Class ${\cal S}_2$ interpretation of a selfdual $SU(4)$ SYM with tensor matter}\label{app:tensant}

Let us take our basic interacting four punctured sphere of $A_1$ class ${\cal S}_2$ and perform
Seiberg duality on one of the gauge nodes. We have discussed this duality in section \ref{sect:basic} and 
here let us writing the index in the Seiberg dual frame,

\be\label{seibdualindex}
&&{\cal I}_{{\mathbf u}\delta{\mathbf v}\alpha}=
\frac{\left[(q;q)(p,p)\right]^4}{4!}\Gamma_e(\frac{t\beta}\gamma u_1^{\pm1}u_2^{\pm1})
\Gamma_e(\frac{t\gamma}\beta v_1^{\pm1}v_2^{\pm1})
\Gamma_e(t\alpha\delta u_2^{\pm1}v_1^{\pm1})
\Gamma_e(\frac{t}{\alpha\delta}v_2^{\pm1}u_1^{\pm1})
\Gamma_e(\frac{p^2q^2}{t^2})^2\nonumber\\
&&\oint \prod_{i=1}^3\frac{d{\frak z}_i}{2\pi i {\frak z}_i}
\frac{\prod_{i=1}^4\Gamma_e(\sqrt\frac{pq}{t}\sqrt{\frac{1}{\alpha\delta\beta^2}}u_2^{\pm1}{\frak z}_i)
\Gamma_e(\sqrt\frac{pq}{t}\sqrt{\frac{\alpha\delta}{\gamma^2}}v_2^{\pm1}{\frak z}_i)
\Gamma_e(\sqrt\frac{pq}{t}\sqrt{{\alpha\delta\gamma^2}}u_1^{\pm1}{\frak z}^{-1}_i)
\Gamma_e(\sqrt\frac{pq}{t}\sqrt{\frac{\beta^2}{\alpha\delta}}v_1^{\pm1}{\frak z}^{-1}_i)}{\prod_{i\neq j}\Gamma_e(\frak z_i/\frak z_j)}\nonumber\\
&&\oint \frac{dz_2}{4\pi i z_2}
\frac{\Gamma_e(\frac{p^2q^2}{t^2}z_2^{\pm2})\prod_{i=1}^4\Gamma_e(\sqrt{\frac{t^2}{pq}}\sqrt{\frac{\delta}{\alpha}}z_2^{\pm1}{\frak z}_i)\Gamma_e(\sqrt{\frac{t^2}{pq}}\sqrt{\frac{\alpha}{\delta}}z_2^{\pm1}{{\frak z}_i}^{-1})}{\Gamma_e(z_2^{\pm2})}\,.
\ee The first line has the singlet mesonic operators appearing after the duality transformation and
the last line is the ${\cal N}=2$ block. Note that we cannot here take Macdonald limit of the index  of the ingredients.  We can perform the S-duality transformation on the ${\cal N}=2$ block exchanging $\alpha$ with $\delta$ which will give us the exchange of the two minimal punctures which we discussed in section \ref{sect:basic}. However,
 we also can act with the other elements of the S-triality. Denoting,

\be\label{abcd}
a=\sqrt{\frac{\frak z_1}{\frak z_2}},\quad
b=\sqrt{\frac\delta\alpha \frak z_1\frak z_2},\quad
c={\frak z_3}\sqrt{{\frak z_1}{\frak z}_2},\quad
d=\sqrt{\frac\alpha\delta {\frak z_1}{\frak z_2}},\quad
\ee the first duality above would exchange $b$ with $d$ but now we will exchange $c$ with $b$ to obtain

\be\label{ASind}
&&{\cal I}_{{\mathbf u}\delta{\mathbf v}\alpha}=
\frac{\left[(q;q)(p,p)\right]^4}{4!}\Gamma_e(\frac{t\beta}\gamma u_1^{\pm1}u_2^{\pm1})
\Gamma_e(\frac{t\gamma}\beta v_1^{\pm1}v_2^{\pm1})
\Gamma_e(t\alpha\delta u_2^{\pm1}v_1^{\pm1})
\Gamma_e(\frac{t}{\alpha\delta}v_2^{\pm1}u_1^{\pm1})
\Gamma_e(\frac{p^2q^2}{t^2})^2\nonumber\\
&&\oint \frac{dz_2}{4\pi i z_2}
\frac{\Gamma_e(\frac{p^2q^2}{t^2}z_2^{\pm2})}{\Gamma_e(z_2^{\pm2})}
\Gamma_e(\sqrt{\frac{t^2}{pq}}z_2^{\pm1}\left(\frac\alpha\delta\right)^{\pm1})\;
\oint \prod_{i=1}^3\frac{d{\frak z}_i}{2\pi i {\frak z}_i}\frac{\prod_{i<j<4}\Gamma_e(\sqrt{\frac{t^2}{pq}}({\frak z}_i{\frak z}_j)^{\pm1}z_2^{\pm1})}{\prod_{i\neq j}\Gamma_e(\frak z_i/\frak z_j)}
\,\\
&&\prod_{i=1}^4\Gamma_e(\sqrt\frac{pq}{t}\sqrt{\frac{1}{\alpha\delta\beta^2}}u_2^{\pm1}{\frak z}_i)
\Gamma_e(\sqrt\frac{pq}{t}\sqrt{\frac{\alpha\delta}{\gamma^2}}v_2^{\pm1}{\frak z}_i)
\Gamma_e(\sqrt\frac{pq}{t}\sqrt{{\alpha\delta\gamma^2}}u_1^{\pm1}{\frak z}^{-1}_i)
\Gamma_e(\sqrt\frac{pq}{t}\sqrt{\frac{\beta^2}{\alpha\delta}}v_1^{\pm1}{\frak z}^{-1}_i)\,.\nonumber
\ee We see that we have here the ${\cal N}=2$ block coupled to ${\cal N}=1$ SQCD with four flavors
in fundamental representation of $SU(4)$ and $2$ flavors in the antisymmetric representation.
The symmetry under exchanging the two minimal punctures, $\alpha\leftrightarrow\delta$, is 
manifest. 
Let us denote the  ${\cal N}=1$ theory with the antisymmetric matter by ${\cal T}_A$. The index of this theory is,

\be\label{Ta}
&&{\cal I}_{{\cal T}_A} = 
\frac{\left[(q;q)(p,p)\right]^3}{4!}\Gamma_e(\frac{t\beta}\gamma u_1^{\pm1}u_2^{\pm1})
\Gamma_e(\frac{t\gamma}\beta v_1^{\pm1}v_2^{\pm1})
\Gamma_e(t\alpha\delta u_2^{\pm1}v_1^{\pm1})
\Gamma_e(\frac{t}{\alpha\delta}v_2^{\pm1}u_1^{\pm1})
\Gamma_e(\frac{p^2q^2}{t^2})\nonumber\\
&&
\Gamma_e(\frac{p^2q^2}{t^2}z_2^{\pm2})\oint \prod_{i=1}^3\frac{d{\frak z}_i}{2\pi i {\frak z}_i}\frac{\prod_{i<j<4}\Gamma_e(\sqrt{\frac{t^2}{pq}}({\frak z}_i{\frak z}_j)^{\pm1}z_2^{\pm1})}{\prod_{i\neq j}\Gamma_e(\frak z_i/\frak z_j)}
\,\\
&&\prod_{i=1}^4\Gamma_e(\sqrt\frac{pq}{t}\sqrt{\frac{1}{\alpha\delta\beta^2}}u_2^{\pm1}{\frak z}_i)
\Gamma_e(\sqrt\frac{pq}{t}\sqrt{\frac{\alpha\delta}{\gamma^2}}v_2^{\pm1}{\frak z}_i)
\Gamma_e(\sqrt\frac{pq}{t}\sqrt{{\alpha\delta\gamma^2}}u_1^{\pm1}{\frak z}^{-1}_i)
\Gamma_e(\sqrt\frac{pq}{t}\sqrt{\frac{\beta^2}{\alpha\delta}}v_1^{\pm1}{\frak z}^{-1}_i)\,.\nonumber
\ee
 The theory ${\cal T}_A$ is actually dual to itself under Seiberg duality \cite{csakiterningshmaltzskiba}.
The index of this duality was discussed in~\cite{Spiridonov:2009za}.
The global non-R symmetry of the theory is as follows,

\be\label{symmantiAS}
(SU(2))^2_{\mathbf u}\times (SU(2))^2_{\mathbf v}\times SU(2)_{z_2}\times U(1)_{\alpha\delta}
\times U(1)_t\times U(1)_\beta\times U(1)_\gamma\,.
\ee The group $SU(2)_{z_2}$ rotates the two quarks in the antisymmetric representation of $SU(4)$
gauge group. Under Seiberg duality we have the following map of the symmetries,

\be\label{seibmapta}
\a\delta \to \a^{-1}\delta^{-1}\,,\qquad 
\b \to \g\,,\qquad  \g\to\b\,.
\ee Alternatively Seiberg duality can be thought of as exchanging $(SU(2))^2_{\mathbf u}$ and $(SU(2))^2_{\mathbf v}$. We would like to understand this theory in terms of class ${\cal S}_2$.
The two symmetries ${\mathbf u}$ and ${\mathbf v}$ live to two maximal punctures of the same color and the symmetries $z_2$ and $\a\delta$ will need to be interpreted.
We can write the index using the eigenfunctions as 

\be\label{eigta}
{\cal I}_{{\cal T}_A} = \sum_\lambda \Upsilon_\lambda(\a\delta,z_2) \psi_\lambda({\mathbf u}) \psi_\lambda({\mathbf v})\,,
\ee
where this equation can be viewed as definition of $\Upsilon_\lambda$. We will derive the relation of $\Upsilon_\lambda$ to the eigenfunctions soon. The fact that \eqref{eigta} can be written in terms of single sum over eigenfunction is a manifestation of S-duality or in this case the Seiberg duality of \cite{csakiterningshmaltzskiba}. We have the following relation,

\be
&&{\cal I}_{\a{\mathbf u}\delta{\mathbf v}} = \sum_\lambda \phi_\lambda(\a)
\phi_\lambda(\delta) \psi_\lambda({\mathbf u})\psi_\lambda({\mathbf v})=\\
&&(p;p)(q;q)
\oint \frac{dz_2}{4\pi i z_2}
\frac1{\Gamma_e(z_2^{\pm2})}
\Gamma_e(\sqrt{\frac{t^2}{pq}}z_2^{\pm1}\left(\frac\alpha\delta\right)^{\pm1}) {\cal I}_{{\cal T}_A} =\nonumber\\
&&\sum_\lambda \psi_\lambda({\mathbf u})\psi_\lambda({\mathbf v})
(p;p)(q;q)
\oint \frac{dz_2}{4\pi i z_2}
\frac{\Gamma_e(\frac{p^2q^2}{t^2})}{\Gamma_e(z_2^{\pm2})}
\Gamma_e(\sqrt{\frac{t^2}{pq}}z_2^{\pm1}\left(\frac\alpha\delta\right)^{\pm1}) \Upsilon_\lambda(z_2,\a\delta)\,.\nonumber
\ee From here we deduce that

\be\label{ccupsilon}
\phi_\lambda(\a)
\phi_\lambda(\delta) =
(p;p)(q;q)
\oint \frac{dz}{4\pi i z}
\frac{\Gamma_e(\frac{p^2q^2}{t^2})}{\Gamma_e(z^{\pm2})}
\Gamma_e(\sqrt{\frac{t^2}{pq}}z^{\pm1}\left(\frac\alpha\delta\right)^{\pm1}) \Upsilon_\lambda(z,\a\delta)\,.\nonumber\\
\ee We already have encountered a similar relation while closing maximal punctures~\eqref{respsi}, 

\be\label{respsiasglory}
&&\phi_\lambda(\a)
\phi_\lambda(\delta) =
C_\lambda \sqrt{C^{(\b,+)}_\lambda C^{(\g,-)}_\lambda} (p;p)(q;q) \Gamma_e\biggl( t\frac\g\b\left(\frac\a\delta\right)^{\pm1} (\a\delta\b\g)^{\pm1}\biggr)\\
&&\qquad\quad
\oint \frac{dz}{4\pi i z}
\frac{\Gamma_e(\frac{pq}{t}(\a\b\g\delta)^{\pm1}z^{\pm1})}{\Gamma_e(z^{\pm2})}
\Gamma_e\biggl(\frac\b\g z^{\pm1}\left(\frac\alpha\delta\right)^{\pm1}\biggr)\;\,\, 
\widetilde \psi_\lambda(z,\a\delta)\,.\nonumber
\ee We thus use the elliptic Fourier transform to write

\be
&&\Upsilon_\lambda(w,\a\delta)=
C_\lambda \sqrt{C^{(\b,+)}_\lambda C^{(\g,-)}_\lambda}  (p;p)^2(q;q)^2
\oint \frac{dz}{4\pi i z}
\frac{\Gamma_e(\frac{pq}{t}(\a\b\g\delta)^{\pm1}z^{\pm1})}{\Gamma_e(\frac{pq}{t^2})\Gamma_e(z^{\pm2})}
\widetilde \psi_\lambda(z,\a\delta)\times\nonumber\\
&&\;\oint_{\cal C} \frac{d\xi}{4\pi i \xi} \frac1{\Gamma_e(\xi^{\pm2})}
\Gamma_e\biggl( t\frac\g\b\xi^{\pm1} (\a\delta\b\g)^{\pm1}\biggr)
\Gamma_e\biggl(\frac\b\g z^{\pm1}\xi^{\pm1}\biggr)
\Gamma_e\biggl(\sqrt{\frac{pq}{t^2}}w^{\pm1}\xi^{\pm1}\biggr)\,.
\ee 
On the second line we have the index of $SU(2)$ SYM with three flavors which can be exactly evaluated 
in terms of the dual mesons,

\be
&&(q;q)(p;p)\oint_{\cal C} \frac{d\xi}{4\pi i \xi} \frac1{\Gamma_e(\xi^{\pm2})}
\Gamma_e\biggl( t\frac\g\b\xi^{\pm1} (\a\delta\b\g)^{\pm1}\biggr)
\Gamma_e\biggl(\frac\b\g z^{\pm1}\xi^{\pm1}\biggr)
\Gamma_e\biggl(\sqrt{\frac{pq}{t^2}}w^{\pm1}\xi^{\pm1}\biggr)=\nonumber\\
&&\Gamma_e\biggl(t^2\frac{\g^2}{\b^2}\biggr)
\Gamma_e\biggl(\frac{\b^2}{\g^2}\biggr)
\Gamma_e\biggl(\frac{pq}{t^2}\biggr)
\Gamma_e\biggl(t (\a\b\g\delta)^{\pm1}z^{\pm1}\biggr)\\
&&\qquad\quad\qquad\Gamma_e\biggl(\sqrt{pq}\frac\g\b(\a\b\g\delta)^{\pm1}w^{\pm1}\biggr)
\Gamma_e\biggl(\sqrt{\frac{pq}{t^2}}\frac\b\g z^{\pm1}w^{\pm1}\biggr)\,.\nonumber
\ee Plugging this back we note that some of the fields become massive and we obtain,

\be\label{upsik}
&&\Upsilon_\lambda(w,\a\delta)=\Gamma_e\biggl(t^2\frac{\g^2}{\b^2}\biggr)
\Gamma_e\biggl(\frac{\b^2}{\g^2}\biggr)
\Gamma_e\biggl(\sqrt{pq}\frac\g\b(\a\b\g\delta)^{\pm1}w^{\pm1}\biggr)\, \nonumber\\
&&\qquad 
C_\lambda \sqrt{C^{(\b,+)}_\lambda C^{(\g,-)}_\lambda} (q;q)(p;p)
\oint \frac{dz}{4\pi i z}
\frac{\Gamma_e\biggl(\sqrt{\frac{pq}{t^2}}\frac\b\g z^{\pm1}w^{\pm1}\biggr)}{\Gamma_e(z^{\pm2})}
\widetilde \psi_\lambda(z,\a\delta)\,.
\ee Writing the index of ${\cal T}_A$ using this expression we finally obtain,

\be\label{tasphere}
&&{\cal I}_{{\cal T}_A}=
\Gamma_e\biggl(t^2\frac{\g^2}{\b^2}\biggr)
\Gamma_e\biggl(\frac{\b^2}{\g^2}\biggr)
\Gamma_e\biggl(\sqrt{pq}\frac\g\b(\a\b\g\delta)^{\pm1}w^{\pm1}\biggr)\, \\
&&\qquad \quad 
 (q;q)(p;p)
\oint \frac{dz}{4\pi i z}
\frac{\Gamma_e\biggl(\sqrt{\frac{pq}{t^2}}\frac\b\g z^{\pm1}w^{\pm1}\biggr)}{\Gamma_e(z^{\pm2})}
\left[\sum_\lambda  C_\lambda \sqrt{C^{(\b,+)}_\lambda C^{(\g,-)}_\lambda} \widetilde \psi_\lambda(z,\a\delta)\psi_\lambda({\mathbf u})\psi_\lambda({\mathbf v})\right]\,=\nonumber\\
&&\Gamma_e\biggl(t^2\frac{\g^2}{\b^2}\biggr)
\Gamma_e\biggl(\sqrt{pq}\frac\g\b(\a\b\g\delta)^{\pm1}w^{\pm1}\biggr) (q;q)(p;p)
\oint \frac{dz}{4\pi i z}
\frac{\Gamma_e\biggl(\sqrt{\frac{pq}{t^2}}\frac\b\g z^{\pm1}w^{\pm1}\biggr)}{\Gamma_e(z^{\pm2})}{{\cal I}^{(z,\a\delta)}}_{{\mathbf u}{\mathbf v}}\,.\nonumber
\ee The index in the square brackets is that of the theory corresponding to a sphere with three maximal punctures of two different colors and appropriate discrete charges, $\Gamma_e(pq \frac{\g^2}{\b^2}){{\cal I}^{(z,\a\delta)}}_{{\mathbf u}{\mathbf v}}$, we encountered in section \ref{sect:closemaxi:index}. We thus deduce that the $SU(4)$ SYM with four quark flavors and two antisymmetric tensors is equivalent to certain $SU(2)$ gauging with extra chiral fields of the strongly-coupled trinion with three maximal punctures. We have a singlet field $M_0$ with fugacities $t^2\frac{\g^2}{\b^2}$ and R-charge zero, fields $q$ with fugacities
$\frac\g\b(\a\b\g\delta)^{\pm1}w^{\pm1}$ and R-charge one, and fundamental quarks $Q$ with R-charge one and fugacities $t^{-1}\frac\b\g w^{\pm1}$. Moreover the SCFT has mesons $M$ associated to the maximal puncture we partially gauge. This mesons are in fundamental of the gauge $SU(2)$ and have fugacities $t (\g\b)^{\pm1}(\a\delta)^{\pm1}$ with R-charge zero. The superpotential we turn on is thus,

\be
M_0 Q Q+ qMQ\,.
\ee  Note also that the gauging is non anomalous with all the symmetries at hand.

We can now go back further to the four-punctured sphere. The duality frame \eqref{ASind} is then

\be\label{foursphagain}
&&{\cal I}_{{\mathbf u}\a{\mathbf v}\delta}=
\Gamma_e\biggl(t^2\frac{\g^2}{\b^2}\biggr)\Gamma_e(\frac{p^2q^2}{t^2})
(q;q)(p;p)\oint \frac{dz}{4\pi i z}\frac1{\Gamma_e(z^{\pm2})}
{{\cal I}^{(z,\a\delta)}}_{{\mathbf u}{\mathbf v}}\times\\
&&(q;q)(p;p)\oint \frac{dw}{4\pi i w} \frac1{\Gamma_e(w^{\pm2})}
\Gamma_e\biggl(\sqrt{pq}\frac\g\b(\a\b\g\delta)^{\pm1}w^{\pm1}\biggr)
\Gamma_e\biggl(\sqrt{\frac{pq}{t^2}}\frac\b\g z^{\pm1}w^{\pm1}\biggr)
\Gamma_e\biggl(\sqrt{\frac{t^2}{pq}}\left(\frac\a\delta\right)^{\pm1}w^{\pm1}\biggr)\,.\nonumber
\ee On the second line we have again $SU(2)$ SYM with three flavors index of which can be evaluated to be,

\be
&&(q;q)(p;p)\oint \frac{dw}{4\pi i w} \frac1{\Gamma_e(w^{\pm2})}
\Gamma_e\biggl(\sqrt{pq}\frac\g\b(\a\b\g\delta)^{\pm1}w^{\pm1}\biggr)
\Gamma_e\biggl(\sqrt{\frac{pq}{t^2}}\frac\b\g z^{\pm1}w^{\pm1}\biggr)
\Gamma_e\biggl(\sqrt{\frac{t^2}{pq}}\left(\frac\a\delta\right)^{\pm1}w^{\pm1}\biggr)=\nonumber\\
&&\Gamma_e\biggl(pq\frac{\g^2}{\b^2}\biggr)
\Gamma_e\biggl(\frac{pq}{t^2}\frac{\b^2}{\g^2}\biggr)
\Gamma_e\biggl(\frac{t^2}{pq}\biggr)
\Gamma_e\biggl(\frac{pq}t (\a\b\g\delta)^{\pm1}z^{\pm1}\biggr)\times\\
&&\qquad\qquad\quad\Gamma_e\biggl(t\frac\g\b \biggl(\frac\a\delta\biggr)^{\pm1}(\a\b\g\delta)^{\pm1}\biggr)
\Gamma_e\biggl(\frac\b\g\biggl(\frac\a\delta\biggr)^{\pm1}z^{\pm1}\biggr)\,.\nonumber
\ee Plugging this back to \eqref{foursphagain} we obtain 

\be
&&{\cal I}_{{\mathbf u}\a{\mathbf v}\delta}=(q;q)(p;p)\Gamma_e\biggl(pq\frac{\g^2}{\b^2}\biggr)
\Gamma_e\biggl(t\frac\g\b \biggl(\frac\a\delta\biggr)^{\pm1}(\a\b\g\delta)^{\pm1}\biggr)\\
&&\qquad\quad\,
\oint \frac{dz}{4\pi i z} \frac1{\Gamma_e(z^{\pm2})}
\Gamma_e\biggl(\frac{pq}t (\a\b\g\delta)^{\pm1}z^{\pm1}\biggr)
\Gamma_e\biggl(\frac\b\g\biggl(\frac\a\delta\biggr)^{\pm1}z^{\pm1}\biggr)
{{\cal I}^{(z,\a\delta)}}_{{\mathbf u}{\mathbf v}}\,.\nonumber
\ee	
 This is the Argyres-Seiberg frame for the four-punctured sphere obtained previously in \eqref{asframe}.

\

\section{Comments on index of theories with $U(1)_t$ discrete charge}\label{app:indext}

Let us outline here how to compute the index of theories with $U(1)_t$ discrete charges.
The index for $k=1$ case was thoroughly discussed in \cite{Beem:2012yn} (see also \cite{Gadde:2013fma,Agarwal:2013uga,Agarwal:2014rua}). The higher $k$ 
follows similar pattern but there are some new features which we will discuss here.
We restrict the explicit discussion as usual to $A_1$ and class ${\cal S}_2$. 

First, as we discussed in section \ref{sect:chargeuonet}, we have to introduce a second set of our theories with charges under the global symmetries aproppriately  flipped and R-charges shifted. 
This means that the indices of the second copy of the theories are built using the eigenfunctions,

\be
\hat \psi_\lambda({\mathbf v};\b,\g,t) \equiv \psi_\lambda({\mathbf v}^\dagger; \b^{-1},\g^{-1},\frac{p\,q}t)\,.
\ee Note that $\hat \psi_\lambda({\mathbf v};\b,\g,t)$ are eigenfunctions of 

\be\label{fopaaa}
{\hat{\frak S}}^{(\b,-)}_{(0,1)}\cdot f(v_1,v_2)=\sum_{a,b=\pm1} \hat {\cal T}(v_1^a,v_2^b;\b,\g,t) f(q^{\frac{a}2} v_1,q^{\frac{b}2} v_2)\,,
\ee where

\be\label{difopppaaa}
\hat {\cal T}(v_1,v_2;\b,\g,t)={\cal T}(v_2,v_1;\b^{-1},\g^{-1},\frac{pq}t)=
\frac{\theta(t v_1 v_2(\frac\g\b)^{\pm1};p)\theta(\frac{t\b}{q\g} \frac{v_1}{  v_2};p)\theta(\frac{t\b^3\g}{q} \frac{v_2}{v_1};p)}{\theta(v_1^{2};p)\theta(v_2^{2};p)}\,.
\ee  
We can derive yet another relation between eigenfunctions. Let us glue as in section \ref{sect:surface}
an interacting trinion to a general theory but now glue it to the puncture of the opposite type, {\it i.e.} glue $\psi$ puncture to $\hat \psi$ one without extra fields but with a quartic superpotential. Next, we close the  minimal puncture residing on the interacting trinion. Assuming dualities to hold, {\it i.e.} that we can freely move around on the Riemann surface different punctures and consider various pair of pants decompositions leading to the same theory, we deduce that $\hat \psi$ and $\psi$ have to be orthonormal under the measure involving only the ${\cal N}=1$ vector multiplets. These two facts translate into the following relation between eigenfunctions,

\be
\hat \psi_\lambda({\mathbf u})= \Gamma_e(\frac{pq}t \biggl(\frac\b\g\biggr)^{\pm1}u_1^{\pm1}u_2^{\pm1}) \psi_\lambda({\mathbf u})\,.
\ee 
Note that this means that $\hat\psi$ are eigenfunctions of the following operator,

\be
{{\widetilde{\frak S}}}^{(\b,-)}_{(0,1)}=\Gamma_e(\frac{pq}t \biggl(\frac\b\g\biggr)^{\pm1}u_1^{\pm1}u_2^{\pm1})
\;\left[{{\frak S}}^{(\b,-)}_{(0,1)}\right]\;
\Gamma_e(t \biggl(\frac\b\g\biggr)^{\pm1}u_1^{\pm1}u_2^{\pm1})\,.
\ee We can compute this operator to be

\be\label{fopbb}
{{\widetilde{\frak S}}}^{(\b,-)}_{(0,1)}\cdot f(v_1,v_2)=\sum_{a,b=\pm1} {\widetilde {\cal T}}(v_1^a,v_2^b;\b,\g,t) f(q^{\frac{a}2} v_1,q^{\frac{b}2} v_2)\,,
\ee and 

\be\label{difopppbb}
{\widetilde {\cal T}}(v_1,v_2;\b,\g,t)=
\frac{\theta(t v_1 v_2(\frac\g\b)^{\pm1};p)\theta(\frac{t\b}{\g} \frac{v_1}{  v_2};p)\theta(t\b^3\g\frac{v_2}{v_1};p)}{\theta(v_1^{2};p)\theta(v_2^{2};p)}\,.
\ee   Note also that the operators ${{\widetilde{\frak S}}}^{(\b,-)}_{(0,1)}$ and ${{\hat{\frak S}}}^{(\b,-)}_{(0,1)}$
are not the same and an analogue of the former we did not encounter till now for theories without discrete charges for $U(1)_t$. The functions $\hat \psi$ have to be eigenfunctions of both operators
if our assumptions are correct and indeed the two operators are selfadjoint under the same measure and commute,

\be
\left[{\widetilde{\frak S}}^{(\b,-)}_{(0,1)},\,{{\hat{\frak S}}}^{(\b,-)}_{(0,1)}\right]=0\,.
\ee In the $k=1$ case discussed in \cite{Beem:2012yn} the two types of difference operators
turn out to coincide but for higher $k$ one derives two different commuting operators. 

\

\bibliography{sdualityMAC}

\bibliographystyle{JHEP}

\end{document}